\documentstyle[epsf,axodraw]{article}

\newcommand{\gsim}{ \mathop{}_{\textstyle \sim}^{\textstyle >} }
\newcommand{\lsim}{ \mathop{}_{\textstyle \sim}^{\textstyle <} }

\begin{document}
\baselineskip 0.7cm

\begin{titlepage}
\begin{center}

\hfill KEK-TH-602, UT-828\\
\hfill hep-ph/9810479\\
\hfill October, 1998

\vskip 0.35cm
  {\large  
       Solar and Atmospheric Neutrino Oscillations  \\              
                         and   \\
              Lepton Flavor Violation \\
                         in \\
    Supersymmetric Models with Right-handed Neutrinos
   }

\vskip 0.5in 

{\large
    J.~Hisano$^{(a)}$ and
    Daisuke~Nomura$^{(b)}$
 }

\vskip 0.4cm 
$(a)$ {\it Theory Group, KEK, Oho 1-1, Tsukuba, Ibaraki 305-0801, Japan}
\\
$(b)$ {\it Department of Physics, University of Tokyo, Tokyo 113-0033, Japan}
\vskip 1.5cm

\abstract{ Taking the solar and the atmospheric neutrino experiments
into account we discuss the lepton flavor violating processes, such as
$\tau\to\mu\gamma$ or $\mu\to e\gamma$, in the minimal supersymmetric
standard model with right-handed neutrinos (MSSMRN) and the
supersymmetric SU(5) GUT with right-handed neutrinos [SU(5)RN].
The predicted branching ratio of $\mu\to e\gamma$ in the MSSMRN with
the Mikheyev-Smirnov-Wolfenstein (MSW) large angle solution is so large 
that it goes beyond the
current experimental bound if the second-generation right-handed
Majorana mass $M_{\nu_2}$ is greater than $\sim 10^{13}(\sim
10^{14})$GeV for $\tan\beta=30(3)$.  When we take the MSW small angle
solution, the $\mu\to e\gamma$ rate is at most about 1/100 of that of
the MSW large angle solution.  The 'just so' solution implies
$10^{-5}$ of that of the MSW large angle solution. Also, in the SU(5)RN
the large $\mu\to e\gamma$ rate naturally follows from the MSW large
angle solution, and the predicted rate is beyond the current
experimental bound if the typical right-handed Majorana mass $M_N$ is
larger than $\sim 10^{13}(\sim 10^{14})$GeV for $\tan\beta=30(3)$,
similarly to the MSSMRN.  We show the multimass insertion formulas
and their applications to $\tau\to\mu\gamma$ and $\mu\to e\gamma$.  }
\end{center}
\end{titlepage}

\setcounter{footnote}{0}

\section{Introduction}

Introduction of supersymmetry (SUSY) to the Standard Model (SM) is a
solution for the naturalness problem on the radiative correction to
the Higgs boson mass. The minimal supersymmetric standard model
(MSSM) is considered as one of the most promising models beyond the
Standard Model. Nowadays the signal of supersymmetry is being
searched for by many experimental ways.

Lepton flavor conservation, lepton number conservation in each
generation, is an exact symmetry in the SM, however, it may be
violated in the MSSM \cite{EN}.  The SUSY breaking mass terms for sleptons have
to be introduced phenomenologically. Then, the mass eigenstates for
sleptons may be different from those for leptons. This leads to the
lepton flavor violating (LFV) rare processes, such as
$\mu \to e \gamma$, $\tau \to \mu \gamma$, and so on. In
fact, the experimental bounds on them have given a constraint on the
slepton mass matrices.

The structure of the SUSY breaking mass matrices for sleptons depends
on the mechanism to generate the SUSY breaking terms in the MSSM. One
of the interesting mechanisms is the minimal supergravity (SUGRA)
scenario. Similar to the slepton masses, arbitrary SUSY breaking masses for
squarks are also strongly constrained from the FCNC processes, such as
$K^0-\overline{K}^0$ mixing. In the minimal SUGRA scenario, the
SUSY breaking masses for squarks, sleptons, and the Higgs bosons are
expected to be given universally at the tree level, and we can escape from
these phenomenological constraints.

However, the universality of the SUSY breaking masses for the scalar
bosons is not stable for the radiative correction. Especially, if the
physics below the gravitational scale $M_{\rm grav}$ ($\sim 10^{18}$GeV) 
has the LFV interaction, the interaction induces radiatively the LFV SUSY
breaking masses for sleptons.  Then, the LFV rare processes are
sensitive to physics beyond the MSSM \cite{HKR}.

Recently, the Super-Kamiokande experiment has given us a convincing 
result \cite{atm}
that the atmospheric neutrino anomaly \cite{atmanomaly} comes from the neutrino
oscillation. From the zenith-angle dependence of  $\nu_e$ and 
$\nu_{\mu}$ fluxes following neutrino mass square difference and  
mixing angle are expected,
\begin{eqnarray}
&\Delta m^2_{\nu_{\mu} \nu_X} \simeq (10^{-3} - 10^{-2}) 
{\rm eV^2}, &
        \nonumber\\
&\sin^2 2\theta_{\nu_{\mu} \nu_X}  \gsim 0.8.&
\end{eqnarray}
From the negative result for $\nu_e$-$\nu_\mu$ oscillation in the CHOOZ 
experiment \cite{CHOOZ} it is natural to consider 
$\nu_X = \nu_{\tau}$ from above results, 
and the tau neutrino mass is given as
\begin{eqnarray}
m_{\nu_\tau} \simeq (3 \times 10^{-2} - 1 \times 10^{-1}){\rm eV},
\label{taumass}
\end{eqnarray}
provided mass hierarchy $m_{\nu_{\tau}}\gg m_{\nu_{\mu}}$.

The simplest model to generate the neutrino masses is the seesaw
mechanism \cite{seesaw}. The neutrino mass Eq.~(\ref{taumass}) leads
to the right-handed neutrino masses below $\sim$ $(10^{14}-10^{15})$
GeV, even if the Yukawa coupling constant for the tau neutrino mass is
of the order of one. This means that a LFV interaction exists below
the gravitational scale. Then it is expected that the LFV large mixing
for sleptons between the second- and the third-generations is
generated radiatively in the minimal SUGRA scenario, and that the LFV
rare processes may occur with rates accessible by future experiments
\cite{BM,HMTYY}. In fact, the branching ratio of
$\tau\to\mu\gamma$ in the MSSM with the right-handed neutrinos can
reach the present experimental bound \cite{HMTY}.

The solar neutrino deficit \cite{solar} may also come
from neutrino oscillation between $\nu_\mu$-$\nu_e$. 
The Mikheyev-Smirnov-Wolfenstein (MSW) solution \cite{MSW} due 
to the matter effect in the sun is natural for its explanation, 
and the observation favors  
\begin{eqnarray}
&\Delta m^2_{\nu_e \nu_Y} \simeq (8\times10^{-6}-3\times10^{-4})
{\rm eV^2}, &
        \nonumber\\
&\sin^2 2\theta_{\nu_e \nu_Y} \gsim 0.5, &
\end{eqnarray}
or
\begin{eqnarray}
&\Delta m^2_{\nu_{e} \nu_Y} \simeq (4\times10^{-6}-1\times10^{-5})
{\rm eV^2}, &
        \nonumber\\
&\sin^2 2\theta_{\nu_e \nu_Y} \simeq (10^{-3}-10^{-2}) .&
\end{eqnarray}
If the solar neutrino anomaly comes from so-called 'just so' solution
\cite{justso}, neutrino oscillation in vacuum, the following mass
square difference and mixing angle are expected \cite{justso2},
\begin{eqnarray}
&\Delta m^2_{\nu_{e} \nu_Y} \simeq (6\times10^{-11}-1\times10^{-10})
{\rm eV^2}, &
        \nonumber\\
&\sin^2 2\theta_{\nu_{e} \nu_Y}  \gsim 0.5 .&
\end{eqnarray}
It is natural to consider $\nu_Y = \nu_{\mu}$, combined with the
atmospheric neutrino observation. If one of the large angle solutions for the
solar neutrino anomaly is true, the large
mixing $\theta_{\nu_{\mu}\nu_e}$ may imply the LFV large mixing for 
sleptons between the first- and the second-generations.

In this article we investigate the LFV processes in the supersymmetric 
models with the right-handed neutrinos, assuming the minimal SUGRA
scenario. We take the above results for the atmospheric and solar neutrinos.
In section 2 after discussing the origin of the observed mixing angles
we calculate the $\tau\to\mu\gamma$ and $\mu\to e\gamma$ branching ratios
under the assumption of the MSSM with the right-handed neutrinos.
It is argued that the $\mu\to e\gamma$ rate depends on the 
solar neutrino solutions, and that 
especially the MSW large angle solution naturally leads to a large 
$\mu \to e\gamma$ rate. 
In section 3 we consider them in the SU(5) SUSY GUT 
with the right-handed neutrinos. 
Here also it is shown that the large $\mu \to e\gamma$ rate 
naturally follows from the MSW large angle solution.
Section 4 is for our conclusion. 
In Appendix A we give our convention used in this article.
In Appendices B and C we show the multimass insertion formulas
and their applications to $\tau\to\mu\gamma$ and $\mu\to e\gamma$, 
which are useful for estimating the LFV amplitudes and 
understanding the qualitative behavior of the LFV rates.
In Appendix D the renormalization group equations (RGE's) relevant for our 
discussion are given.

\section{The Lepton Flavor Violation in the MSSM with 
Right-handed Neutrinos}

Before starting to investigate the LFV rare processes in the MSSM
with the right-handed neutrinos (MSSMRN),
we discuss the origin of the large mixing of neutrino between
$\nu_\tau$-$\nu_\mu$ or $\nu_\mu$-$\nu_e$. The
MSSMRN is the simplest supersymmetric
model to explain the neutrino masses, and following discussion is
valid to the extension. The superpotential of the Higgs and lepton sector is
given as
\begin{eqnarray}
W_{\rm MSSMRN}&=& f_{\nu_{ij}} H_2 \overline{N}_i L_j 
                 +f_{e_{ij}} H_1 \overline{E}_i L_j 
                 + \frac12 M_{\nu_i\nu_j} \overline{N}_i \overline{N}_j 
                 +\mu H_1 H_2,
\label{yukawacoupling}
\end{eqnarray}
where $L$ is a chiral superfield for the left-handed lepton, and 
$\overline{N}$ and $\overline{E}$ are for the right-handed neutrino 
and the charged lepton.
$H_1$ and $H_2$ are for the Higgs
doublets in the MSSM. Here, $i$ and $j$ are generation indices.  
After redefinition of the fields, the Yukawa coupling constants 
and the Majorana masses 
can be taken as 
\begin{eqnarray}
f_{\nu_{ij}}&=&f_{\nu_i} V_{D ij},\nonumber\\
f_{e_{ij}}&=&f_{e_i}\delta_{ij}, \nonumber\\
M_{\nu_i\nu_j} &=& U_{ik}^{\ast} M_{\nu_{k}} U^\dagger_{kj},
\end{eqnarray}
where $V_D$ and $U$ are unitary matrices. 
In this model the mass matrix for the left-handed neutrinos
$(m_{\nu})$ becomes
\begin{eqnarray}
(m_{\nu})_{ij} &=& 
V_{D ik}^\top (\overline{m}_{\nu})_{kl} V_{D lj},
\end{eqnarray}
where 
\begin{eqnarray}
(\overline{m}_{\nu})_{ij} &=& 
m_{{\nu_i}D} \left[M^{-1}\right]_{ij} m_{{\nu_j}D} 
\nonumber\\
&\equiv& V_{Mik}^\top m_{\nu_k}  V_{Mkj}.
\end{eqnarray}
Here, $m_{{\nu_i}D}=f_{\nu_i}v\sin\beta/\sqrt{2}$
and $V_M$ is a unitary matrix.\footnote{ 
$\langle h_1 \rangle=(v\cos\beta/\sqrt{2},0)^\top$ and
$\langle h_2 \rangle=(0,v\sin\beta/\sqrt{2})^\top$ with $v\simeq 246$GeV.}
We assume
$f_{\nu_3}\gsim  f_{\nu_2}\gsim f_{\nu_1}$, similar to the quark
sector,  
and $m_{\nu_\tau} \gg m_{\nu_\mu} \gg m_{\nu_e}$.
Also, we take the
Yukawa coupling and the Majorana masses for the right-handed neutrinos
real for simplicity. 

When we consider only the tau and the mu neutrino masses,
we parameterize two unitary matrices as
\begin{eqnarray}
V_D= 
\left(
\begin{array}{cc} 
\cos\theta_D& \sin\theta_D \\
-\sin\theta_D& \cos\theta_D
\end{array}
\right),&&
V_M= 
\left(
\begin{array}{cc} 
\cos\theta_M&  \sin\theta_M \\
-\sin\theta_M& \cos\theta_M
\end{array}
\right).
\end{eqnarray}
The observed large angle $\theta_{\nu_{\mu} \nu_\tau}$ is a sum of
$\theta_D$ and $\theta_M$. However, in order to derive large
$\theta_M$ we need to fine-tune the independent Yukawa coupling
constants and the mass parameters.  The neutrino mass matrix
$(\overline{m}_{\nu})$ in the second and the third generations is written
explicitly as
\begin{eqnarray}
(\overline{m}_{\nu}) &=& 
\frac{1}{1-\frac{M_{\nu_2 \nu_3}^2}{M_{\nu_2 \nu_2}M_{\nu_3\nu_3}}}
\left(
\begin{array}{cc}
\frac{m_{\nu_2 D}^2}{M_{\nu_2 \nu_2}}&
-\frac{m_{\nu_2 D}m_{\nu_3 D}}{M_{\nu_2\nu_3}} \frac{M_{\nu_2\nu_3}^2}{M_{\nu_2\nu_2}M_{\nu_3\nu_3}}\\
-\frac{m_{\nu_2 D}m_{\nu_3 D}}{M_{\nu_2\nu_3}} \frac{M_{\nu_2\nu_3}^2}{M_{\nu_2\nu_2}M_{\nu_3\nu_3}}&
\frac{m_{\nu_3 D}^2}{M_{\nu_3\nu_3}} 
\end{array}
\right).
\end{eqnarray}
If the following relations are imposed, 
\begin{equation}
\frac{m_{\nu_3 D}^2}{M_{\nu_3\nu_3}} 
\simeq \frac{m_{\nu_2 D}^2}{M_{\nu_2\nu_2}}
\simeq \frac{m_{\nu_2 D} m_{\nu_3 D}}{M_{\nu_2\nu_3}},
\label{relation}
\end{equation}
the neutrino mass hierarchy
$m_{\nu_\tau} \gg m_{\nu_\mu}$ and $\theta_M\simeq \pi/4$ can be
derived. 
However, it is difficult to explain the relation among the
independent coupling constants and masses without some mechanism or symmetry.
Also, the hierarchy 
$m_{\nu_3 D}\gg m_{\nu_2 D}$ suppresses the mixing angle
$\theta_M$ as
\begin{equation}
\tan 2 \theta_M \simeq
2 \left(\frac{m_{\nu_2 D}}{m_{\nu_3 D}} \right)
  \left(\frac{M_{\nu_2\nu_3}}{M_{\nu_2\nu_2}}     \right),
\end{equation}
as far as the Majorana masses for the right-handed neutrinos do not have
stringent hierarchical structure as Eq.~(\ref{relation}).
Therefore, in the following discussion we assume that the large mixing
angle between $\nu_\tau$ and $\nu_\mu$ comes from
$\theta_D$ and that $U$ is a unit matrix. Similarly, it is natural to 
consider that the large mixing
angle between $\nu_\mu$ and $\nu_e$ in the MSW solution or the 'just
so' solution for the solar neutrino anomaly comes from $V_D$. 

The existence of the large mixing angles in $V_D$ may lead to
radiative generation of sizable LFV masses for the 
sleptons in the minimal SUGRA scenario. Though the SUSY breaking
masses for the left-handed slepton are flavor-independent at tree
level, the Yukawa interaction for the neutrino masses induces radiatively 
the LFV
off-diagonal components in the left-handed slepton mass matrix.

The SUSY breaking terms for the Higgs and lepton sector in the MSSMRN
are in general given as
\begin{eqnarray}
-{\cal L}_{\rm \ SUSY \ breaking}&=& \phantom{+}
 (m_{\tilde L}^2)_{ij} \tilde{l}_{Li}^{\dagger} \tilde{l}_{Lj} 
+(m_{\tilde e}^2)_{ij} \tilde{e}_{Ri}^\ast \tilde{e}_{Rj}  
+(m_{\tilde \nu}^2)_{ij} \tilde{\nu}_{Ri}^\ast \tilde{\nu}_{Rj}  
\nonumber\\
&& +\tilde{m}_{h1}^2 h_1^\dagger h_1  
+\tilde{m}_{h2}^2 h_2^\dagger h_2  
\nonumber \\
& &
 + (A_\nu^{ij} h_2 \tilde{\nu}^\ast_{Ri} \tilde{l}_{Lj}
   +A_e^{ij} h_1 \tilde{e}^\ast_{Ri} \tilde{l}_{Lj} 
   +\frac12 B_{\nu}^{ij}  \tilde{\nu}^\ast_{Ri} \tilde{\nu}^\ast_{Rj} 
   +B_{h} h_1h_2 + h.c.),
\label{MSSMsoft}
\end{eqnarray}
where $\tilde{l}_L$, $\tilde{e}_R$, and $\tilde{\nu}_R$ represent the
left-handed slepton, and the right-handed charged slepton, and the
right-handed sneutrino. Also, $h_1$ and $h_2$ are the doublet Higgs
bosons. In the minimal SUGRA scenario 
at the gravitational scale the SUSY breaking masses
for sleptons, squarks, and the Higgs bosons are universal, and the
SUSY breaking parameters associated with the supersymmetric Yukawa
couplings or masses (A or B parameters)  are proportional to the 
Yukawa coupling constants or masses.
Then, the SUSY breaking parameters in Eq.~(\ref{MSSMsoft}) are given as
\begin{eqnarray}
&(m_{\tilde L}^2)_{ij}=(m_{\tilde e}^2)_{ij}= (m_{\tilde \nu}^2)_{ij}
=\delta_{ij} m_0^2 ,& \nonumber\\ 
&\tilde{m}^2_{h1}=\tilde{m}^2_{h2}=m_0^2,& \nonumber\\ 
&A_\nu^{ij} = f_{\nu_{ij}}a_0,~~~A_e^{ij} = f_{e_{ij}} a_0,& \nonumber\\ 
&B_{\nu}^{ij} = M_{\nu_i \nu_j} b_0,~~~B_{h} = \mu b_0.&
\end{eqnarray}

In order to know the values of the SUSY breaking parameters at the 
low energy, we have to include the radiative corrections to them. We
can  evaluate them by the RGE's.
We present them in Appendix~\ref{section:RGE}, 
and here we discuss only the qualitative
behavior of the solution using the logarithmic approximation.
The SUSY breaking masses of squarks, sleptons, and the  Higgs bosons 
at the low energy are enhanced by gauge interactions, and the 
corrections are flavor-independent and proportional to square of 
the gaugino masses. On the other hand, Yukawa interactions reduce
the SUSY breaking masses. If the Yukawa coupling is LFV, the radiative
correction to the SUSY breaking parameters is LFV.
The LFV off-diagonal components for $(m_{\tilde L}^2)$, $(m_{\tilde e}^2)$, 
and $A_e^{ij}$ are given at the low energy  as
\begin{eqnarray}
(m_{\tilde L}^2)_{ij}&\simeq&
-\frac1{8\pi^2} 
(3m_0^2+a_0^2) 
V_{D ki}^\ast V_{D kj} f_{\nu_k}^2
\log \frac{M_{\rm grav}}{M_{\nu_k}},
\nonumber\\
(m^2_{\tilde e})_{ij} &\simeq&   0,
\nonumber\\
A_e^{ij} &\simeq& 
-\frac{3}{8\pi^2} a_0
f_{e_i} V_{D ki}^\ast V_{D kj} f_{\nu_k}^2 
\log \frac{M_{\rm grav}}{M_{\nu_k}},
\nonumber
\end{eqnarray}
where $i\ne j$. In these equations, the off-diagonal components of
$(m_{\tilde L}^2)$ and $A_e$ are generated radiatively while those of
$(m_{\tilde e}^2)$ are not. This is because the right-handed leptons
have only one kind of the Yukawa interaction $f_e$ and we can 
always take a basis where $f_e$ is diagonal. 
The magnitudes of the off-diagonal
components of $(m_{\tilde L}^2)$ and $A_e$ are sensitive to $f_{\nu_i}$
and $V_D$.\footnote{
If $U$ is not a unit matrix, the off-diagonal components 
for $(m_{\tilde L}^2)$ and $A_e$ become
\begin{eqnarray}
(m_{\tilde L}^2)_{ij}&\simeq&
-\frac1{8\pi^2} 
(3m_0^2+a_0^2) 
V_{D ki}^\ast V_{D lj} f_{\nu_k} f_{\nu_l} U_{km}^\ast U_{lm}
\log \frac{M_{\rm grav}}{M_{\nu_m}},
\nonumber\\
A_e^{ij} &\simeq& 
-\frac{3}{8\pi^2} a_0
f_{e_i} V_{D ki}^\ast V_{D lj} f_{\nu_k} f_{\nu_l} U_{km}^\ast U_{lm}
\log \frac{M_{\rm grav}}{M_{\nu_m}}.
\label{LFVinMSSMRN}
\end{eqnarray}
Then they are insensitive to the detail of $U$ 
since the dependence on ${M_{\nu_i}}$ is logarithmic. }

As shown above, $V_{D 32}$ is expected to be of the order of one 
from the atmospheric neutrino observation. This leads to the
non-vanishing $(m_{\tilde L}^2)_{32}$ and $A_e^{32}$,
which result in a finite $\tau\to\mu\gamma$ decay rate via diagrams
involving them.
The dominant contributions are proportional to 
\begin{eqnarray}
(m_{\tilde L}^2)_{32}&\simeq&
-\frac1{8\pi^2} (3m_0^2+a_0^2) 
  V_{D 33}^\ast V_{D 32} f_{\nu_3}^2  \log \frac{M_{\rm grav}}{M_{\nu_3}} .
\end{eqnarray}
As will be shown, if $f_{\nu_3}$ is of the order of one, 
the branching ratio of
$\tau\to \mu\gamma$ may reach the present experimental bound.

Moreover if $V_{D 31}$ is finite, 
$(m_{\tilde L}^2)_{31}$ and $(m_{\tilde L}^2)_{21}$ are also large.
They are approximately
\begin{eqnarray}
(m_{\tilde L}^2)_{31}&\simeq&
-\frac1{8\pi^2} (3m_0^2+a_0^2) 
  V_{D 33}^\ast V_{D 31} f_{\nu_3}^2  \log \frac{M_{\rm grav}}{M_{\nu_3}} ,
\nonumber \\
(m_{\tilde L}^2)_{21}&\simeq&
-\frac1{8\pi^2} (3m_0^2+a_0^2) 
  V_{D 32}^\ast V_{D 31} f_{\nu_3}^2  \log \frac{M_{\rm grav}}{M_{\nu_3}} .
\end{eqnarray}
This fact implies a sizable $\mu\to e\gamma$ rate
because the amplitudes proportional to 
$(m^2_{\tilde L})_{23}(m^2_{\tilde L})_{31}$ or
$(m^2_{\tilde L})_{21}$ are dominant.
When $V_{D 21}$ is also of the order of one to explain the solar neutrino
anomaly, an extra contribution to $(m_{\tilde L}^2)_{21}$
has to be taken into account as
\begin{equation}
(m_{\tilde L}^2)_{21} \simeq
-\frac1{8\pi^2} (3m_0^2+a_0^2) 
 \left(
  V_{D 32}^\ast V_{D 31} f_{\nu_3}^2  \log \frac{M_{\rm grav}}{M_{\nu_3}} 
+ V_{D 22}^\ast V_{D 21} f_{\nu_2}^2  \log \frac{M_{\rm grav}}{M_{\nu_2}} 
\right) .
\end{equation}
The experimental upper bound on the branching ratio of
$\mu\to e\gamma$ is so severe that the predicted branching ratio may reach it
even if $f_{\nu_2}$ is ${\cal O}(10^{-1})$.

\subsection{The Branching Ratio of $\tau\to\mu\gamma$}

\begin{figure}[p]
\begin{center} 
\begin{picture}(455,140)(30,-20)
\ArrowArcn(135,25)(75,180,135)
\ArrowArc(135,25)(75,90,135)
\ArrowArcn(135,25)(75,90,45)
\ArrowArc(135,25)(75,0,45)
\Vertex(188,78){3}
\Vertex(82,78){3}
\Vertex(135,100){3}
\Text(135,110)[]{$v\sin\beta$}
\Text(55,60)[]{$\tilde{H}_1^-$}
\Text(78,88)[]{$\mu$}
\Text(105,105)[]{$\tilde{H}_2^-$}
\Text(160,80)[]{$\tilde{W}^-$}
\Text(200,90)[]{$M_2$}
\Text(225,60)[]{$\tilde{W}^-$}

\ArrowLine(60,25)(30,25)
\DashArrowLine(60,25)(135,25){3}             \Vertex(135,25){3}
\DashArrowLine(135,25)(210,25){3}        
\ArrowLine(210,25)(240,25)

\Text(45,15)[]{$\tau_R$}
\Text(97,15)[]{$\tilde{\nu}_\tau$}
\Text(172,15)[]{$\tilde{\nu}_\mu$}
\Text(225,15)[]{$\mu_L$}

\Text(135,38)[]{$(m^2_{\tilde{L}})_{32}$}

\Photon(172,110)(195,135){2}{5}
\Text(203,145)[]{$\gamma$}

\Text(135,-10)[]{(a)}

\SetOffset(245,0)
\ArrowArcn(135,25)(75,180,135)
\ArrowArc(135,25)(75,90,135)
\ArrowArcn(135,25)(75,90,45)
\ArrowArc(135,25)(75,0,45)
\Vertex(188,78){3}
\Vertex(82,78){3}
\Vertex(135,100){3}
\Text(135,110)[]{$v\sin\beta$}
\Text(55,60)[]{$\tilde{H}_1^0$}
\Text(78,88)[]{$\mu$}
\Text(105,105)[]{$\tilde{H}_2^0$}
\Text(160,80)[]{$\tilde{W}^0$}
\Text(200,90)[]{$M_2$}
\Text(225,60)[]{$\tilde{W}^0$}

\ArrowLine(60,25)(30,25)
\DashArrowLine(60,25)(135,25){3}             \Vertex(135,25){3}
\DashArrowLine(135,25)(210,25){3}        
\ArrowLine(210,25)(240,25)

\Text(45,15)[]{$\tau_R$}
\Text(97,15)[]{$\tilde{\tau}_L$}
\Text(172,15)[]{$\tilde{\mu}_L$}
\Text(225,15)[]{$\mu_L$}

\Text(135,38)[]{$(m^2_{\tilde{L}})_{32}$}

\Photon(172,110)(195,135){2}{5}
\Text(203,145)[]{$\gamma$}

\Text(135,-10)[]{(b)}

\end{picture} 

\caption{The Feynman diagrams which give
dominant contributions to $\tau^+ \to \mu^+ \gamma$ 
when $\tan\beta\gsim1$ and the off-diagonal elements of 
the right-handed slepton mass matrix are negligible, 
as in the MSSM with the right-handed neutrinos.
In the diagrams, $(m^2_{\tilde{L}})_{32}$ is the $(3,2)$ element of 
the left-handed slepton soft mass matrix.
$\tilde{\tau}_{L(R)}$ and $\tilde{\mu}_{L(R)}$
are the left-handed (right-handed) stau and smuon, respectively,
and $\tilde{\nu}_{\tau}$ and $\tilde{\nu}_{\mu}$ 
the tau sneutrino and the mu sneutrino.
$\tilde{H}_{1}$ and $\tilde{H}_{2}$ are Higgsino,
$\tilde{W}$ wino.
The symbol $\mu$ is the Higgsino mass. 
The arrows represent the chirality.}

\label{fig:MSSMtmdiag}
\end{center}
\end{figure}
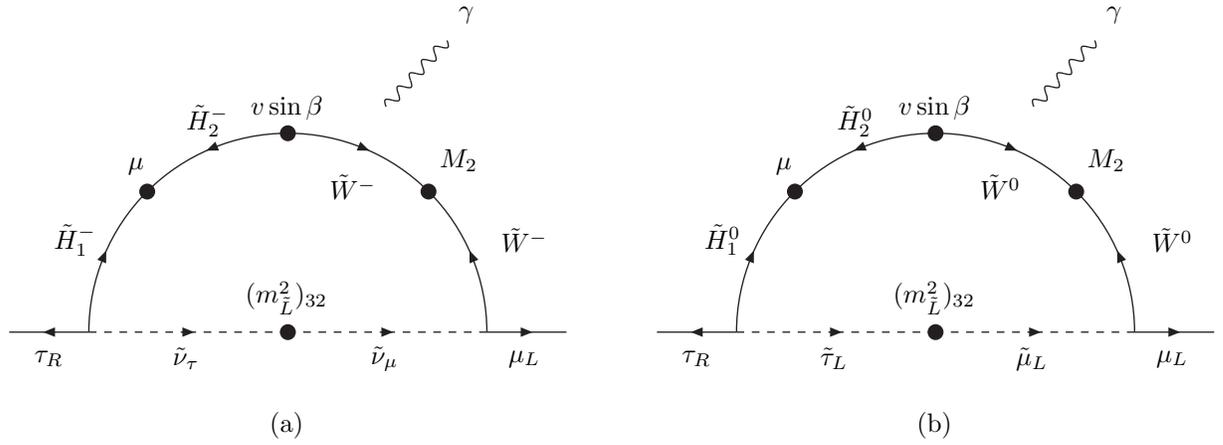

Let us discuss the branching ratios of the LFV rare processes in the
MSSMRN. First, $\tau\to\mu\gamma$. The amplitude of the
$e_i^+\to e_j^+\gamma$ ($i>j$) takes a form
\begin{eqnarray}
T=e \epsilon^{\alpha*}(q) \bar{v}_{i} (p) 
i \sigma_{\alpha \beta} q^\beta (A^{(ij)}_L P_L + A^{(ij)}_R P_R)
v_{j}(p-q),
\label{Penguin}
\end{eqnarray}
where $p$ and $q$ are momenta of $e_i$ and photon, and the decay rate
is given by
\begin{eqnarray}
\Gamma(e_i \to e_j\gamma)
= \frac{e^2}{16 \pi} m_{e_i}^3 (|A^{(ij)}_L|^2+|A^{(ij)}_R|^2).
\label{eventrate}
\end{eqnarray}
Here, we neglect the mass of $e_j$.  The amplitude is not invariant 
for the SU(2)$_L$ and U(1)$_Y$ symmetry and the chiral symmetry of
leptons. Then the coefficients $A^{(ij)}_L$ and $A^{(ij)}_R$ are 
proportional to the charged
lepton masses. Since in the MSSMRN the mismatch between the
left-handed slepton and the charged lepton mass eigenstates is
induced, $A^{(ij)}_L$ is much larger than $A^{(ij)}_R$ since $A^{(ij)}_R$ is
suppressed by $m_{e_j}/m_{e_i}$ compared with $A^{(ij)}_L$. Also, when 
$\tan\beta(\equiv v_2/v_1)$ is large, the contribution to $A^{(ij)}_L$ 
proportional to $f_{e_i} v_2 (= - \sqrt{2} m_{e_i} \tan\beta)$ 
becomes dominant. 
In the MSSMRN, 
the dominant contribution to $\tau\to\mu\gamma$ is from the diagram
of Figs.~(\ref{fig:MSSMtmdiag})(a) and (b) and its expression is 
\begin{eqnarray}
A_L^{(\tau\mu)} &\simeq & m_\tau \frac{\alpha_2}{4\pi} 
                       \mu M_2 \tan\beta (m^2_{\tilde L})_{32} \nonumber \\
&& \times D \left[ D \left[ \frac{1}{m^2} 
     \left\{ 
                         f_{c2}\left(\frac{M^2}{m^2}\right)
              - \frac14  f_{n2}\left(\frac{M^2}{m^2}\right)
     \right\} ; M^2 
             \right] (M_2^2, \mu^2) 
      ; m^2 \right] (m^2_{\tilde{\nu}_\mu},m^2_{\tilde{\nu}_\tau}) , 
\label{al_tau}
\end{eqnarray}
which comes from the SU(2)$_L$ interaction.
The functions $f_{n2}(x)$ and $f_{c2}(x)$ are defined 
in Appendix~\ref{section:applimassins} and the operator $D[f(x);x](x_1,x_2)$
to a function $f(x)$ is defined by
\begin{equation}
D[f(x);x](x_1,x_2) \equiv \frac{1}{x_1-x_2}(f(x_1)-f(x_2)) .
\end{equation}
Here, for a demonstrational purpose, we take a limit where the SUSY
breaking scale is much larger than the $W$ and $Z$ gauge boson masses
and $\tan\beta\gsim 1$. This equation can be derived from the 
mass-insertion formula represented in Appendix~\ref{section:applimassins}. 
The LFV A term 
can not give a dominant contribution when $\tan\beta\gsim 1$.

\begin{figure}[t]
\epsfxsize=12cm
\centerline{\epsfbox{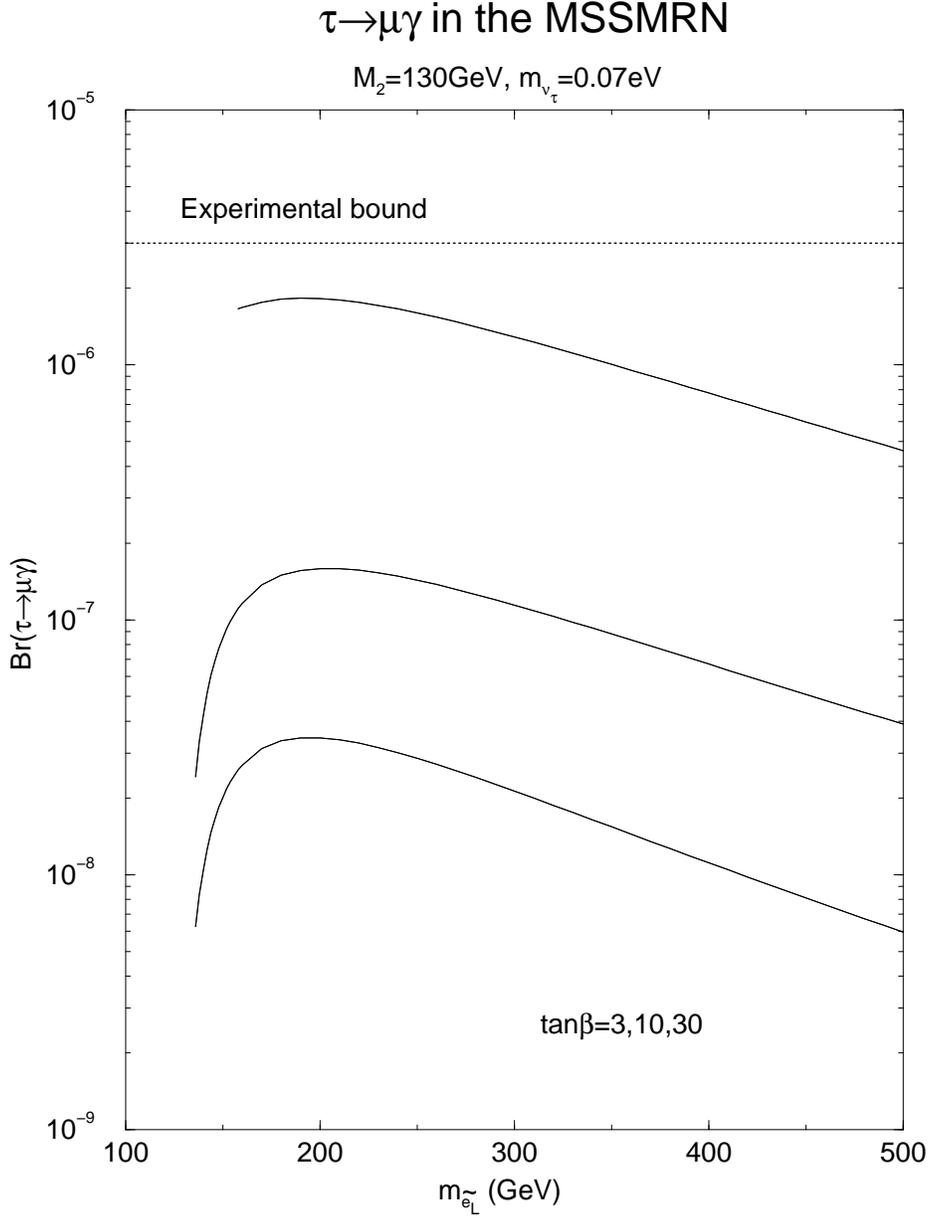}}

\caption{Dependence of the branching ratio of $\tau\to\mu\gamma$ 
on the left-handed selectron mass $m_{\tilde{e}_L}$ in the MSSM
with the right-handed neutrinos. 
We take the tau neutrino mass 0.07eV and $V_{D 32}=-0.71$,
which are suggested by the atmospheric neutrino result.
$M_{\nu_3}$ is fixed at $\simeq 10^{14}$GeV by imposing a condition
$f_{u_3}=f_{\nu_3}$ at the gravitational scale.
The dotted line shown in the figure is the present experimental bound.
We set the wino mass $M_2$ 130GeV, 
and the Higgsino mass parameter $\mu$ positive.
The mu neutrino mass is neglected.
We take $\tan\beta=3,10$, and 30.
The larger $\tan\beta$ corresponds to the upper curve.}
\label{fig:MSSMmlvsbrtm}

\end{figure}

In Fig.~(\ref{fig:MSSMmlvsbrtm}) we show the branching ratio of
$\tau\to\mu\gamma$ as a function of the left-handed selectron
mass ($m_{\tilde{e}_L}$).  Here, $m_{\nu_\tau}=0.07$eV,
$V_{D 33}=V_{D 22}=-V_{D 32}=V_{D 23}=1/\sqrt{2}$, and we assume that
$f_{\nu_3}$ is as large as the Yukawa coupling constant for the
top quark at the gravitational scale. This corresponds to
$M_{\nu_3}\sim 10^{14}$GeV. Also, we impose the radiative breaking
condition of the SU(2)$_L\times$U(1)$_Y$ gauge symmetry with 
$\tan\beta=3,10,30$ and the Higgsino mass parameter $\mu$ positive.
In our calculation we considered the experimental constraints from the
negative results of the SUSY particle search.
Though we do not assume the GUT's, 
we take the wino mass ($M_2$) 130GeV and 
determine the other gaugino masses 
by the GUT relation for the gaugino masses. We use the formula for
$\tau\to\mu\gamma$ in Ref. \cite{HMTY} for the numerical
calculation.

The branching ratio is reduced where the left-handed selectron mass is
comparable to the wino mass. This is because the slepton masses are almost
determined by the radiative correction from the gaugino masses, and
$m^2_0$, which $(m_{\tilde L}^2)_{32}$ is proportional to, is negligible in
the region.  As mentioned above, the branching ratio is proportional to
$\tan^2\beta$ (see Eq.~(\ref{al_tau})), and the line for
$\tan\beta=30$ is close to the experimental bound, 
${\rm Br}(\tau\to\mu\gamma) < 3.0\times 10^{-6}$ \cite{PDG}.

In Fig.~(\ref{fig:MSSMMNvsbrtm}) we present the dependence of 
the branching ratio of $\tau\to\mu\gamma$ on $M_{\nu_3}$. Here, we take 
$m_{\tilde{e}_L} =170$GeV, and the other SUSY breaking parameters are 
the same as 
in Fig.~(\ref{fig:MSSMmlvsbrtm}). The branching ratio is proportional to 
$M_{\nu_3}^2$
since we fix $m_{\nu_\tau}=0.07$eV. If $10^{-8}$ can be reached
in the future experiments, we can probe $M_{\nu_3}> 10^{13}(10^{14})$GeV
for $\tan\beta=30(3)$.\footnote{
An alternative way to prove $(m_{\tilde L}^2)_{32}$ is to search for the 
slepton oscillation \cite{ACHF,HNST}.}

\subsection{The Branching Ratio of $\mu\to e\gamma$}

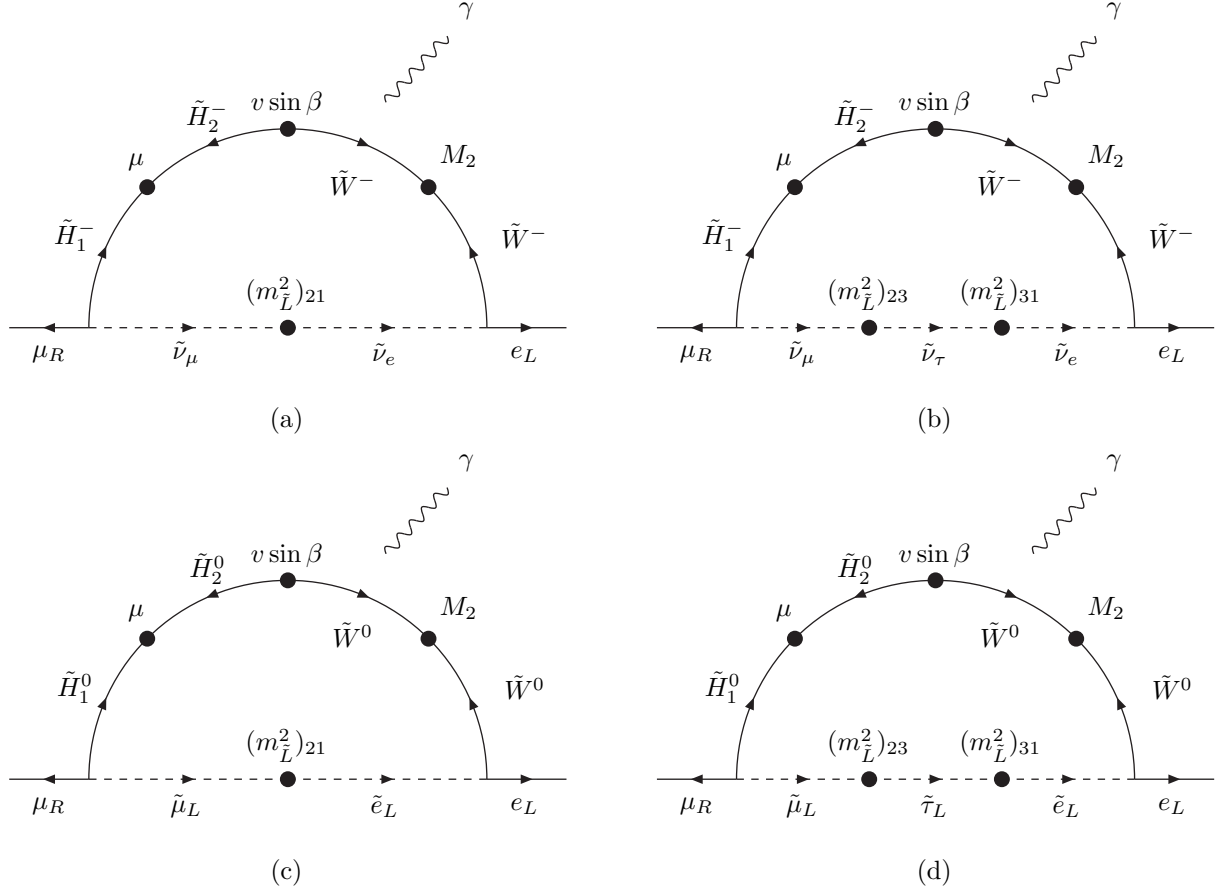
\begin{figure}
\begin{center} 
\begin{picture}(455,140)(30,-20)
\ArrowArcn(135,25)(75,180,135)
\ArrowArc(135,25)(75,90,135)
\ArrowArcn(135,25)(75,90,45)
\ArrowArc(135,25)(75,0,45)
\Vertex(188,78){3}
\Vertex(82,78){3}
\Vertex(135,100){3}
\Text(135,110)[]{$v\sin\beta$}
\Text(55,60)[]{$\tilde{H}_1^-$}
\Text(78,88)[]{$\mu$}
\Text(105,105)[]{$\tilde{H}_2^-$}
\Text(160,80)[]{$\tilde{W}^-$}
\Text(200,90)[]{$M_2$}
\Text(225,60)[]{$\tilde{W}^-$}

\ArrowLine(60,25)(30,25)
\DashArrowLine(60,25)(135,25){3}             \Vertex(135,25){3}
\DashArrowLine(135,25)(210,25){3}        
\ArrowLine(210,25)(240,25)

\Text(45,15)[]{$\mu_R$}
\Text(97,15)[]{$\tilde{\nu}_{\mu}$}
\Text(172,15)[]{$\tilde{\nu}_e$}
\Text(225,15)[]{$e_L$}

\Text(135,38)[]{$(m^2_{\tilde{L}})_{21}$}

\Photon(172,110)(195,135){2}{5}
\Text(203,145)[]{$\gamma$}

\Text(135,-10)[]{(a)}
\SetOffset(245,0)
\ArrowArcn(135,25)(75,180,135)
\ArrowArc(135,25)(75,90,135)
\ArrowArcn(135,25)(75,90,45)
\ArrowArc(135,25)(75,0,45)
\Vertex(188,78){3}
\Vertex(82,78){3}
\Vertex(135,100){3}
\Text(135,110)[]{$v\sin\beta$}
\Text(55,60)[]{$\tilde{H}_1^-$}
\Text(78,88)[]{$\mu$}
\Text(105,105)[]{$\tilde{H}_2^-$}
\Text(160,80)[]{$\tilde{W}^-$}
\Text(200,90)[]{$M_2$}
\Text(225,60)[]{$\tilde{W}^-$}

\ArrowLine(60,25)(30,25)
\DashArrowLine(60,25)(110,25){3}             \Vertex(110,25){3}
\DashArrowLine(110,25)(160,25){3}            \Vertex(160,25){3}
\DashArrowLine(160,25)(210,25){3}        
\ArrowLine(210,25)(240,25)

\Text(45,15)[]{$\mu_R$}
\Text(85,15)[]{$\tilde{\nu}_{\mu}$}
\Text(135,15)[]{$\tilde{\nu}_{\tau}$}
\Text(185,15)[]{$\tilde{\nu}_e$}
\Text(225,15)[]{$e_L$}

\Text(110,38)[]{$(m^2_{\tilde{L}})_{23}$}
\Text(160,38)[]{$(m^2_{\tilde{L}})_{31}$}

\Photon(172,110)(195,135){2}{5}
\Text(203,145)[]{$\gamma$}

\Text(135,-10)[]{(b)}

\end{picture} 

\begin{picture}(455,170)(30,-20)
\ArrowArcn(135,25)(75,180,135)
\ArrowArc(135,25)(75,90,135)
\ArrowArcn(135,25)(75,90,45)
\ArrowArc(135,25)(75,0,45)
\Vertex(188,78){3}
\Vertex(82,78){3}
\Vertex(135,100){3}
\Text(135,110)[]{$v\sin\beta$}
\Text(55,60)[]{$\tilde{H}_1^0$}
\Text(78,88)[]{$\mu$}
\Text(105,105)[]{$\tilde{H}_2^0$}
\Text(160,80)[]{$\tilde{W}^0$}
\Text(200,90)[]{$M_2$}
\Text(225,60)[]{$\tilde{W}^0$}

\ArrowLine(60,25)(30,25)
\DashArrowLine(60,25)(135,25){3}             \Vertex(135,25){3}
\DashArrowLine(135,25)(210,25){3}        
\ArrowLine(210,25)(240,25)

\Text(45,15)[]{$\mu_R$}
\Text(97,15)[]{$\tilde{\mu}_L$}
\Text(172,15)[]{$\tilde{e}_L$}
\Text(225,15)[]{$e_L$}

\Text(135,38)[]{$(m^2_{\tilde{L}})_{21}$}

\Photon(172,110)(195,135){2}{5}
\Text(203,145)[]{$\gamma$}

\Text(135,-10)[]{(c)}

\SetOffset(245,0)
\ArrowArcn(135,25)(75,180,135)
\ArrowArc(135,25)(75,90,135)
\ArrowArcn(135,25)(75,90,45)
\ArrowArc(135,25)(75,0,45)
\Vertex(188,78){3}
\Vertex(82,78){3}
\Vertex(135,100){3}
\Text(135,110)[]{$v\sin\beta$}
\Text(55,60)[]{$\tilde{H}_1^0$}
\Text(78,88)[]{$\mu$}
\Text(105,105)[]{$\tilde{H}_2^0$}
\Text(160,80)[]{$\tilde{W}^0$}
\Text(200,90)[]{$M_2$}
\Text(225,60)[]{$\tilde{W}^0$}

\ArrowLine(60,25)(30,25)
\DashArrowLine(60,25)(110,25){3}             \Vertex(110,25){3}
\DashArrowLine(110,25)(160,25){3}            \Vertex(160,25){3}
\DashArrowLine(160,25)(210,25){3}        
\ArrowLine(210,25)(240,25)

\Text(45,15)[]{$\mu_R$}
\Text(85,15)[]{$\tilde{\mu}_L$}
\Text(135,15)[]{$\tilde{\tau}_L$}
\Text(185,15)[]{$\tilde{e}_L$}
\Text(225,15)[]{$e_L$}

\Text(110,38)[]{$(m^2_{\tilde{L}})_{23}$}
\Text(160,38)[]{$(m^2_{\tilde{L}})_{31}$}

\Photon(172,110)(195,135){2}{5}
\Text(203,145)[]{$\gamma$}

\Text(135,-10)[]{(d)}

\end{picture} 

\caption{Candidates of the Feynman diagrams which give 
dominant contributions to $\mu^+ \to e^+ \gamma$
when $\tan\beta\gsim1$ and 
the off-diagonal elements of the right-handed soft mass matrix
are negligible, as in the MSSM with the right-handed neutrinos.
In the diagrams, $(m^2_{\tilde{L}})_{ij}$ is the $(i,j)$ element of
the left-handed slepton mass matrix.
$\tilde{e}_{L(R)}$ is the left-handed (right-handed) selectron and 
$\tilde{\nu}_e$ is the electron sneutrino.
Other symbols are the same as those in Fig.~(1).}

\label{fig:MSSMmediag}

\end{center}
\end{figure}

Next, we discuss $\mu\to e\gamma$ in the MSSMRN. The forms of
the amplitude and the event rate are the same as those of
$\tau\to\mu\gamma$ (Eqs.~(\ref{Penguin},\ref{eventrate})). 
This process has two types of the
contribution, depending on the structure of the Yukawa coupling for the
neutrino masses.  One is the diagrams where $(m_{\tilde L}^2)_{21}$ or
$A_e^{21}$ is inserted, 
and another is those that
$(m_{\tilde L}^2)_{32}$ or $A_e^{32}$ and 
$(m_{\tilde L}^2)_{13}$ or $A_e^{13}$ are inserted. 
Then the dominant contributions (Figs.~(\ref{fig:MSSMmediag})(a)-(d)) 
are following,
\begin{eqnarray}
A_L^{(\mu e)}  &=& -m_\mu \frac{\alpha_2}{4\pi}M_2 \mu \tan\beta \nonumber \\
&& 
\times D\left[ 
\left\{ \phantom{-}
        \left[  (m^2_{\tilde{L}})_{21} 
              + \frac{(m^2_{\tilde{L}})_{23}(m^2_{\tilde{L}})_{31}}
                     {m^2_{\tilde{\nu}}-m^2_{\tilde{\nu}_{\tau}}}
        \right]
        \frac{1}{m^4_{\tilde{\nu}}}
        \left\{ 
                    g_{c2}\left( \frac{M^2}{m^2_{\tilde{\nu}}}\right)
           -\frac14 g_{n2}\left( \frac{M^2}{m^2_{\tilde{\nu}}}\right)
        \right\}  \right. \right.
\nonumber \\
&& \phantom{ \times D\left[ \left\{ \right. \right.} \left. \left.
  -     \left[ 
               \frac{(m^2_{\tilde{L}})_{23}(m^2_{\tilde{L}})_{31}}
                    {(m^2_{\tilde{\nu}}-m^2_{\tilde{\nu}_{\tau}})^2}
        \right]
        \frac{1}{m^2_{\tilde{\nu}_{\tau}}}
        \left\{ 
                    f_{c2}\left( \frac{M^2}{m^2_{\tilde{\nu}_{\tau}}}\right)
           -\frac14 f_{n2}\left( \frac{M^2}{m^2_{\tilde{\nu}_{\tau}}}\right)
        \right\} 
\right\}  ;  M^2 \right] (M^2_2, \mu^2)
\label{al_mu}
\end{eqnarray}
Here, we take a limit where the SUSY breaking scale is much larger
than the $W$ and $Z$ gauge boson masses and $\tan\beta\gsim 1$, again.
We also assumed the mass degeneracy between the first- and 
the second-generation left-handed sleptons as
\begin{equation}
  m^2_{\tilde{e}_L}   = m^2_{\tilde{\mu}_L} 
= m^2_{\tilde{\nu}_e} = m^2_{\tilde{\nu}_\mu} \equiv m^2_{\tilde{\nu}}.
\end{equation}
The functions $f_{c2,n2}(x)$ and $g_{c2,n2}(x)$ are defined in 
Appendix~\ref{section:applimassins}.

\begin{figure}[t]
\epsfxsize=12cm
\centerline{\epsfbox{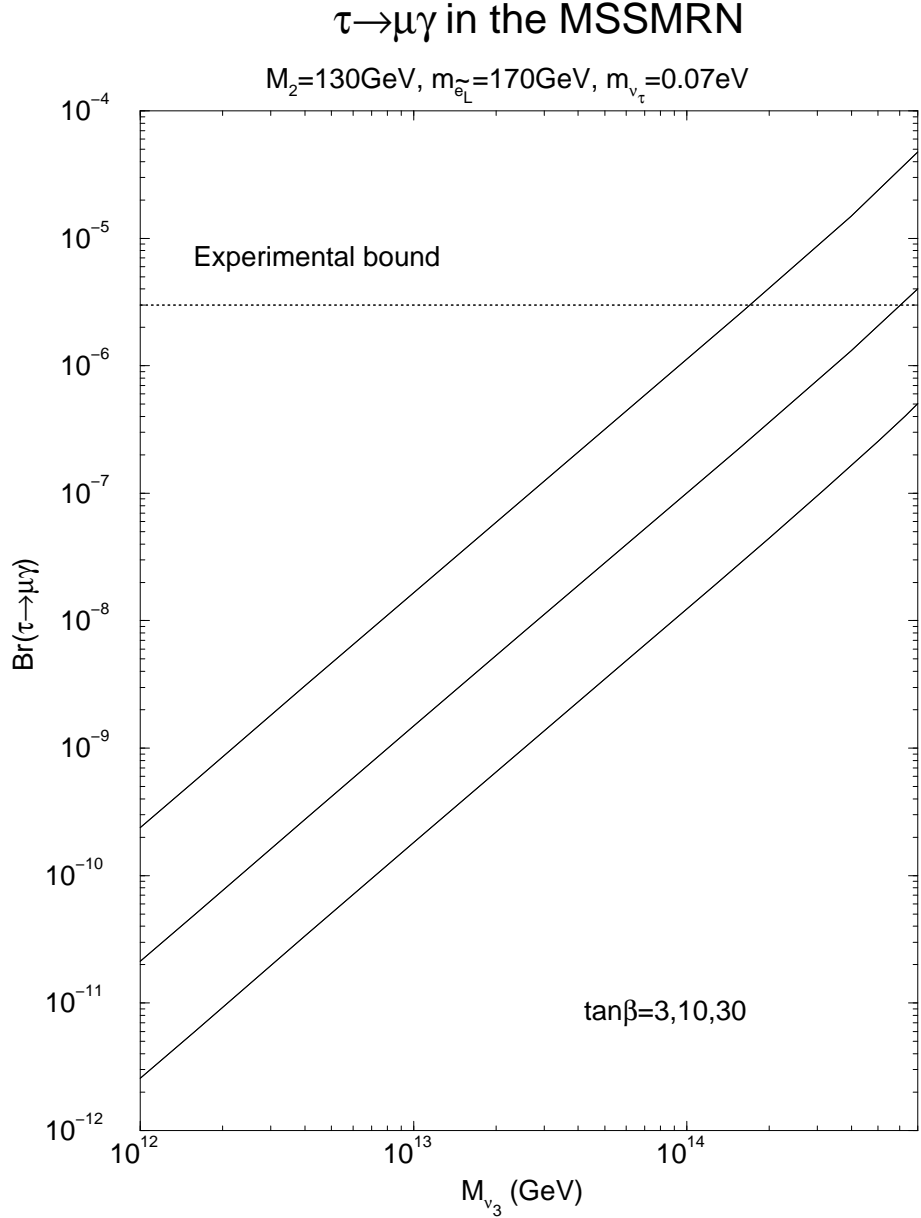}}

\caption{Dependence of the branching ratio of 
$\tau\to\mu\gamma$ on the third-generation right-handed neutrino 
Majorana mass $M_{\nu_3}$ in the MSSM with the right-handed neutrinos.
The input parameters are the same as those of Fig.~(2)
except that in this figure we take $m_{\tilde{e}_L}=170$GeV and 
that we do not impose the condition 
$f_{u_3}=f_{\nu_3}$ but treat $M_{\nu_3}$ as an independent variable.
The dotted line shown in the figure is the present experimental bound.
Here also the larger $\tan\beta$ corresponds to the upper curve.}
\label{fig:MSSMMNvsbrtm}

\end{figure}

\begin{figure}[t]
\epsfxsize=12cm
\centerline{\epsfbox{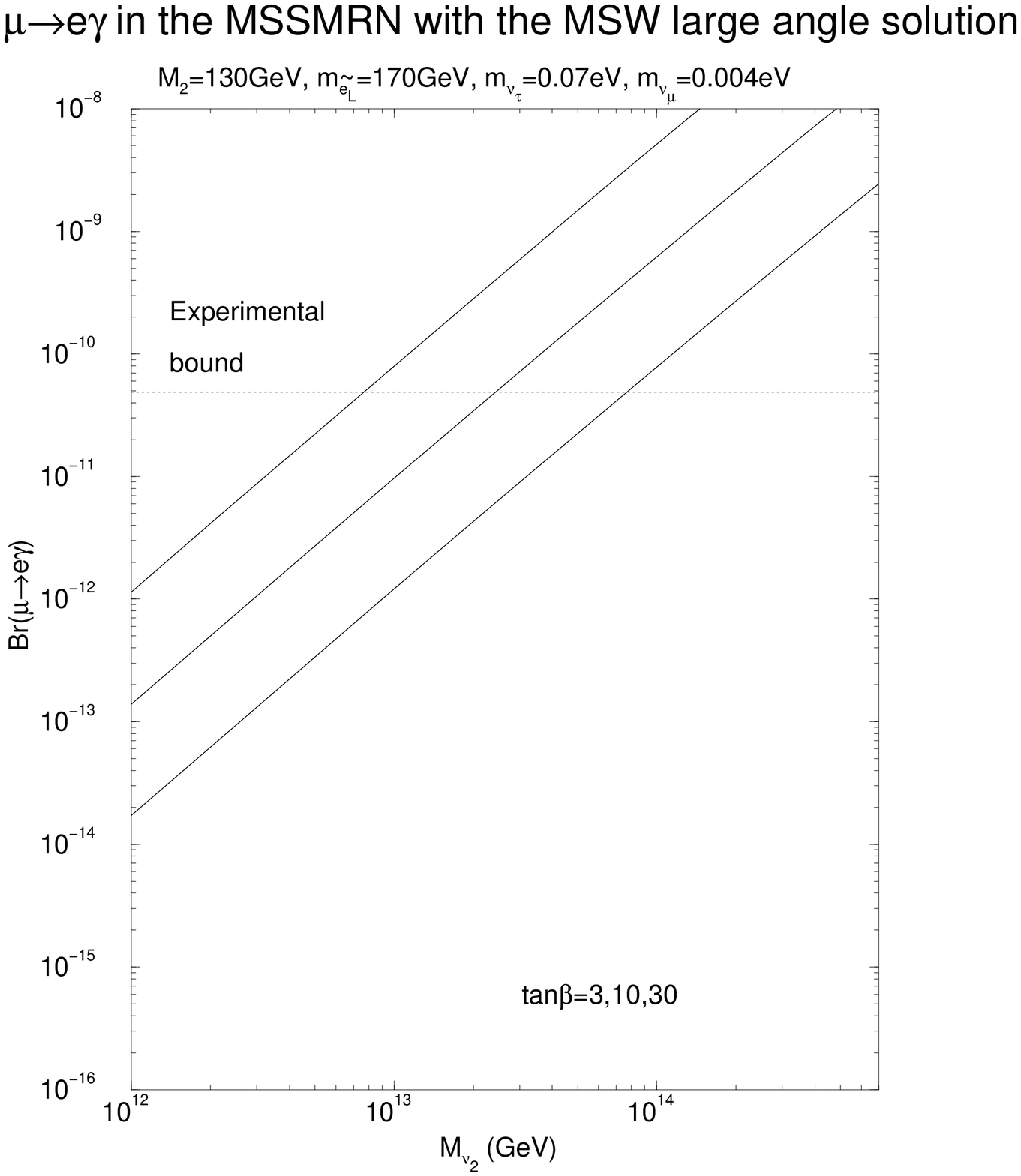}}

\caption{Dependence of the branching ratio of $\mu \to e \gamma$ 
on the second-generation right-handed neutrino Majorana mass $M_{\nu_2}$ 
in the MSSM with the right-handed neutrinos
under the assumption of the MSW large angle solution
with $V_{D31}=0$. 
We take $V_{D 21}=-0.42$ and the mu neutrino mass as 0.004eV,
as suggested by the MSW large angle solution.
The dotted line shown in the figure is the present experimental bound.
Other input parameters are 
the wino mass 130GeV, the left-handed selectron 170GeV, 
the tau neutrino mass 0.07eV, and $\tan\beta=3,10$, and 30.
The larger $\tan\beta$ corresponds to the upper curve.}

\label{fig:MSSMmeMl}
\end{figure}

As mentioned above, if the solar neutrino anomaly comes from the 
MSW effect or the vacuum oscillation  with the large angle, $V_{D 21}$
is expected to be large. This leads to non-vanishing $(m_{\tilde L}^2)_{21}$.
In Fig.~(\ref{fig:MSSMmeMl}), under the condition that 
\begin{equation}
V_{D} =\left( \begin{array}{ccc}  
  0.91 &   0.35  & 0.24            \\
- 0.42 &   0.72  & 0.55            \\
  0    & - 0.60  & 0.80
\end{array} \right)  \label{eq:MSWLargeMixing} 
\end{equation}
and $m_{\nu_\mu}=0.004$eV \cite{BG}
we show the branching ratio of $\mu\to e\gamma$ as a function
of $M_{\nu_2}$.  This corresponds to the MSW solution with the large
mixing. Here we take $V_{D31}=0$ and we will discuss a case with
finite $V_{D31}$ later. The input parameters are taken to be the same
as in Fig.~(\ref{fig:MSSMMNvsbrtm}). 
For $\tan\beta=30(3)$, the branching ratio reaches
the experimental bound 
(${\rm Br}(\mu\to e\gamma) < 4.9\times10^{-11}$ \cite{PDG}) when 
$M_{\nu_2} \simeq 8 \times 10^{12} (8 \times 10^{13})$GeV. This 
corresponds to $f_{\nu_2} \simeq 0.03(0.11)$. Future experiments are expected
to reach 10$^{-14}$ \cite{kuno}. 
This corresponds to $M_{\nu_2} \simeq 10^{11}$ $(10^{12})$GeV. 

\begin{figure}[t]
\epsfxsize=12cm

\centerline{\epsfbox{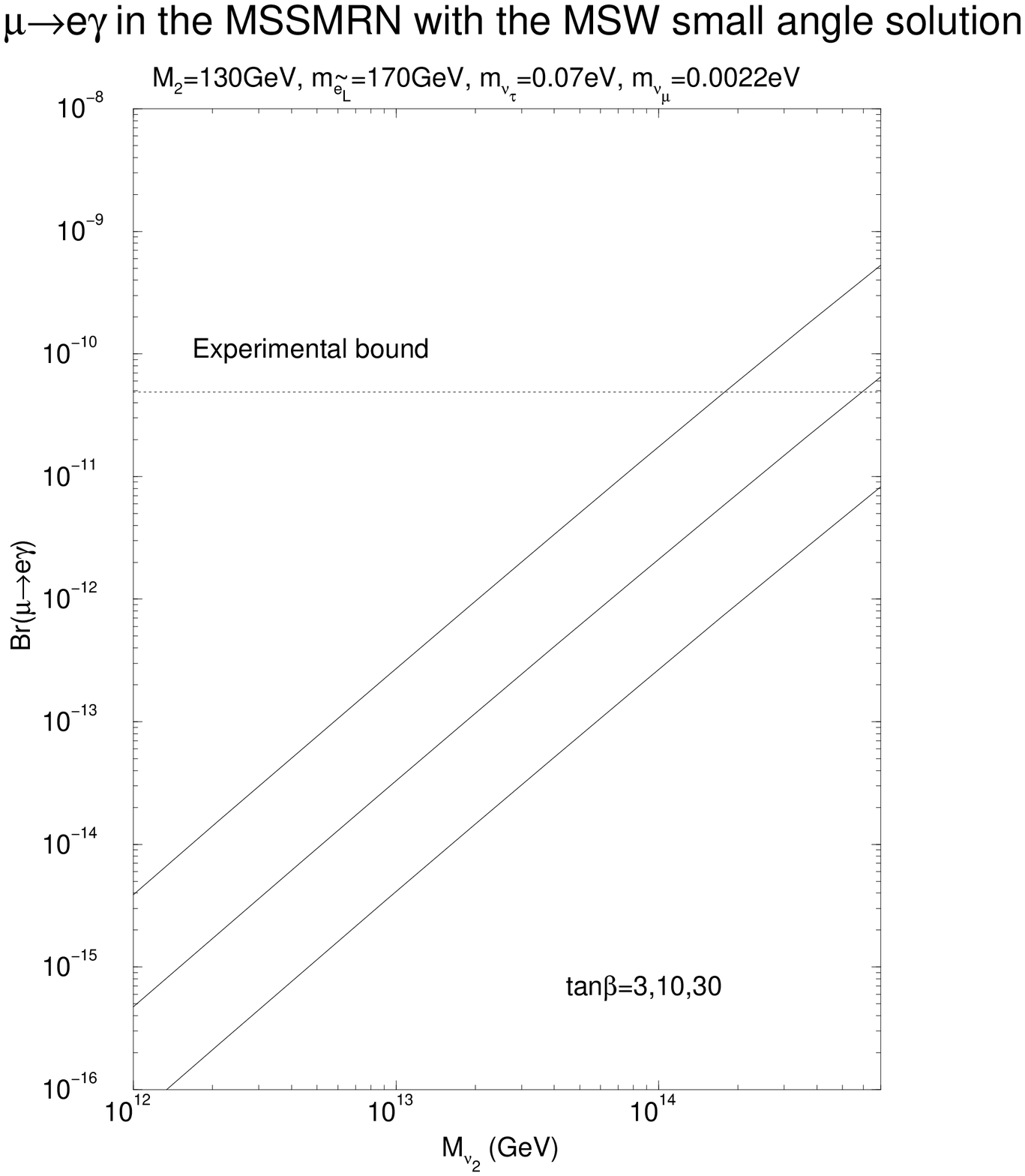}}

\caption{Dependence of the branching ratio of $\mu \to e \gamma$ 
on the second-generation right-handed neutrino Majorana mass $M_{\nu_2}$ 
in the MSSM with the right-handed neutrinos
under the assumption of the MSW small angle solution with $V_{D 31}=0$.
We take $V_{D 21}=-0.04$ and the mu neutrino mass as 0.0022eV,
which are suggested by the MSW small angle solution
if the mixing comes from $V_{D}$.
Other input parameters are the same as those in Fig.~(5).
The dotted line shown in the figure is the present experimental bound.
$\tan\beta=3,10$, and 30, and 
the larger $\tan\beta$ corresponds to the upper curve.}

\label{fig:MSSMmeMs}
\end{figure}

If the solar neutrino anomaly comes from the MSW solution with the
small mixing, we cannot distinguish whether the mixing comes from 
$V_D$ or $V_M$.
If it comes from $V_D$, the branching ratio is smaller by about 
1/100 compared with that in the MSW solution with the large mixing, 
as shown in Fig.~(\ref{fig:MSSMmeMs}).
In Fig.~(\ref{fig:MSSMmeMs}) we assume that 
\begin{equation}
V_{D} =\left( \begin{array}{ccc}  
  1    &   0.04  & 0.03            \\
- 0.04 &   0.79  & 0.59            \\
  0    & - 0.60  & 0.80
\end{array} \right)  \label{eq:MSWSmallMixing} 
\end{equation}
and $m_{\nu_\mu}=0.0022$eV \cite{BG}.
Other input parameters are the same as Fig.~(\ref{fig:MSSMmeMl}).

\begin{figure}[t]
\epsfxsize=12cm

\centerline{\epsfbox{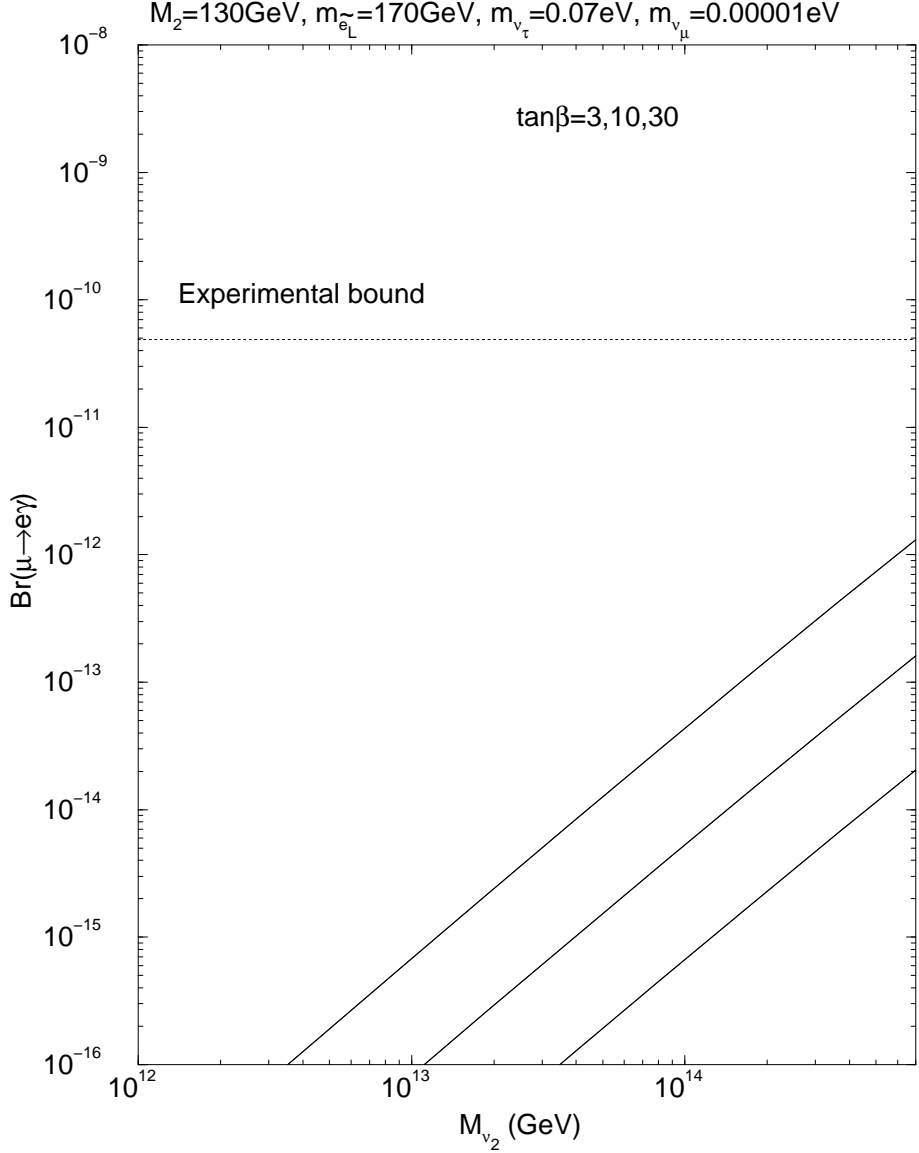}}

\caption{Dependence of the branching ratio of $\mu \to e \gamma$ 
on the second-generation right-handed neutrino Majorana mass $M_{\nu_2}$ 
in the MSSM with the right-handed neutrinos 
under the assumption of the 'just so' solution
with $V_{D 31}=0$.
We take $V_{D 21}=-0.71$ and the mu neutrino mass as 
$1.0\times10^{-5}$eV,
as suggested by the 'just so' solution.
Other input parameters are the same as those in Fig.~(5).
The dotted line shown in the figure is the present experimental bound.
$\tan\beta=3,10$, and 30, and 
the larger $\tan\beta$ corresponds to the upper curve.}

\label{fig:MSSMmeJs}
\end{figure}

In Fig.~(\ref{fig:MSSMmeJs}) we take 
\begin{equation}
V_{D} =\left( \begin{array}{ccc}  
 \frac{1}{\sqrt{2}}   &         \frac12      &   \frac12            \\
-\frac{1}{\sqrt{2}}   &         \frac12      &   \frac12            \\
            0         & -\frac{1}{\sqrt{2}}  &  \frac{1}{\sqrt{2}} 
\end{array} \right) \label{eq:BiMax} 
\end{equation}
and $m_{\nu_\mu}=1.0 \times 10^{-5}$eV \cite{BiMax}.
Other parameters are 
the same as in Fig.~(\ref{fig:MSSMmeMl}). 
This corresponds to the 'just so' solution
for the solar neutrino anomaly. Since the mu neutrino mass is smaller,
the branching ratio is suppressed by $10^{-5}$ compared with that in 
the MSW solution with the large mixing. 

\begin{figure}[t]
\epsfxsize=12cm

\centerline{\epsfbox{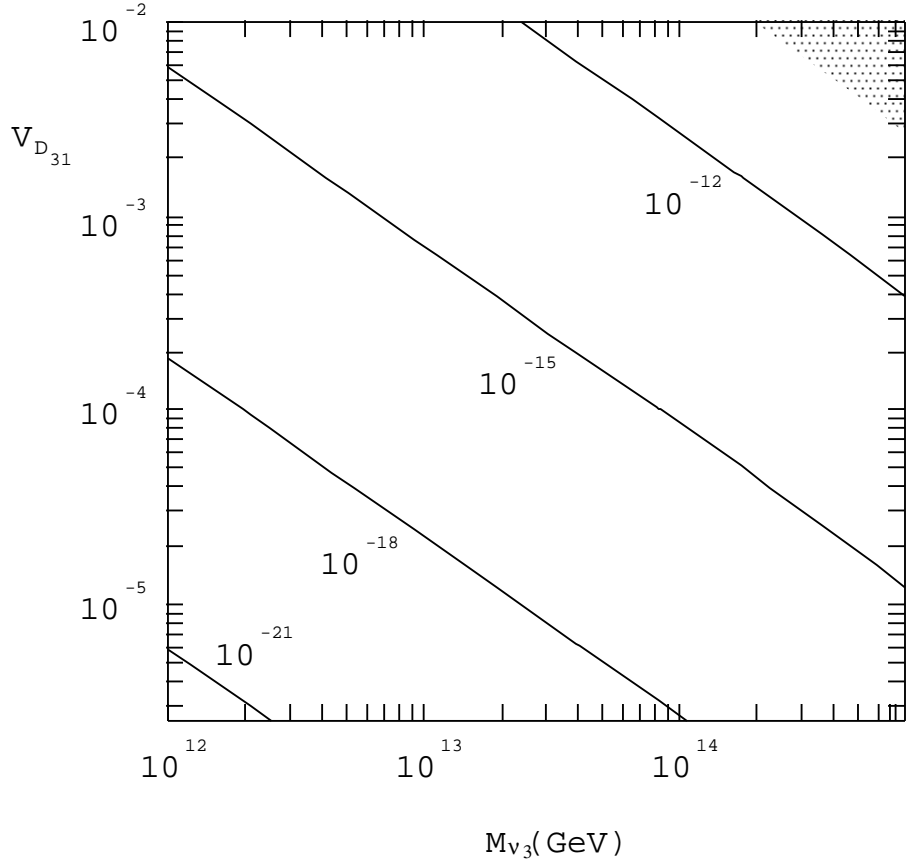}}

\caption{Dependence of the branching ratio of $\mu\to e \gamma$ 
on the third-generation right-handed neutrino Majorana mass $M_{\nu_3}$ 
and $V_{D 31}$ in the MSSM with the right-handed neutrinos. 
Here the tau neutrino mass is 0.07eV and 
$V_{D32}=-0.71$, as suggested by the atmospheric neutrino result.
We neglect $f_{\nu_2}$ here.
The curves mean the contours on which 
the branching ratio of $\mu \to e \gamma$
is $10^{-21},10^{-18},10^{-15}$, and $10^{-12}$, respectively.
The shaded region is already excluded experimentally.
$\tan\beta$ is set to be 3.
The wino mass is 130GeV, the left-handed selectron mass 170GeV,
and the Higgsino mass parameter $\mu$ positive.
}

\label{fig:MSSMMNV31me3}
\end{figure}

\begin{figure}[t]
\epsfxsize=12cm

\centerline{\epsfbox{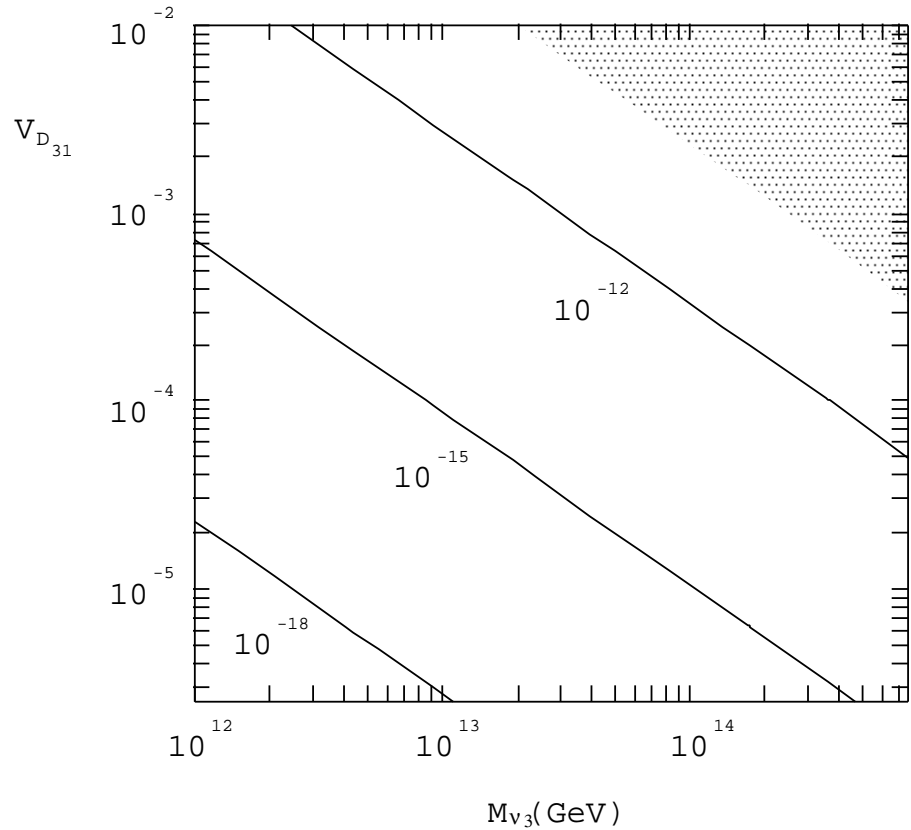}}

\caption{Dependence of the branching ratio of $\mu \to e \gamma$ 
on the third-generation right-handed neutrino Majorana mass $M_{\nu_3}$ 
and $V_{D 31}$ in the MSSM with the right-handed neutrinos. 
The input parameters are the same as those in
Fig.~(8) except that we take $\tan\beta=30$ here.
The curves mean the contours on which 
the branching ratio of $\mu \to e \gamma$
is $10^{-18},10^{-15}$, and $10^{-12}$, respectively.
The shaded region is already excluded experimentally.
}

\label{fig:MSSMMNV31me30}
\end{figure}

Next we discuss the branching ratio of $\mu\to e\gamma$
when $V_{D31}$ is finite. 
In Figs.~(\ref{fig:MSSMMNV31me3},\ref{fig:MSSMMNV31me30})
we show the branching ratio
as a function of $V_{D31}$ and $M_{\nu_3}$ for $\tan\beta=3$ and
30.  Here we assume that $f_{\nu_2}$ is negligibly small.  The other
parameters are the same as in Fig.~(\ref{fig:MSSMMNvsbrtm}). 
The branching ratio is
almost proportional to $V_{D31}^2 M_{\nu_3}^2$. Compared with this
figure to Fig.~(\ref{fig:MSSMmeMl}), when $M_{\nu_2} = M_{\nu_3}$, the
contribution from $V_{D31}$ is negligible in the MSW solution with
the large mixing angle unless
$V_{D31}$ is larger than $10^{-2.5}$. On the other hand, it can be dominant
in the 'just so' solution even if $V_{D 31} \sim 10^{-4}$.

Finally we consider the $\mu^+ \to e^+ e^- e^+$ process 
and the $\mu$-$e$ conversion on ${}^{48}_{22}$Ti.
For these processes the penguin type diagrams dominate over the others,
so the behavior of the decay rate is similar to that of $\mu \to e
\gamma$. For the $\mu \to 3e$ process the following approximate relation 
holds between the branching ratios of the two processes,
\begin{eqnarray}
{\rm Br}(\mu \to 3e) &\simeq& \frac{\alpha}{8\pi} \frac83  
        \left(  \log \frac{m_\mu^2}{m_e^2} - \frac{11}{4}\right)
       {\rm Br}(\mu \to e \gamma) \\
 &\simeq& 7 \times 10^{-3} {\rm Br}(\mu \to e \gamma) .
\end{eqnarray}
For the $\mu$-$e$ conversion rate $\Gamma(\mu \to e)$
a similar relation holds at $\tan\beta>1$ region,
\begin{equation}
\Gamma (\mu \to e) \simeq 16\alpha^4 Z_{\rm eff}^4 Z |F(q^2)|^2
                           {\rm Br}(\mu \to e \gamma) .
\end{equation}
Here $Z$ is the proton number in the nucleus,
and $Z_{\rm eff}$ is the effective charge, 
$F(q^2)$ the nuclear form factor at the momentum transfer $q$. 
For ${}^{48}_{22}$Ti, $Z_{\rm eff}=17.6$
and $F(q^2\simeq - m_\mu^2) \simeq 0.54$ \cite{Zeff,BNT}.
We express the magnitude of the  $\mu$-$e$ conversion 
with the normalization the muon capture rate in Ti nucleus. 
Then the normalized conversion rate 
$R(\mu^- \to e^- ;{}^{48}_{22}{\rm Ti})$
is approximately
\begin{equation}
R(\mu^- \to e^- ;{}^{48}_{22}{\rm Ti}) 
\simeq 6 \times 10^{-3} {\rm Br}(\mu \to e \gamma) .
\end{equation}
The future experiment for the $\mu$-$e$ conversion is planed to reach
$R(\mu^- \to e^- ;{}^{48}_{22}{\rm Ti})<10^{-18}$ \cite{mu-e}.

\section{The Lepton Flavor Violation in the SU(5) SUSY GUT with 
Right-handed Neutrinos}

In the SUSY GUT the gauge coupling unification is predicted, and the
predicted weak mixing angle is
consistent with the experimental data at the 1 \% level of accuracy.
Moreover if the unified gauge group is SO(10),
the right-handed neutrinos are introduced 
automatically into the matter multiplet. 
However, in order to accommodate the observed
large mixing angle in the framework of the SO(10) SUSY GUT,
one needs unnatural extension
of the simplest version of the SO(10) SUSY GUT.
Hence in this article we do not discuss the SO(10) SUSY GUT.
Here we investigate the SU(5) SUSY GUT 
with the right-handed neutrinos
as one of the extension of the MSSMRN 
in which the small neutrino mass is naturally obtained 
and the large neutrino mixing angle is possible
without unnatural fine-tuning. 
We here call this model as SU(5)RN, for brevity.
After introducing the model we estimate the off-diagonal elements
of the slepton soft mass matrices using 
the one-loop level RGE's under an assumption of the minimal SUGRA scenario.
With them we study the LFV processes 
$\tau \to \mu \gamma$ and $\mu \to e \gamma$.
After that we comment on the $b \to s \gamma$ branching ratio.
We show that the LFV rates in this model 
is larger in general than those in the MSSMRN model,
due to the fact that in this model the right-handed slepton mass 
matrix also can have non-negligible off-diagonal elements,
in addition to the left-handed one \cite{BH,BHS,su5}.

First we introduce the model. 
This model has three families of matter multiplets $\psi_i$, $\phi_i$, 
and $\eta_i$, which are ${\bf 10}$, ${\bf 5}^\ast$, and ${\bf 1}$ 
dimension representations of SU(5), respectively.
$\psi_i$ contains the quark doublet, the charged lepton singlet, and 
the up-type quark singlet, while $\phi_i$ the down-type quark singlet
and the lepton doublet and $\eta_i$ the right-handed neutrino, respectively. 
The model has ${\bf 5}$ and ${\bf 5}^\ast$
dimension representation Higgs multiplets, $H$ and $\overline{H}$. 
$H$ consists of the MSSM Higgs multiplet $H_2$ and a colored Higgs
multiplet $H_C$, and $\overline{H}$ another MSSM Higgs multiplet $H_1$ and
another colored Higgs multiplet $\overline{H}_C$.
The GUT gauge symmetry is spontaneously broken into the SM one 
at the GUT scale $M_{\rm GUT} \simeq 2\times 10^{16}$GeV. 
Above the GUT scale 
the superpotential $W$ of the matter sector of this model is
\begin{eqnarray}
W&=&\frac14 f_{u_{ij}} \psi_i^{AB} \psi_j^{CD} H^E \epsilon_{ABCDE} 
  + \sqrt{2} f_{d_{ij}} \psi_i^{AB} \phi_{jA} \overline{H}_B  \nonumber  \\
 &&+ f_{\nu_{ij}} \eta_i \phi_{jA} H^A 
  + \frac12 M_{\eta_i \eta_j} \eta_i \eta_j , \nonumber 
\end{eqnarray}
where $A,B,...$ are indices of SU(5) and run from 1 to 5.
We also introduce the soft SUSY breaking terms associated with
the GUT multiplets. The relevant part of them is
\footnote{For simplicity we neglect the Yukawa coupling $\lambda H \Sigma
\bar{H}$ and the soft SUSY-breaking parameters associated with it, 
where $\Sigma$ is an adjoint representation Higgs multiplet causing the
breaking SU(5)$_{\rm GUT} \to$ SU(3)$_c \times$ SU(2)$_L \times$ U(1)$_Y$.}
\begin{eqnarray}
-{\cal L}_{\rm SUSY~breaking} 
&=&
 (m_{\psi}^2)_{ij}  \tilde{\psi}_i^{\dagger} \tilde{\psi}_j
+(m_{\phi}^2)_{ij}  \tilde{\phi}_i^{\dagger} \tilde{\phi}_{j} 
+(m_{\eta}^2)_{ij}  \tilde{\eta}_i^{\dagger} \tilde{\eta}_{j} 
+m_{h}^2 h^\dagger h +  m_{\bar{h}}^2 \bar{h}^\dagger \bar{h} 
\nonumber\\
&&
+\left\{
 \frac14  A_{u_{ij}} \tilde{\psi}_i \tilde{\psi}_j h
+ \sqrt{2}A_{d_{ij}} \tilde{\psi}_i \tilde{\phi}_{j} \bar{h}
+ A_{\nu_{ij}} \tilde{\eta}_i \tilde{\phi}_{j} h +  h.c.
\right\},
\end{eqnarray} 
where $\tilde{\psi}_i$, $\tilde{\phi}_i$, and 
$\tilde{\eta}_i$ are the scalar components of the ${\psi}_i$, 
${\phi}_i$, and ${\eta}_i$ chiral multiplets, respectively, and 
$h$ and $\bar{h}$ are the Higgs bosons. In the minimal SUGRA scenario
these coefficients are given at the gravitational scale as 
\begin{eqnarray}
&(m_{\psi}^2)_{ij}=(m_{\phi}^2)_{ij}=(m_{\eta}^2)_{ij}=\delta_{ij} m_0^2,&
\nonumber\\
&m_{h}^2 = m_{\bar{h}}^2= m_0^2,&
\nonumber\\
& A_{u_{ij}}=f_{u_{ij}}a_0, \,  A_{d_{ij}}=f_{d_{ij}}a_0, \, 
 A_{\nu_{ij}}=f_{\nu_{ij}}a_0. &
\label{InitCondAtGravScale}
\end{eqnarray}

At the GUT scale we choose a basis where 
the up-type quark and the neutrino Yukawa coupling 
matrices are diagonalized as     
\begin{eqnarray}
f_{u_{ij}}&=&f_{u_i} e^{i\phi_{u_i}} \delta_{ij}, \nonumber \\
f_{d_{ij}}&=&(V^{\ast}_{\rm KM})_{ik} f_{d_k} (V_D^{\dagger})_{kj}, 
\nonumber \\
f_{\nu_{ij}}&=& f_{\nu_i} e^{i\phi_{\nu_i}} \delta_{ij} , \label{eq:basis}
\end{eqnarray}
where $f_{\psi_i}$ ($\psi = u,d,\nu$) are the eigenvalues of 
$f_{\psi_{ij}}$, respectively, $V_{\rm KM}$ the 
Kobayashi-Maskawa matrix at the GUT scale, and $V_D$ a unitary matrix 
which describes the generation mixing in the lepton sector.
$\phi_{\psi_i} (\psi=u,\nu)$ are phase factors which satisfy
$\phi_{ u_1 }+\phi_{ u_2 }+\phi_{ u_3 }=0$ and
$\phi_{\nu_1}+\phi_{\nu_2}+\phi_{\nu_3}=0$. 
However these phases are completely irrelevant for our below discussion.

At the GUT scale the Yukawa coupling constants responsible to
the down-type quark masses and those responsible to the charged lepton masses 
are supposed to unify as
\begin{equation}
f_{d_{i}}=f_{e_{i}}. \label{eq:unification}
\end{equation}
This relation is consistent with the particle spectrum at the low energy
only for the third-generation.
In order to explain the fermion masses
of the first- and the second-generations, one has to consider
the effect of the nonrenormalizable terms also. At that time 
those terms can be another source of LFV \cite{ACH,HNOST}, 
but we do not take them into account for simplicity.

Below the GUT scale we take the basis in which the Yukawa coupling 
constant matrix responsible for charged lepton masses is diagonalized.
The basis we take at low energy region is related to
that of GUT multiplets by the following embedding:
\begin{eqnarray}
\psi_i&=&\{Q_i, e^{-i\phi_{u_i}} \overline{U}_i, (V_{\rm KM})_{ij}
           \overline{E}_j \}, \nonumber \\
\phi_i&=&\{V_{D ij}\overline{D}_j,V_{D ij}L_j\}, \nonumber \\
\eta_i&=&\{e^{-i\phi_{\nu_i}} \overline{N}_i \} . \label{eq:embed}
\end{eqnarray}
Then the superpotential $W$ is expanded in terms of the MSSM fields as
\begin{eqnarray}
\label{SU5RNdecomp}
W&=& \phantom{+}f_{u_i} Q_i \overline{U}_i H_2
  + (V_{\rm KM}^{\ast})_{ij} f_{d_j} Q_i \overline{D}_{j} H_1 \nonumber \\
&&  +  f_{d_i} \overline{E}_i L_i H_1
    - f_{\nu_i} V_{D ij} \overline{N}_i L_j H_2 \nonumber \\
&& + f_{u_j} (V_{\rm KM})_{ji} \overline{E}_i \overline{U}_j H_C 
   -\frac12 f_{u_i}e^{i\phi_{u_i}} Q_i Q_i H_C  \nonumber \\
&& + (V_{\rm KM}^{\ast})_{ij} f_{d_j} e^{-i\phi_{u_i}} 
\overline{U}_i \overline{D}_j \overline{H}_C
 - (V_{\rm KM}^{\ast})_{ij} f_{d_j} Q_i L_j \overline{H}_C \nonumber \\
&& + f_{\nu_i} V_{D ij} \overline{N}_i \overline{D}_j H_C \nonumber \\
&& + \frac12 M_{\nu_i \nu_j} \overline{N}_i \overline{N}_j .
\end{eqnarray}
Here we should notice that the fifth term of the right-hand side of the above
equation is no longer generation-diagonal. This is nothing but a 
direct consequence of the GUT unification, that is, 
one of the central goal of the grand unification is to embed the leptons
and the quarks into the same multiplet, which forces the mixing 
in the quark sector related to that of the lepton sector. 
No redefinition of $\overline{E}$ in generation-space can eliminate this
mixing, as can be seen from Eq.~(\ref{SU5RNdecomp}). This mixing causes 
the off-diagonal elements of $(m^2_{\tilde{e}})$ via radiative corrections.

As for the origin of the observed mixing angle between the left-handed 
neutrinos a parallel discussion to that of the previous section applies. 
The Majorana mass matrix in Eq.~(\ref{SU5RNdecomp}) 
has an inter-generational mixing as
\begin{equation} 
M_{\nu_i\nu_j}=U^\ast_{ik}M_{\nu_k}U^\dagger_{kj}.
\end{equation}
The mass matrix of the left-handed neutrinos $(m_\nu)$ is then
\begin{equation} 
(m_\nu)_{ij}= V^\top_{D ik} (\overline{m}_\nu)_{kl} V_{D lj},
\end{equation}
where 
\begin{eqnarray}
(\overline{m}_{\nu})_{ij} &=& 
m_{{\nu_i}D} \left[M^{-1}\right]_{ij} m_{{\nu_j}D}     \nonumber\\
&\equiv& V_{Mik}^\top m_{\nu_k}  V_{Mkj}.
\end{eqnarray}
Here also $m_{{\nu_i}D}=f_{\nu_i}v\sin\beta/\sqrt{2}$, the same notation
as that of the previous section.
The discussion in the previous section shows that the
large mixing angle from $V_M$ requires a fine-tuning between 
the elements of $M_{\nu_i\nu_j}$ (Eq.~(\ref{relation})).
Therefore
as for the large mixing angle between neutrinos it is natural that 
its origin is in the mixing matrix $V_D$ in Eq.~(\ref{SU5RNdecomp}).
Here we assume $U_{ij}=\delta_{ij}$, for simplicity,
which means that the mixing comes only from
$V_D$.

Now we evaluate the off-diagonal elements of the slepton mass matrices
at the low energy. 
As stated above, both the left- and right-handed slepton's ones have
non-negligible off-diagonal elements at one-loop level.
Assuming $m_{\nu_e} \ll m_{\nu_\mu} \ll m_{\nu_\tau}$ we neglect
$f_{\nu_1}$, and also $f_{u_1}$ and $f_{u_2}$
to obtain approximate formulas for the off-diagonal elements of 
the slepton mass matrices as
\begin{eqnarray}
(m^2_{\tilde{e}})_{ij} 
&\simeq& - \frac{3}{8\pi^2}
  f_{u_3}^2 (V_{\rm KM})_{3i} (V^{\ast}_{\rm KM})_{3j} 
 (3 m_0^2 +a_0^2)  \log  \frac{M_{\rm grav}}{M_{\rm GUT}} ,
\label{offdiagofR} \\
(m^2_{\tilde{L}})_{ij} 
&\simeq& - \frac{1}{8\pi^2} \left( 
  f_{\nu_3}^2 V^{\ast}_{D 3i} V_{D 3j} 
 \log \frac{M_{\rm grav}}{M_{\nu_3}}
+ f_{\nu_2}^2 V^{\ast}_{D 2i} V_{D 2j} 
 \log \frac{M_{\rm grav}}{M_{\nu_2}}
\right)  (3 m_0^2 +a^2_0)  ,
\label{offdiagofL} \\
A_e^{ij}&\simeq&
-\frac{3}{8\pi^2} a_0 \left( 
  f_{e_i} V_{D 3i}^\ast V_{D 3j} f_{\nu_3}^2 
       \log \frac{M_{\rm grav}}{M_{\nu_3}}  
+ f_{e_i} V_{D 2i}^\ast V_{D 2j} f_{\nu_2}^2 
       \log \frac{M_{\rm grav}}{M_{\nu_2}}  
\right. \nonumber \\
&&
\phantom{-\frac{3}{8\pi^2} a_0 }
\left. + 3 f_{e_j} V_{{\rm KM}3j}^\ast V_{{\rm KM}3i} f_{u_3}^2 
\log \frac{M_{\rm grav}}{M_{\rm GUT}} \right) ,
\end{eqnarray}
for $i \ne j$.
These formulas are obtained by a logarithmic approximation 
from the RGE's (given in Appendix~\ref{section:RGE}). 

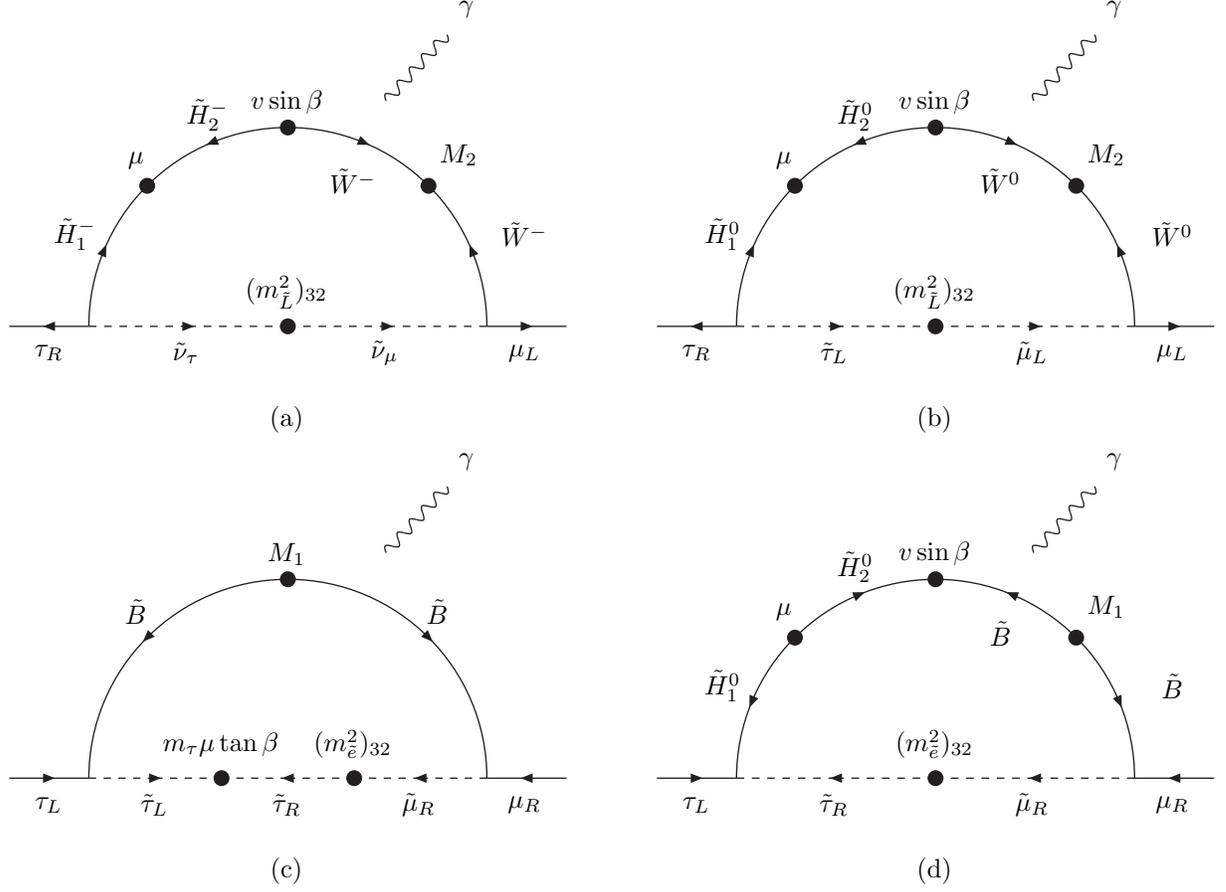
\begin{figure}[p]
\begin{center} 
\begin{picture}(455,140)(30,-20)
\ArrowArcn(135,25)(75,180,135)
\ArrowArc(135,25)(75,90,135)
\ArrowArcn(135,25)(75,90,45)
\ArrowArc(135,25)(75,0,45)
\Vertex(188,78){3}
\Vertex(82,78){3}
\Vertex(135,100){3}
\Text(135,110)[]{$v\sin\beta$}
\Text(55,60)[]{$\tilde{H}_1^-$}
\Text(78,88)[]{$\mu$}
\Text(105,105)[]{$\tilde{H}_2^-$}
\Text(160,80)[]{$\tilde{W}^-$}
\Text(200,90)[]{$M_2$}
\Text(225,60)[]{$\tilde{W}^-$}

\ArrowLine(60,25)(30,25)
\DashArrowLine(60,25)(135,25){3}             \Vertex(135,25){3}
\DashArrowLine(135,25)(210,25){3}        
\ArrowLine(210,25)(240,25)

\Text(45,15)[]{$\tau_R$}
\Text(97,15)[]{$\tilde{\nu}_\tau$}
\Text(172,15)[]{$\tilde{\nu}_\mu$}
\Text(225,15)[]{$\mu_L$}

\Text(135,38)[]{$(m^2_{\tilde{L}})_{32}$}

\Photon(172,110)(195,135){2}{5}
\Text(203,145)[]{$\gamma$}

\Text(135,-10)[]{(a)}
\SetOffset(245,0)
\ArrowArcn(135,25)(75,180,135)
\ArrowArc(135,25)(75,90,135)
\ArrowArcn(135,25)(75,90,45)
\ArrowArc(135,25)(75,0,45)
\Vertex(188,78){3}
\Vertex(82,78){3}
\Vertex(135,100){3}
\Text(135,110)[]{$v\sin\beta$}
\Text(55,60)[]{$\tilde{H}_1^0$}
\Text(78,88)[]{$\mu$}
\Text(105,105)[]{$\tilde{H}_2^0$}
\Text(160,80)[]{$\tilde{W}^0$}
\Text(200,90)[]{$M_2$}
\Text(225,60)[]{$\tilde{W}^0$}

\ArrowLine(60,25)(30,25)
\DashArrowLine(60,25)(135,25){3}             \Vertex(135,25){3}
\DashArrowLine(135,25)(210,25){3}        
\ArrowLine(210,25)(240,25)

\Text(45,15)[]{$\tau_R$}
\Text(97,15)[]{$\tilde{\tau}_L$}
\Text(172,15)[]{$\tilde{\mu}_L$}
\Text(225,15)[]{$\mu_L$}

\Text(135,38)[]{$(m^2_{\tilde{L}})_{32}$}

\Photon(172,110)(195,135){2}{5}
\Text(203,145)[]{$\gamma$}

\Text(135,-10)[]{(b)}

\end{picture} 

\begin{picture}(455,200)(30,-50)
\ArrowArcn(135,25)(75,90,0)
\ArrowArc(135,25)(75,90,180)
\Vertex(135,100){3}
\Text(135,110)[]{$M_1$}
\Text(78,88)[]{$\tilde{B}$}
\Text(192,88)[]{$\tilde{B}$}

\ArrowLine(30,25)(60,25)
\DashArrowLine(60,25)(110,25){3}             \Vertex(110,25){3}
\DashArrowLine(160,25)(110,25){3}            \Vertex(160,25){3}
\DashArrowLine(210,25)(160,25){3}
\ArrowLine(240,25)(210,25)

\Text(45,15)[]{$\tau_L$}
\Text(85,15)[]{$\tilde{\tau}_L$}
\Text(135,15)[]{$\tilde{\tau}_R$}
\Text(185,15)[]{$\tilde{\mu}_R$}
\Text(225,15)[]{$\mu_R$}

\Text(110,38)[]{$m_{\tau} \mu \tan\beta$}
\Text(160,38)[]{$(m^2_{\tilde{e}})_{32}$}

\Photon(172,110)(195,135){2}{5}
\Text(203,145)[]{$\gamma$}

\Text(135,-10)[]{(c)}
\SetOffset(245,0) 
\ArrowArc(135,25)(75,135,180)
\ArrowArcn(135,25)(75,135,90)
\ArrowArc(135,25)(75,45,90)
\ArrowArcn(135,25)(75,45,0)
\Vertex(188,78){3}
\Vertex(82,78){3}
\Vertex(135,100){3}
\Text(135,110)[]{$v\sin\beta$}
\Text(55,60)[]{$\tilde{H}_1^0$}
\Text(78,88)[]{$\mu$}
\Text(105,105)[]{$\tilde{H}_2^0$}
\Text(160,80)[]{$\tilde{B}$}
\Text(200,90)[]{$M_1$}
\Text(225,60)[]{$\tilde{B}$}

\ArrowLine(30,25)(60,25)
\DashArrowLine(135,25)(60,25){3}             \Vertex(135,25){3}
\DashArrowLine(210,25)(135,25){3}        
\ArrowLine(240,25)(210,25)

\Text(45,15)[]{$\tau_L$}
\Text(97,15)[]{$\tilde{\tau}_R$}
\Text(172,15)[]{$\tilde{\mu}_R$}
\Text(225,15)[]{$\mu_R$}

\Text(135,38)[]{$(m^2_{\tilde{e}})_{32}$}

\Photon(172,110)(195,135){2}{5}
\Text(203,145)[]{$\gamma$}

\Text(135,-10)[]{(d)}

\end{picture} 

\caption{Candidates of the Feynman diagrams which give
dominant contributions to $\tau^+ \to \mu^+ \gamma$ when 
$\tan\beta\gsim1$ and $(m^2_{\tilde{e}})_{32}$ is not negligible.
The arrows represent the chirality.}

\label{fig:SU5tmdiag}
\end{center}
\end{figure}

\begin{figure}[p]
\epsfxsize=12cm

\centerline{\epsfbox{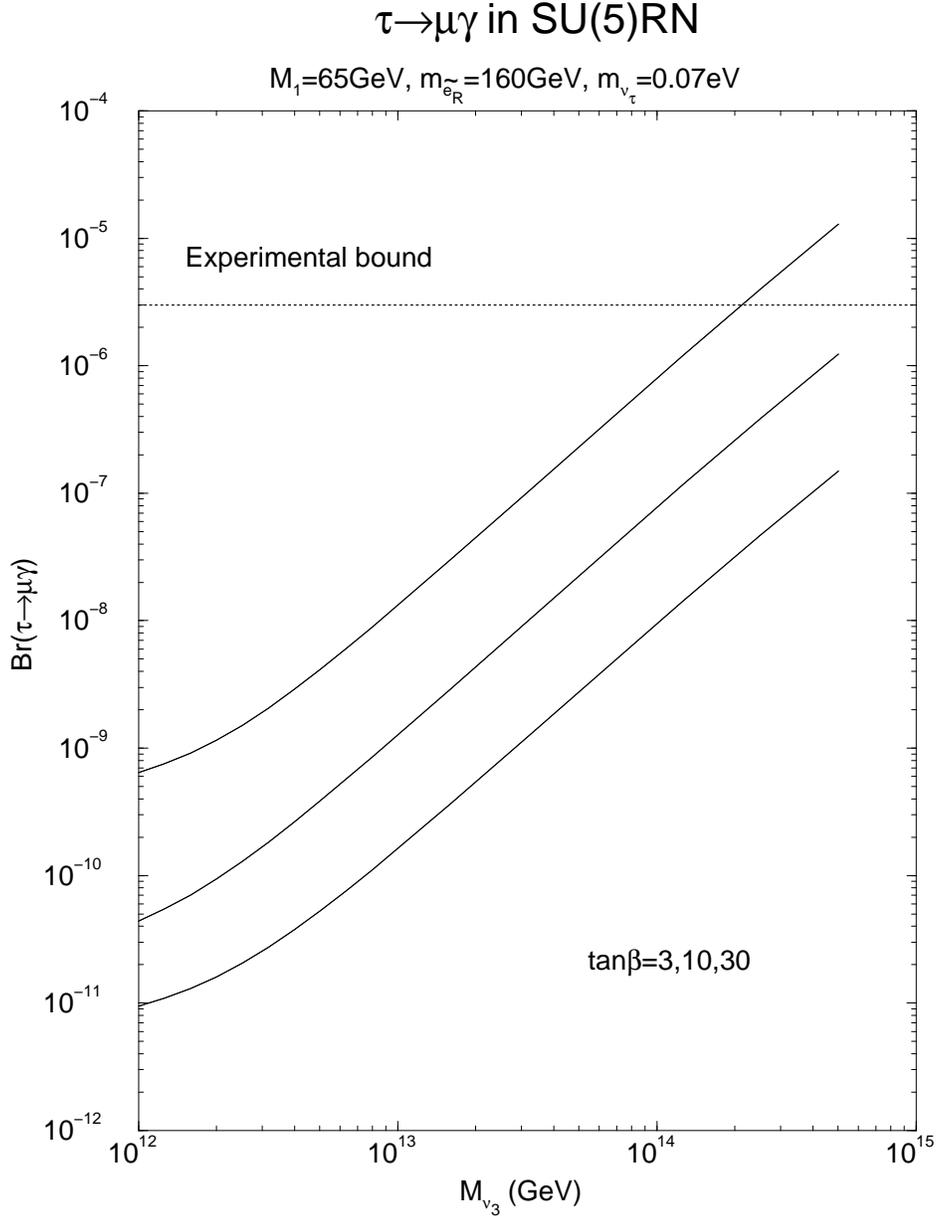}}

\caption{Dependence of the branching ratio of $\tau \to \mu \gamma$ 
on the third-generation right-handed neutrino Majorana mass $M_{\nu_3}$ 
in the SU(5) SUSY GUT with the right-handed neutrinos. 
Here the tau neutrino mass $m_{\nu_\tau}$ is 0.07eV 
and $V_{D 32}=-0.71$, 
as suggested by the atmospheric neutrino result.
We take the bino mass $M_1$ 65GeV, the right-handed selectron mass
$m_{\tilde{e}_R}$ 160GeV. 
The three curves correspond to the case where $\tan\beta=3,10$, and 30,
respectively.
The branching ratio becomes larger for larger $\tan\beta$ value.}

\label{fig:SU5MNvsbrtm}
\end{figure}

Now we study the individual LFV processes.
First we concentrate on the $\tau \to \mu \gamma$ decay.
For $\tau \to \mu \gamma$ the most important contribution is from
diagrams which involves $(m^2_{\tilde{L}})_{32}$ and winos (that is,
diagrams of Figs.~(\ref{fig:SU5tmdiag}) (a) and (b)),
which is the common feature with the MSSMRN case. 
These diagrams dominate because the large $V_{D 32}$ element,
suggested by the atmospheric neutrino anomaly, 
enhances $(m^2_{\tilde{L}})_{32}$ as
\begin{equation}
(m_{\tilde L}^2)_{32} \simeq
-\frac1{8\pi^2} (3m_0^2+a_0^2) 
  V_{D 33}^\ast V_{D 32} f_{\nu_3}^2  \log \frac{M_{\rm grav}}{M_{\nu_3}} ,
\end{equation}
which is the same situation as in the MSSMRN case.
The main difference from the MSSMRN case is a presence of 
$(m_{\tilde{e}})_{32}$, but the contribution to the $\tau\to\mu\gamma$
is too small at the broad parameter region 
to be comparable to those from Figs.~(\ref{fig:SU5tmdiag})(a) and (b),
because $(m_{\tilde{e}})_{32}$ is suppressed by small 
$(V_{\rm KM})_{32}$ \cite{HNY}.  
Our result of numerical calculation, Fig.~(\ref{fig:SU5MNvsbrtm}), indeed
shows that almost the same situation as in the MSSMRN case is realized.
In the figure 
we plot the dependence of the branching ratio of $\tau\to\mu\gamma$ 
on the third generation right-handed Majorana mass $M_{\nu_3}$
for $\tan\beta=3$, 10, and 30. 
The upper curve corresponds to larger $\tan\beta$.
We take the bino mass as 65GeV, the right-handed selectron
mass 160GeV, and the tau neutrino mass 0.07eV, as expected from the
atmospheric neutrino result. We take $a_0=0$ for simplicity.
The figure shows us that the branching ratio of $\tau \to \mu \gamma$
is nearly proportional to the square of $M_{\nu_3}$.
At the right-hand side of each curve the Yukawa coupling 
constant $f_{\nu_3}$ blows up below the gravitational scale, so 
the perturbative treatment is no longer valid in this region. 
In the region near $M_{\nu_3} \simeq 10^{14}$GeV 
the branching ratio is close to
or even beyond the current experimental bound, 
${\rm Br}(\tau\to\mu\gamma)<3.0 \times 10^{-6}$ \cite{PDG}.
At relatively small $M_{\nu_3}$ region 
($M_{\nu_3} \lsim 4\times10^{12}$GeV) 
the contribution from the right-handed slepton mass matrix
(the diagrams shown in Figs.~(\ref{fig:SU5tmdiag})(c) and (d)) 
tends to dominate, and the curves of the branching ratio show deviation from 
simple straight lines. 

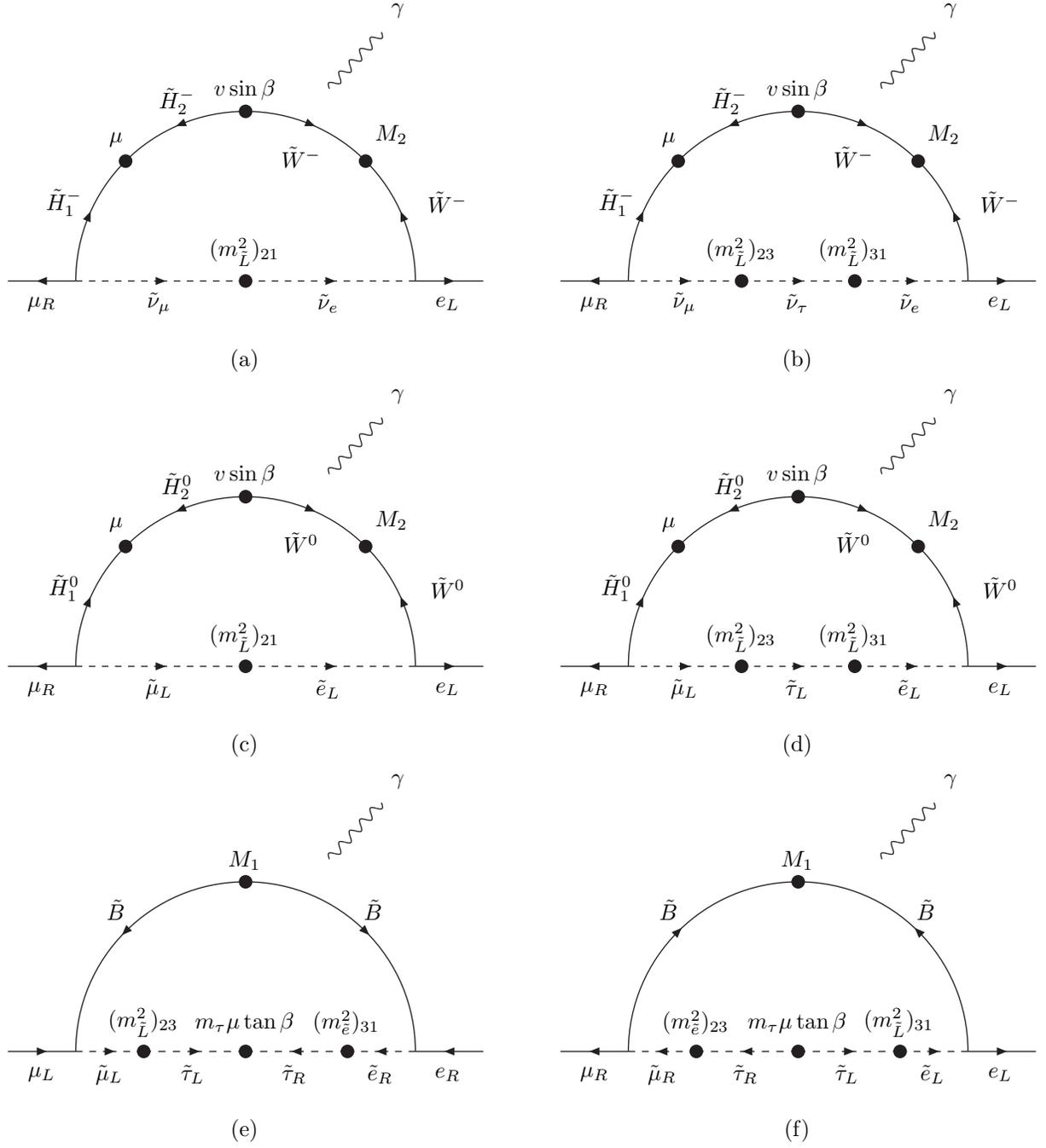
\begin{figure}
\begin{center} 
\begin{picture}(455,140)(30,-20)
\ArrowArcn(135,25)(75,180,135)
\ArrowArc(135,25)(75,90,135)
\ArrowArcn(135,25)(75,90,45)
\ArrowArc(135,25)(75,0,45)
\Vertex(188,78){3}
\Vertex(82,78){3}
\Vertex(135,100){3}
\Text(135,110)[]{$v\sin\beta$}
\Text(55,60)[]{$\tilde{H}_1^-$}
\Text(78,88)[]{$\mu$}
\Text(105,105)[]{$\tilde{H}_2^-$}
\Text(160,80)[]{$\tilde{W}^-$}
\Text(200,90)[]{$M_2$}
\Text(225,60)[]{$\tilde{W}^-$}

\ArrowLine(60,25)(30,25)
\DashArrowLine(60,25)(135,25){3}             \Vertex(135,25){3}
\DashArrowLine(135,25)(210,25){3}        
\ArrowLine(210,25)(240,25)

\Text(45,15)[]{$\mu_R$}
\Text(97,15)[]{$\tilde{\nu}_{\mu}$}
\Text(172,15)[]{$\tilde{\nu}_e$}
\Text(225,15)[]{$e_L$}

\Text(135,38)[]{$(m^2_{\tilde{L}})_{21}$}

\Photon(172,110)(195,135){2}{5}
\Text(203,145)[]{$\gamma$}

\Text(135,-10)[]{(a)}
\SetOffset(245,0)
\ArrowArcn(135,25)(75,180,135)
\ArrowArc(135,25)(75,90,135)
\ArrowArcn(135,25)(75,90,45)
\ArrowArc(135,25)(75,0,45)
\Vertex(188,78){3}
\Vertex(82,78){3}
\Vertex(135,100){3}
\Text(135,110)[]{$v\sin\beta$}
\Text(55,60)[]{$\tilde{H}_1^-$}
\Text(78,88)[]{$\mu$}
\Text(105,105)[]{$\tilde{H}_2^-$}
\Text(160,80)[]{$\tilde{W}^-$}
\Text(200,90)[]{$M_2$}
\Text(225,60)[]{$\tilde{W}^-$}

\ArrowLine(60,25)(30,25)
\DashArrowLine(60,25)(110,25){3}             \Vertex(110,25){3}
\DashArrowLine(110,25)(160,25){3}            \Vertex(160,25){3}
\DashArrowLine(160,25)(210,25){3}        
\ArrowLine(210,25)(240,25)

\Text(45,15)[]{$\mu_R$}
\Text(85,15)[]{$\tilde{\nu}_{\mu}$}
\Text(135,15)[]{$\tilde{\nu}_{\tau}$}
\Text(185,15)[]{$\tilde{\nu}_e$}
\Text(225,15)[]{$e_L$}

\Text(110,38)[]{$(m^2_{\tilde{L}})_{23}$}
\Text(160,38)[]{$(m^2_{\tilde{L}})_{31}$}

\Photon(172,110)(195,135){2}{5}
\Text(203,145)[]{$\gamma$}

\Text(135,-10)[]{(b)}

\end{picture} 

\begin{picture}(455,170)(30,-20)
\ArrowArcn(135,25)(75,180,135)
\ArrowArc(135,25)(75,90,135)
\ArrowArcn(135,25)(75,90,45)
\ArrowArc(135,25)(75,0,45)
\Vertex(188,78){3}
\Vertex(82,78){3}
\Vertex(135,100){3}
\Text(135,110)[]{$v\sin\beta$}
\Text(55,60)[]{$\tilde{H}_1^0$}
\Text(78,88)[]{$\mu$}
\Text(105,105)[]{$\tilde{H}_2^0$}
\Text(160,80)[]{$\tilde{W}^0$}
\Text(200,90)[]{$M_2$}
\Text(225,60)[]{$\tilde{W}^0$}

\ArrowLine(60,25)(30,25)
\DashArrowLine(60,25)(135,25){3}             \Vertex(135,25){3}
\DashArrowLine(135,25)(210,25){3}        
\ArrowLine(210,25)(240,25)

\Text(45,15)[]{$\mu_R$}
\Text(97,15)[]{$\tilde{\mu}_L$}
\Text(172,15)[]{$\tilde{e}_L$}
\Text(225,15)[]{$e_L$}

\Text(135,38)[]{$(m^2_{\tilde{L}})_{21}$}

\Photon(172,110)(195,135){2}{5}
\Text(203,145)[]{$\gamma$}

\Text(135,-10)[]{(c)}
\SetOffset(245,0)
\ArrowArcn(135,25)(75,180,135)
\ArrowArc(135,25)(75,90,135)
\ArrowArcn(135,25)(75,90,45)
\ArrowArc(135,25)(75,0,45)
\Vertex(188,78){3}
\Vertex(82,78){3}
\Vertex(135,100){3}
\Text(135,110)[]{$v\sin\beta$}
\Text(55,60)[]{$\tilde{H}_1^0$}
\Text(78,88)[]{$\mu$}
\Text(105,105)[]{$\tilde{H}_2^0$}
\Text(160,80)[]{$\tilde{W}^0$}
\Text(200,90)[]{$M_2$}
\Text(225,60)[]{$\tilde{W}^0$}

\ArrowLine(60,25)(30,25)
\DashArrowLine(60,25)(110,25){3}             \Vertex(110,25){3}
\DashArrowLine(110,25)(160,25){3}            \Vertex(160,25){3}
\DashArrowLine(160,25)(210,25){3}        
\ArrowLine(210,25)(240,25)

\Text(45,15)[]{$\mu_R$}
\Text(85,15)[]{$\tilde{\mu}_L$}
\Text(135,15)[]{$\tilde{\tau}_L$}
\Text(185,15)[]{$\tilde{e}_L$}
\Text(225,15)[]{$e_L$}

\Text(110,38)[]{$(m^2_{\tilde{L}})_{23}$}
\Text(160,38)[]{$(m^2_{\tilde{L}})_{31}$}

\Photon(172,110)(195,135){2}{5}
\Text(203,145)[]{$\gamma$}

\Text(135,-10)[]{(d)}

\end{picture} 

\begin{picture}(455,200)(30,-50)
\ArrowArcn(135,25)(75,90,0)
\ArrowArc(135,25)(75,90,180)
\Vertex(135,100){3}
\Text(135,110)[]{$M_1$}
\Text(78,88)[]{$\tilde{B}$}
\Text(192,88)[]{$\tilde{B}$}

\ArrowLine(30,25)(60,25)
\DashArrowLine(60,25)(90,25){3}             \Vertex(90,25){3}
\DashArrowLine(90,25)(135,25){3}            \Vertex(135,25){3}
\DashArrowLine(180,25)(135,25){3}           \Vertex(180,25){3}
\DashArrowLine(210,25)(180,25){3}
\ArrowLine(240,25)(210,25)

\Text(45,15)[]{$\mu_L$}
\Text(75,15)[]{$\tilde{\mu}_L$}
\Text(112,15)[]{$\tilde{\tau}_L$}
\Text(157,15)[]{$\tilde{\tau}_R$}
\Text(195,15)[]{$\tilde{e}_R$}
\Text(225,15)[]{$e_R$}

\Text(135,38)[]{$m_\tau \mu \tan \beta$}
\Text(90,38)[]{$(m^2_{\tilde{L}})_{23}$}
\Text(180,38)[]{$(m^2_{\tilde{e}})_{31}$}

\Photon(172,110)(195,135){2}{5}
\Text(203,145)[]{$\gamma$}

\Text(135,-10)[]{(e)}

\SetOffset(245,0)
\ArrowArc(135,25)(75,0,90)
\ArrowArcn(135,25)(75,180,90)
\Vertex(135,100){3}
\Text(135,110)[]{$M_1$}
\Text(78,88)[]{$\tilde{B}$}
\Text(192,88)[]{$\tilde{B}$}

\ArrowLine(60,25)(30,25)
\DashArrowLine(90,25)(60,25){3}             \Vertex(90,25){3}
\DashArrowLine(135,25)(90,25){3}            \Vertex(135,25){3}
\DashArrowLine(135,25)(180,25){3}           \Vertex(180,25){3}
\DashArrowLine(180,25)(210,25){3}
\ArrowLine(210,25)(240,25)

\Text(45,15)[]{$\mu_R$}
\Text(75,15)[]{$\tilde{\mu}_R$}
\Text(112,15)[]{$\tilde{\tau}_R$}
\Text(157,15)[]{$\tilde{\tau}_L$}
\Text(195,15)[]{$\tilde{e}_L$}
\Text(225,15)[]{$e_L$}

\Text(135,38)[]{$m_\tau \mu \tan \beta$}
\Text(90,38)[]{$(m^2_{\tilde{e}})_{23}$}
\Text(180,38)[]{$(m^2_{\tilde{L}})_{31}$}

\Photon(172,110)(195,135){2}{5}
\Text(203,145)[]{$\gamma$}

\Text(135,-10)[]{(f)}

\end{picture} 

\caption{Candidates of the Feynman diagrams which give 
dominant contributions to $\mu^+ \to e^+ \gamma$
when $\tan\beta\gsim1$ and the off-diagonal elements of 
$(m^2_{\tilde{e}})$ are non-negligible.
The arrows represent the chirality.}

\label{fig:SU5mediag}

\end{center}
\end{figure}

Next the $\mu \to e \gamma$ process. 
Since we know $V_{D 32}={\cal O}(1)$ from the atmospheric neutrino result 
we can calculate $(m^2_{\tilde L})_{23}$ as
\begin{equation}
(m^2_{\tilde L})_{23} \simeq
 - \frac{1}{8\pi^2} f_{\nu_3}^2 V^{\ast}_{D 32}  V_{D 33}
   (3m^2_0 +a^2_0) \log \frac{M_{\rm grav}}{M_{\nu_3}} .
\end{equation}
On the other hand $(m^2_{\tilde e})_{31}$ is determined from the GUT 
symmetry as
\begin{equation}
(m^2_{\tilde e})_{31} \simeq
 - \frac{3}{8\pi^2} f_{u_3}^2   (V_{\rm KM})_{33} (V^\ast_{\rm KM})_{31}
   (3m^2_0 +a^2_0) \log \frac{M_{\rm grav}}{M_{\rm GUT}}.
\end{equation}
Then we can definitely calculate the diagram shown 
in Fig.~(\ref{fig:SU5mediag})(e), which contributes to 
$A^{(\mu e)}_R$ defined in Eq.~(\ref{Penguin}).
This contribution is proportional to $m_\tau$, 
and it tends to dominate over
the other contribution to $A_R^{(\mu e)}$.
We need to know $V_{D31}$ and $V_{D21}$ to calculate $A_L^{(\mu e)}$.
However, we can evaluate only the lower bound of the branching ratio
from the structure given in Eq.~(\ref{eventrate}) even if we do not know
them.
The contribution from the diagram (e) 
can be so large that it reaches
the present experimental bound at some parameter regions \cite{HNY}.

We can imagine some cases where much larger rate than this lower bound 
is predicted by the contribution from $A^{(\mu e)}_L$. 
One of such cases is that $V_{D31}$ is large.
In this case, the diagrams in Figs.~(\ref{fig:SU5mediag})
(a)-(d) and (f) are enhanced since 
$(m^2_{\tilde L})_{31}$ and $(m^2_{\tilde L})_{21}$ 
are proportional to $V_{D31}$ as
\begin{eqnarray}
(m^2_{\tilde{L}})_{31} 
&\simeq& - \frac{1}{8\pi^2} 
  (3m_0^2+a_0^2) 
  f_{\nu_3}^2 V^{\ast}_{D 33} V_{D 31} 
 \log \frac{M_{\rm grav}}{M_{\nu_3}} , \\
(m^2_{\tilde{L}})_{21} 
&\simeq& - \frac{1}{8\pi^2}
  (3m_0^2+a_0^2) 
  f_{\nu_3}^2 V^{\ast}_{D 32} V_{D 31} 
 \log \frac{M_{\rm grav}}{M_{\nu_3}} .
\label{21elementofmL}
\end{eqnarray}
While the diagram (f) is enhanced by $m_\tau$, the contribution is
suppressed by $(m_{\tilde{e}}^2)_{32}$, which is proportional to
$(V_{\rm KM})_{32}$. 
As a result, the diagrams (a)-(d) dominate since
$V_{D32}$ is large.  Also, if $f_{\nu_2}$ and $V_{D21}$ are
non-negligibly large, $(m_{\tilde{e}}^2)_{21}$ is enhanced as 
\begin{equation}
(m^2_{\tilde{L}})_{21} 
\simeq - \frac{1}{8\pi^2}
 \left(
  f_{\nu_3}^2 V^{\ast}_{D 32} V_{D 31} 
 \log \frac{M_{\rm grav}}{M_{\nu_3}} 
+ f_{\nu_2}^2 V^{\ast}_{D 22} V_{D 21} 
 \log \frac{M_{\rm grav}}{M_{\nu_2}} 
 \right)
(3m_0^2+a_0^2), 
\label{21elementofmL2}
\end{equation}
and the diagrams (a) and (c) dominate over the other contributions.

\begin{figure}[p]
\epsfxsize=12cm
\centerline{\epsfbox{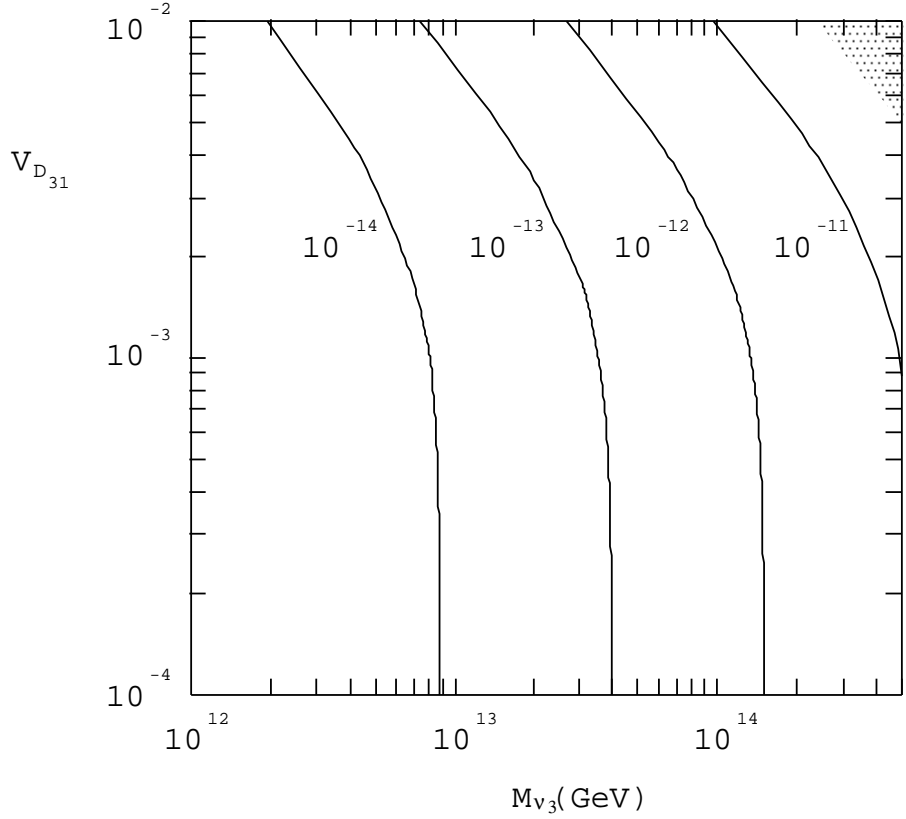}}

\caption{Dependence of the branching ratio of $\mu \to e \gamma$ 
on the third-generation right-handed neutrino Majorana mass 
$M_{\nu_3}$ and $V_{D 31}$
in the SU(5) SUSY GUT with the right-handed neutrinos.  
Here we take the tau neutrino mass $m_{\nu_\tau}$ 0.07eV
and the mu neutrino mass is neglected.
The curves mean the contours on which
the branching ratio of $\mu\to e\gamma$
is $10^{-14}, 10^{-13}, 10^{-12}$, and $10^{-11}$, respectively.
The shaded region is already excluded experimentally.
We take the bino mass $M_1$ 65GeV, the right-handed selectron mass
$m_{\tilde{e}_R}$ 160GeV, and $\tan\beta=3$.}

\label{fig:SU5con3f20}
\end{figure}

\begin{figure}[t]
\epsfxsize=12cm
\centerline{\epsfbox{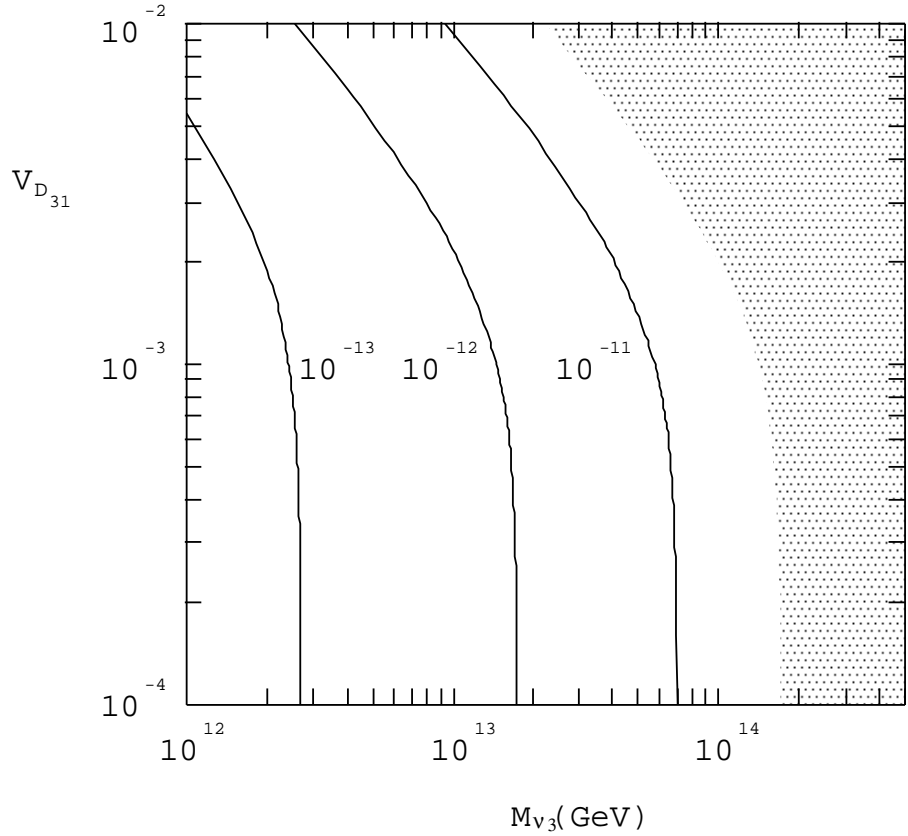}}

\caption{Dependence of the branching ratio of $\mu \to e \gamma$ 
on the third-generation right-handed neutrino Majorana mass 
$M_{\nu_3}$ and $V_{D 31}$
in the SU(5) SUSY GUT with the right-handed neutrinos.   
The curves mean the contours on which
the branching ratio of $\mu\to e\gamma$
is $10^{-13}, 10^{-12}$, and $10^{-11}$, respectively.
The shaded region is already excluded experimentally.
The input parameters are the same as those in Fig.~(13)
except that in this figure $\tan\beta=30$.}

\label{fig:SU5con30f20}
\end{figure}

Now we examine these expectations numerically.
First we set $f_{\nu_2}$ to zero 
and later investigate the non-zero $f_{\nu_2}$ case.
We calculate the dependence of the branching ratio of $\mu \to e \gamma$ 
on $M_{\nu_3}$ and $V_{D 31}$ in the Figs.~(\ref{fig:SU5con3f20},
\ref{fig:SU5con30f20}).
In the figures we take $V_{D 32}=-1/\sqrt{2}$ and
the tau neutrino mass 0.07eV, as suggested 
by the atmospheric neutrino result.
We neglect $m_{\nu_\mu}$ here.
Other parameters are the bino mass 65GeV, 
the right-handed selectron mass 160GeV,
$\tan\beta=3$ and 30, and the Higgsino mass $\mu >0$.
From the Figs.~(\ref{fig:SU5con3f20},\ref{fig:SU5con30f20}) we can see that 
for $V_{D 31} \lsim 10^{-3}$
the diagram (e) dominates over the others. 
At relatively larger $V_{D 31}$ region ($V_{D 31} \gsim 10^{-2.5}$)
the diagrams (a)-(d) become dominant, and the predicted rate is large enough
to reach the experimental bound for $V_{D 31} \simeq 10^{-2}$ 
if $\tan\beta=30(3)$ and $M_{\nu_3} \gsim 10^{13.5}(10^{14.5})$GeV.

\begin{figure}[t]
\epsfxsize=12cm
\centerline{\epsfbox{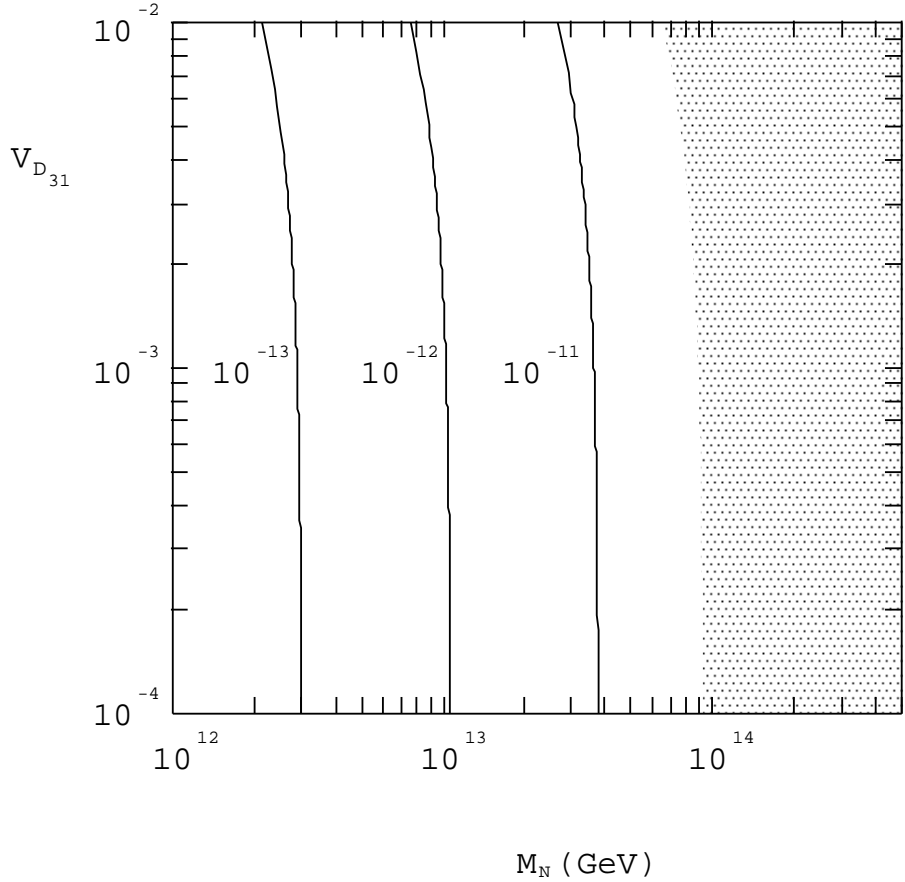}}

\caption{Dependence of the branching ratio of $\mu \to e \gamma$
on the typical right-handed neutrino Majorana mass $M_N$ and $V_{D 31}$
in the SU(5) SUSY GUT with the right-handed neutrinos.   
We assume the MSW large angle solution, 
which suggests $m_{\nu_\mu}$ to be 0.004eV and $V_{D 21}=-0.42$. 
We take the tau neutrino mass $m_{\nu_\tau}$ 0.07eV 
and $V_{D 32}=-0.60$, as suggested by the atmospheric neutrino result.
We assume the universality of the right-handed Majorana masses
$M_{\nu 1}=M_{\nu 2}=M_{\nu 3}\equiv M_N$, for simplicity.
The curves mean the contours on which
the branching ratio of $\mu\to e\gamma$
is $10^{-13}, 10^{-12}$, and $10^{-11}$, respectively.
The shaded region is already excluded experimentally.
We take the bino mass $M_1$ 65GeV, the right-handed selectron mass
$m_{\tilde{e}_R}$ 160GeV. 
In this figure $\tan\beta=3$.}

\label{fig:SU5con3Ml}
\end{figure}

\begin{figure}[t]
\epsfxsize=12cm
\centerline{\epsfbox{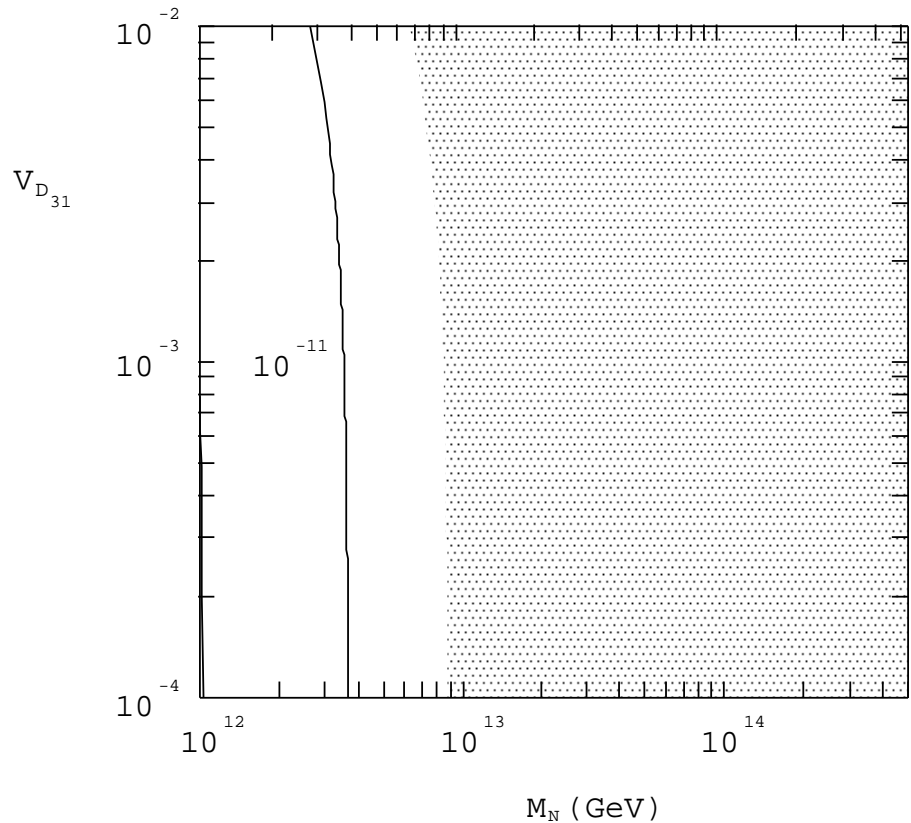}}

\caption{Dependence of the branching ratio of $\mu \to e \gamma$ 
on the typical right-handed neutrino Majorana mass $M_N$ and $V_{D 31}$ 
in the SU(5) SUSY GUT with the right-handed neutrinos. 
We assume the MSW large angle solution and the atmospheric neutrino result.  
All the input parameters are the same as those in Fig.~(15)
except that we take $\tan\beta=30$ in this figure.
The curve means the contour on which  
the branching ratio of $\mu\to e\gamma$ is $10^{-11}$.
The shaded region is already excluded experimentally.}

\label{fig:SU5con30Ml}

\end{figure}

Next let us consider the finite $f_{\nu_2}$ case.
We show in Figs.~(\ref{fig:SU5con3Ml}, \ref{fig:SU5con30Ml})
the dependence of the $\mu\to e\gamma$ branching ratio on $V_{D 31}$
and the typical right-handed neutrino Majorana mass $M_N$.
In the figures the parameters that describe 
the MSW large angle solution are taken as the same as those we used in the
MSSMRN case, and other parameters are taken as the same 
as in Figs.~(\ref{fig:SU5con3f20}, \ref{fig:SU5con30f20}).
We assume in the figure the universality of the right-handed Majorana
mass, that is, $M_{\nu_1}=M_{\nu_2}=M_{\nu_3}(\equiv M_N)$.
We can see from the figures that 
the enhancement due to the MSW large angle solution is so large that
the dependence of the rate on $V_{D 31}$ is small,
and that the almost same situation 
as in the MSSMRN is reproduced here again,
as can be seen when comparing these figures with Fig.~(\ref{fig:MSSMmeMl}).
Since the dominant contribution is determined by the
second term of Eq.~(\ref{21elementofmL2}),
the rate is almost determined from the value of $M_{\nu_2}$
because $f^2_{\nu_2} \propto M_{\nu_2}$ for fixed $m_{\nu_\mu}$.
Hence we can conclude from the Figs.~(\ref{fig:SU5con3Ml},\ref{fig:SU5con30Ml})
that for $\tan\beta=30(3)$ 
the excluded region of $M_{\nu_2}$ 
extends to $9\times10^{12}(1\times10^{14})$GeV, at least 
for the parameters chosen in our calculation. 

We also calculated for the parameters suggested from 
the MSW small angle solution and the 'just so' solution.
For the MSW small angle solution the difference from the $f_{\nu_2}=0$ case
is at most the factor 2 enhancement, and for the 'just so' solution 
no difference from the $f_{\nu_2}=0$ case can be seen. 

Finally we comment on the $b \to s \gamma$ process.
This model has a characteristic feature that 
the right-handed neutrinos couple to 
the right-handed down-type quarks with the large mixing
through the ninth term of Eq.~(\ref{SU5RNdecomp}).
This coupling induces the off-diagonal elements in
the right-handed down-type squark soft mass matrix
via the radiative corrections, which are as large as
\begin{equation}
(m^2_{\tilde{d}_R})_{ij} \simeq - \frac{1}{8\pi^2} f_{\nu_k}^2 
  V^{\ast}_{D kj} V_{D ki} (3 m_0^2 + a_0^2) 
  \log \frac{M_{\rm grav}}{M_{\rm GUT}} .
\end{equation}
This fact causes us an expectation
that the $b \to s \gamma$ rate may become larger by the effect of 
the diagram which contains $(m^2_{\tilde{d}_R})_{23}$,  
by an exact analogy to that of the lepton sector.
We examined whether this is true or not, and concluded 
that a tiny enhancement indeed occurs but it is too small to 
experimentally distinguishable.
This is because the heavy gluino suppresses the contribution.

\section{Conclusion}
In this article taking the solar and the atmospheric neutrino 
experiment results into account
we investigated the lepton flavor
violating decay processes, such as $\mu \to e \gamma$ or $\tau\to \mu\gamma$
in the MSSM with the right-handed neutrinos (MSSMRN)
and in the SU(5) SUSY GUT with the right-handed neutrinos (SU(5)RN). 

In the MSSMRN we first studied
the branching ratio of $\tau\to\mu\gamma$.
It gets larger for larger $\tan\beta$ and
almost proportional to $M_{\nu_3}^2$ for fixed $m_{\nu_\tau}$.
A large rate naturally follows from
the large mixing between $\nu_\tau$ and $\nu_\mu$,
suggested by the atmospheric neutrino result.
If $\tan\beta$=30(10) the rate reaches the current experimental bound
for $M_{\nu_3} \sim 2\times 10^{14}(6\times 10^{14})$GeV.
We investigated the $\mu\to e\gamma$ rate 
under three kinds of the solar neutrino solutions,
the MSW large and small angle solutions and the 'just so' solution.
We argued that the $\mu\to e\gamma$ depends on 
$f_{\nu_2}$, and that especially 
the large $V_{D21}$ and the large $f_{\nu_2}$, 
which the MSW large angle solution suggests,
naturally results in such a large rate that 
the predicted rate is beyond the current experimental bound
if $M_{\nu_2}$ is larger than $8 \times 10^{12}(8 \times 10^{13})$GeV 
for $\tan\beta=30(3)$.
We also investigated the dependence of $\mu\to e\gamma$ on 
$V_{D31}$. For $\tan\beta=30$ and $V_{D31}=10^{-2}$,
$M_{\nu_3}\gsim 3\times 10^{13}$GeV is excluded for our input parameters.

In the SU(5)RN we calculated 
the $\tau\to\mu\gamma$ rate.
Here also the large mixing between $\nu_\tau$ and $\nu_\mu$ 
leads to a large rate, 
which is almost the same situation as in the MSSMRN.
For $\mu\to e\gamma$ in this model, we can predict the lower bound 
of the branching ratio of it.
This lower bound is calculable 
from the value of $m_{\nu_\tau}$ and 
the large mixing angle between $\nu_\tau$-$\nu_\mu$,
suggested from the atmospheric neutrino result, 
and it turns out to be within accessible region by near future experiments.
They are supposed to probe the $\mu \to e\gamma$ 
branching ratio to $10^{-14}$ level \cite{kuno}, and then 
the region $M_{\nu_3}>10^{13}(10^{12})$GeV can be probed for 
$\tan\beta=3(30)$ and $m_{\nu_\tau}=0.07$eV.
We considered the relation between $\mu\to e\gamma$ and 
three kinds of the solar neutrino solutions.
The large rate naturally follows from the MSW large angle solution,
similarly to the MSSMRN case,
while in the MSW small angle solution and in the 'just so' solution 
there is only a little difference from the $f_{\nu_2}=0$ case.

If the LFV processes are discovered or the experimental bounds are improved by near future experiments,
the interesting insight on the lepton sector will be obtained. 
The best effort to implement it is strongly desired.

\section*{Acknowledgments}
The authors would like to thank K.~Tobe for his participation to the
preparation of Appendices B and C at the early stage.  One of the
authors (D.~N.) would like to express his gratitude to T.~Yanagida for
his continuous support, and thanks the members of the theory group of
KEK for the hospitality extended to him during his stay at KEK.  This
work is partially supported by Grants-in-aid for Science and Culture
Research for Monbusho, No.10740133 and "Priority Area: Supersymmetry
and Unified Theory of Elementary Particles (\#707)".

\newpage
\appendix

\section{Definitions and Conventions in the MSSM}
\label{mssm}
In this appendix we collect our notations of the MSSM used in our article.
The superpotential of the MSSM $W_{\rm MSSM}$ is defined as \footnote{
Supersymmetric couplings and masses of chiral multiplets are given as
\begin{equation}
{-\cal L}= \int d^2 \theta ~W + h.c.. \nonumber 
\end{equation}
Our convention is unusual in order to keep the chargino mass matrix
in accordance with the Haber-Kane convention \cite{HaberKane}. }
\footnote{
We implicitly assume the contraction convention over 
SU(2)$_L$ doublet indices $(a,b,..=1,2)$ 
of two doublets $A$ and $B$ 
\begin{equation}
A = \left( \begin{array}{c} A^1 \\ A^2 \end{array} \right) , \, \, 
B = \left( \begin{array}{c} B^1 \\ B^2 \end{array} \right) 
\end{equation}
as
\begin{equation}
AB \equiv \epsilon_{ab} A^a B^b ,
\end{equation}
where $\epsilon_{ab}$ is an antisymmetric tensor with 
$\epsilon_{12}=-\epsilon_{21}=1$. 
}
\begin{equation}
W_{\rm MSSM} \equiv f_{e_{ij}} H_1 \overline{E}_iL_j
  +f_{d_{ij}} H_1 Q_i\overline{D}_j
  +f_{u_{ij}} H_2 Q_i\overline{U}_j 
  +{\mu} H_1 H_2 . \label{MSSMsuppot}
\end{equation}
We use $i$ and $j$ as the generation indices running from 1 to 3.
$\overline{E},\overline{D}$, and $\overline{U}$ are the superfields
associated with the right-handed electron $e_R$, 
the right-handed down-type quark $d_R$,
and the right-handed up-type quark $u_R$, respectively. 
$Q$ and $L$ are ones associated with the quark doublet $q_L$ and 
the lepton doublet $l_L$ defined by
\begin{equation}
q_L=\left( \begin{array}{c} u_L \\ d_L \end{array} \right) , \, \, 
l_L=\left( \begin{array}{c} \nu \\ e_L \end{array} \right) ,
\end{equation}
and $H_1$ and $H_2$ the Higgs doublets whose SU(2)$_L$ components 
we denote as
\begin{equation}
H_1=\left( \begin{array}{c} 
H^0_1 \\ H^-_1 \end{array} \right) , \, \, 
H_2=\left( \begin{array}{c} 
H^+_2 \\ H^0_2 \end{array} \right) .
\end{equation}
The scalar components of $H_1^0$ and $H_2^0$ develop the vacuum expectation
values (vev's) as
\begin{equation}
\langle H_1^0 \rangle \equiv \frac{v_1}{\sqrt{2}} , \ \ \ 
\langle H_2^0 \rangle \equiv \frac{v_2}{\sqrt{2}} ,
\end{equation}
which satisfy $v_1^2+v_2^2 = v^2$ with $v \simeq$ 246 GeV. 
We define the ratio of these vev's as $\tan\beta$,
\begin{equation}
\tan\beta \equiv \frac{v_2}{v_1} .
\end{equation}
The Yukawa couplings are given by
the fermion masses and the Kobayashi-Maskawa matrix as
\begin{eqnarray}
f_{e_{ij}} &=& -\sqrt{2} \frac{m_{e_i}}{v \cos\beta} \delta_{ij},
\nonumber\\
f_{d_{ij}} &=& -\sqrt{2} \frac{m_{d_i}}{v \cos\beta} \delta_{ij},
\nonumber\\
f_{u_{ij}} &=&  \sqrt{2} \frac{m_{u_i}}{v \sin\beta} (V_{\rm KM})_{ij}.
\end{eqnarray}

We also introduce the soft SUSY breaking terms in the Lagrangian,
\begin{eqnarray}
-{\cal L}_{\rm \ SUSY \ breaking}&=& \phantom{+}
 (m_{\tilde L}^2)_{ij} \tilde{l}_{Li}^{\dagger} \tilde{l}_{Lj} 
+(m_{\tilde e}^2)_{ij} \tilde{e}_{Ri}^\ast \tilde{e}_{Rj}  
\nonumber\\
&&
+(m_{\tilde Q}^2)_{ij} \tilde{q}_{Li}^{\dagger} \tilde{q}_{Lj} 
+(m_{\tilde u}^2)_{ij} \tilde{u}_{Ri}^\ast \tilde{u}_{Rj}  
+(m_{\tilde d}^2)_{ij} \tilde{d}_{Ri}^\ast \tilde{d}_{Rj}  
\nonumber\\
&& +\tilde{m}_{h1}^2 h_1^\dagger h_1  
   +\tilde{m}_{h2}^2 h_2^\dagger h_2  
\nonumber \\
&&
 + (A_u^{ij} h_2  \tilde{q}_{Li} \tilde{u}^\ast_{Rj}
  + A_d^{ij} h_1  \tilde{q}_{Li} \tilde{d}^\ast_{Rj}
  + A_e^{ij} h_1 \tilde{e}^\ast_{Ri} \tilde{l}_{Lj} \nonumber \\
&&
  \phantom{+}   +B_{h} h_1h_2 + h.c.) \nonumber \\
&& 
 +(\frac12 M_1 \tilde{B}_L \tilde{B}_L
 + \frac12 M_2 \tilde{W}^a_L \tilde{W}^a_L 
 + \frac12 M_3 \tilde{g}^a_L \tilde{g}^a_L
   + h.c.) .
\end{eqnarray}
Here the first seven terms are the soft SUSY breaking masses for 
the doublet-slepton $\tilde{l}_L$, the right-handed charged slepton 
$\tilde{e}_R$, the doublet-squark $\tilde{q}_L$, 
the right-handed up-type (down-type) squark $\tilde{u}_R \ (\tilde{d}_R)$, 
and the Higgs bosons.
$\tilde{B}$, $\tilde{W}$, and $\tilde{g}$ stand
for bino, wino, and gluino respectively, and the superscript $a$ is
the gauge group index for each corresponding gauge group.

Now we discuss the slepton mass matrices.
Let $\tilde{e}_{Li}$ and $\tilde{e}_{Ri}$ be the superpartners of 
the left-handed electron $e_{Li}$ and 
the right-handed electron $e_{Ri}$, respectively.
Then the slepton mass matrix $(\overline{m}^2_{\tilde{e}})_{ij}$ is 
\begin{equation}
-{\cal L}=
\left( \tilde{e}_{Li}^\dagger ,  \tilde{e}_{Ri}^\dagger  \right)
(\overline{m}^2_{\tilde{e}})_{ij}
\left( \begin{array}{c} \tilde{e}_{Lj} \\  \tilde{e}_{Rj} \end{array} \right) 
\equiv
\left( \tilde{e}_{Li}^\dagger ,  \tilde{e}_{Ri}^\dagger  \right)
\left( 
\begin{array}{cc} 
(m^2_L)_{ij}    & (m^{2\top}_{LR})_{ij} \\
(m^2_{LR})_{ij} & (m^2_{R})_{ij} 
\end{array} \right)
\left( \begin{array}{c} \tilde{e}_{Lj} \\  \tilde{e}_{Rj} \end{array} \right) .
\end{equation}
Here $m^2_L$ and $m^2_R$ are $3\times 3$ hermitian matrices and 
$m^2_{LR}$ is a $3\times 3$ matrix.
They are given as 
\begin{eqnarray}
(m_L^2)_{ij} &=&
(m_{\tilde L}^2)_{ij} + m_{e_i}^2 \delta_{ij}
+ m_Z^2 \delta_{ij} \cos 2 \beta (-\frac12 + \sin^2 \theta_W) ,  \\
(m_R^2)_{ij} &=&
(m_{\tilde e}^2)_{ij}+ m_{e_i}^2 \delta_{ij} 
- m_Z^2 \delta_{ij} \cos 2 \beta  \sin^2 \theta_W  , \\
(m_{LR}^2)_{ij} &=&
 A^{ij}_e v \cos\beta /\sqrt{2} - m_{e_i} \mu \delta_{ij} \tan\beta  .
\end{eqnarray}

As for the neutrinos we should notice that there is no
right-handed sneutrino in the MSSM.  Let $\tilde \nu_{i}$ be the
superpartner of the left-handed neutrino $\nu_{i}$.  The mass matrix of 
sneutrino $(\overline{m}^2_{\tilde{\nu}})_{ij}$ is 
\begin{eqnarray}
-{\cal L}&=& 
\tilde{\nu}^\dagger_i (\overline{m}^2_{\tilde{\nu}})_{ij} 
\tilde{\nu}_j                 ,                           \nonumber \\
(\overline{m}^2_{\tilde{\nu}})_{ij} &=&
(m_{\tilde L}^2 )_{ij}
+ \frac12 m_Z^2 \delta_{ij} \cos 2 \beta   .
\end{eqnarray}

Next we discuss the mass matrix of gauginos. First we consider chargino
mass matrix ${\cal M}_C$. It is a $2\times 2$ matrix that appears in  
the chargino mass terms,
\begin{equation}
  -{\cal L}_m =
     \left ( \overline{\tilde W^-_R}~ \overline{\tilde H^-_{2R}} \right )
     \left (  \begin{array}{cc}
                M_2            & \sqrt{2} m_W \cos \beta \\
                \sqrt{2}m_W \sin \beta &  \mu
              \end{array}                                 \right)
     \left (  \begin{array}{c}
              \tilde W^-_L    \\ \tilde H_{1L}^-
              \end{array}                                 \right) +h.c..
\end{equation}
${\cal M}_C$ is diagonalized by $2 \times 2$ real 
orthogonal matrices $O_L$ and $O_R$ as
\begin{equation}
     O_R {\cal M}_C O_L^\top =
{\rm diag}(M_{\tilde{\chi}_1^-},M_{\tilde{\chi}_2^-}).
\end{equation}
Define the mass eigenstates $\tilde{\chi}_{A L(R)}$  $(A=1,2)$ by
\begin{equation}
   \left( \begin{array}{c}
           \tilde \chi^-_{1L} \\
           \tilde \chi^-_{2L}
           \end{array}                 \right)
  =O_L   \left( \begin{array}{c}
                \tilde W^-_L  \\
                \tilde H^-_{1L}
                \end{array}            \right),
\hspace{1.5cm}
    \left( \begin{array}{c}
           \tilde \chi^-_{1R} \\
           \tilde \chi^-_{2R}
           \end{array}                 \right)
  =O_R   \left( \begin{array}{c}
                \tilde W^-_R  \\
                \tilde H^-_{2R}
                \end{array}            \right).
\end{equation}
Then
\begin{equation}
    \tilde \chi^-_A  =\tilde \chi^-_{AL} + \tilde \chi^-_{AR}
\hspace{1.5cm}
    (A=1,2)
\end{equation}
forms a Dirac fermion with mass $M_{\tilde \chi^-_A}$.

Finally we consider neutralinos.  The mass matrix of the neutralino
sector is given by
\begin{equation}
 -{\cal L}_m =
  \frac{1}{2}
  \left( \tilde B_L \tilde W^0_L \tilde H^0_{1L} \tilde H^0_{2L}
  \right)
   {\cal M}_N
   \left(   \begin{array}{c}
            \tilde B_L  \\
            \tilde W^0_L \\
            \tilde H^0_{1L} \\
            \tilde H^0_{2L}
            \end{array}                  \right)  +h.c.,
\end{equation}
where
\begin{equation}
   {\cal M}_N=
   \left(
   \begin{array}{cccc}
     M_1    & 0 & -m_Z\sin\theta_W\cos\beta & m_Z\sin\theta_W\sin\beta \\
     0 & M_2 & m_Z\cos\theta_W\cos\beta & -m_Z\cos\theta_W\sin\beta \\
     -m_Z\sin\theta_W\cos\beta & m_Z\cos\theta_W\cos\beta & 0 & -\mu
     \\
     m_Z\sin\theta_W\sin\beta & -m_Z\cos\theta_W\sin\beta & -\mu & 0
   \end{array}            \right).
\end{equation}
The diagonalization is done by a real orthogonal matrix $O_N$,
\begin{equation}
    O_N {\cal M}_N O_N^\top = 
{\rm diag}(M_{\tilde{\chi}^0_1},...,M_{\tilde{\chi}^0_4}).
\end{equation}
The mass eigenstates are given by
\begin{equation}
         \tilde \chi^0_{AL} =(O_N)_{AB} \tilde X^0_{BL}
\hspace{1.5cm}
           (A,B=1, \cdots ,4)
\end{equation}
where
\begin{equation}
   \tilde X^0_{AL} = ( \tilde B_L, \tilde W^0_L, \tilde H^0_{1L},
   \tilde H^0_{2L}).
\end{equation}
We have thus Majorana spinors
\begin{equation}
   \tilde \chi^0_A = \tilde \chi^0_{AL} + \tilde \chi^0_{AR},
{}~~~~(A=1, \cdots ,4)
\end{equation}
with mass $M_{\tilde \chi^0_A}$.

Now we give the interaction Lagrangian 
of lepton-slepton-chargino (-neutralino) in a basis
of the slepton weak-eigenstate and the chargino (neutralino) 
mass-eigenstate.
By writing the interactions in this basis
we can get transparent view in the discussion 
of the multimass insertion technique 
discussed in Appendices B and C.
The Lagrangian is
\begin{eqnarray}
   - {\cal L}_{\rm int}& = & \phantom{+}
     \tilde \nu_i^{\dagger}~  \overline{\tilde \chi^-_A}
     (C^{A(i)}_{LR} P_R+ C^{A(i)}_{LL} P_L)~e_i 
\nonumber \\
& &
 +   \tilde e_{Li}^{\dagger}~ \overline{\tilde \chi^0_A}  
     (N^{A(i)}_{LR} P_R +N^{A(i)}_{LL} P_L)~e_i  
\nonumber \\
& &
 +   \tilde e_{Ri}^{\dagger}~ \overline{\tilde \chi^0_A}  
     (N^{A(i)}_{RR} P_R +N^{A(i)}_{RL} P_L)~e_i
   +h.c.,
\end{eqnarray}
where the coefficients are
\begin{eqnarray}
 C^{A(i)}_{LL}& =& g_2(O_R)_{A1},
\nonumber \\
 C^{A(i)}_{LR}& = &- \frac{\sqrt{2} m_{e_i}}{v \cos\beta}(O_L)_{A2},
\nonumber \\
 N^{A(i)}_{LL}&=&  \frac{g_2}{\sqrt{2}} 
       [-(O_N)_{A2} - (O_N)_{A1} \tan \theta_W],
\nonumber \\
 N^{A(i)}_{RL}&=& 
         \frac{\sqrt{2} m_{e_i}}{v \cos\beta} (O_N)_{A3},
\nonumber \\
 N^{A(i)}_{LR} &=&  
         \frac{\sqrt{2} m_{e_i}}{v \cos\beta} (O_N)_{A3}, 
\nonumber \\
 N^{A(i)}_{RR} &=& 
         \sqrt{2} g_2(O_N)_{A1} \tan \theta_W .
\end{eqnarray}

\section{Multimass Insertion Technique}

Mass insertion technique is useful to understand lepton flavor violating
processes in the MSSM since only off-diagonal elements of the 
slepton mass matrices are sources of the flavor violation. 
In this appendix we
introduce the multimass insertion technique with the nondegenerate masses.
The multimass inserted diagrams may not be necessarily suppressed 
than the single-mass inserted ones when the
slepton mass matrix has several small flavor-violating elements. Also,
though in the previous mass insertion formulas the degeneracy of the 
slepton masses is sometimes assumed, such a degeneracy is not
necessarily maintained at low energy even if the universal scalar mass
hypothesis is assumed in the higher energy scale. Our rule to derive
the mass-inserted amplitudes is very simple. We can derive them by
taking the finite difference on amplitudes which have
no mass-insertion. In this section we derive general formulas of
$e^+_i \to e^+_j \gamma$ keeping in mind that we apply 
them to more realistic cases of $\mu^+\to e^+ \gamma$ and
$\tau^+\to \mu^+ \gamma$ in the next section.

We refer to the slepton mass matrices as $(\overline{m}^2_{\tilde{l}})$,
\begin{equation}
-{\cal L}= \sum_{\tilde{l}=\tilde{e},\tilde{\nu}}
 (\overline{m}^2_{\tilde{l}})_{ij} \tilde{l}_i^{\dagger} \tilde{l}_j
\end{equation}
where 
\begin{equation}
\begin{array}{cclc}
\tilde{e}_i &=&
(\tilde{e}_{L}, \tilde{\mu}_{L},\tilde{\tau}_{L},
 \tilde{e}_{R}, \tilde{\mu}_{R},\tilde{\tau}_{R})^\top &
( {\rm for} ~~ \tilde{l}_i=\tilde{e}_i ~~ (i=1-6)),   \\
\tilde{\nu}_i &=& 
(\tilde{\nu}_{e},\tilde{\nu}_{\mu},\tilde{\nu}_{\tau})^\top &
( {\rm for} ~~ \tilde{l}_i=\tilde{\nu}_i ~~(i=1-3)).  
\end{array}
\end{equation}
The explicit forms of these matrices are given in Appendix~\ref{mssm}. Here 
we assume that all the off-diagonal elements of the slepton mass
matrices, both 
flavor violating and conserving ones, are much smaller than the diagonal 
elements, 
($(\overline{m}^2_{\tilde{l}})_{ii} \gg (\overline{m}^2_{\tilde{l}})_{jk}$ 
($j\ne k$)). 

In the mass-insertion technique the internal slepton lines are 
classified to two types at one loop level. 
First type is a slepton line on which the momentum of slepton
is not changed as
\begin{center} 
\begin{picture}(120,125)(185,80)
\LongArrow(130,100)(160,100)
\Text(145,80)[b]{$k$}
\DashLine(120,150)(250,150){5}
\DashLine(255,150)(300,150){2}
\DashLine(305,150)(350,150){5}
\Text(125,170)[t]{$\tilde{l}_1$}
\Vertex(150,150){3}  
\Text(150,120)[b]{$(\overline{m}_{\tilde{l}}^2)_{l_1 l_2}$}  
\Text(175,170)[t]{$\tilde{l}_2$}
\Vertex(200,150){3}  
\Text(200,120)[b]{$(\overline{m}_{\tilde{l}}^2)_{l_2 l_3}$}
\Text(225,170)[t]{$\tilde{l}_3$}
\Vertex(250,150){3}  
\Text(250,120)[b]{$(\overline{m}_{\tilde{l}}^2)_{l_3 l_4}$}
\Vertex(300,150){3}  
\Text(310,120)[b]{$(\overline{m}_{\tilde{l}}^2)_{l_{N-1} l_{N}}$}
\Text(325,170)[t]{$\tilde{l}_N$}
\end{picture}
\end{center}
\begin{equation}
=
\frac1{k^2-\overline{m}_{\tilde{l}_1}^2}(\overline{m}_{\tilde{l}}^2)_{l_1 l_2} 
\frac1{k^2-\overline{m}_{\tilde{l}_2}^2}(\overline{m}_{\tilde{l}}^2)_{l_2 l_3}
\frac1{k^2-\overline{m}_{\tilde{l}_3}^2}(\overline{m}_{\tilde{l}}^2)_{l_3 l_4}
                          \cdots (\overline{m}_{\tilde{l}}^2)_{l_{N-1} l_{N}} 
\frac1{k^2-\overline{m}_{\tilde{l}_N}^2},
\end{equation}
where we call the diagonal components of the slepton mass matrices, 
$(\overline{m}^2_{\tilde{l}})_{ii}$, as $\overline{m}^2_{\tilde{l}_i}$.   
We would like to consider the $e^+_i \to e^+_j \gamma$ process, and so
in the above figure we mean that an anti-slepton is going from left to
right with a momentum $k$. 
The product of propagators in above equation is referred to as $F_N$,
\begin{eqnarray}
F_N
(\overline{m}_{\tilde{l}_1}^2,\overline{m}_{\tilde{l}_2}^2,\cdots,\overline{m}_{\tilde{l}_N}^2) 
&\equiv& 
\frac1{k^2-\overline{m}_{\tilde{l}_1}^2}
\frac1{k^2-\overline{m}_{\tilde{l}_2}^2}
  \cdots 
\frac1{k^2-\overline{m}_{\tilde{l}_N}^2}. 
\end{eqnarray}
The second type is a slepton line where the momentum is changed,
by an emission of a photon with an outgoing momentum $p$, as
\begin{center} 
\begin{picture}(120,125)(185,80)
\LongArrow(130,100)(160,100)
\Text(145,80)[b]{$p+k$}
\LongArrow(310,100)(340,100)
\Text(325,80)[b]{$k$}
\DashLine(120,150)(250,150){5}
\DashLine(255,150)(300,150){2}
\DashLine(305,150)(350,150){5}
\Text(125,170)[t]{$\tilde{l}_1$}
\Vertex(150,150){3}  
\Text(150,120)[b]{$(\overline{m}_{\tilde{l}}^2)_{l_1 l_2}$}  
\Text(175,170)[t]{$\tilde{l}_2$}
\Vertex(200,150){3}  
\Text(200,120)[b]{$(\overline{m}_{\tilde{l}}^2)_{l_2 l_3}$}
\Text(225,170)[t]{$\tilde{l}_3$}
\Vertex(250,150){3}  
\Text(250,120)[b]{$(\overline{m}_{\tilde{l}}^2)_{l_3 l_4}$}
\Vertex(300,150){3}  
\Text(310,120)[b]{$(\overline{m}_{\tilde{l}}^2)_{l_{N-1} l_{N}}$}
\Text(325,170)[t]{$\tilde{l}_N$}
\Photon(270,180)(300,210){4}{4}
\end{picture}
\end{center}
\begin{eqnarray}
&=&
\frac1{(k+p)^2-\overline{m}_{l_1}^2} 
\frac1{k^2-\overline{m}_{l_1}^2}(\overline{m}_{\tilde{l}}^2)_{l_1 l_2} 
\frac1{k^2-\overline{m}_{l_2}^2} (\overline{m}_{\tilde{l}}^2)_{l_2 l_3}
                   \cdots (\overline{m}_{\tilde{l}}^2)_{l_{N-1} l_{N}}
\frac1{k^2-\overline{m}_{l_N}^2}
\nonumber\\
&+& 
\frac1{(k+p)^2-\overline{m}_{l_1}^2} (\overline{m}_{\tilde{l}}^2)_{l_1 l_2}
\frac1{(k+p)^2-\overline{m}_{l_2}^2} 
\frac1{k^2-\overline{m}_{l_2}^2} (\overline{m}_{\tilde{l}}^2)_{l_2 l_3}
                    \cdots(\overline{m}_{\tilde{l}}^2)_{l_{N-1} l_{N}} 
\frac1{k^2-\overline{m}_{l_N}^2}
\nonumber\\
&+& \cdots 
\nonumber\\
&+& 
\frac1{(k+p)^2-\overline{m}_{l_1}^2}(\overline{m}_{\tilde{l}}^2)_{l_1 l_2} 
\frac1{(k+p)^2-\overline{m}_{l_2}^2} (\overline{m}_{\tilde{l}}^2)_{l_2 l_3}
   \cdots(\overline{m}_{\tilde{l}}^2)_{l_{N-1} l_{N}} 
\frac1{(k+p)^2-\overline{m}_{l_N}^2} \frac1{k^2-\overline{m}_{l_N}^2}.
\nonumber\\
\end{eqnarray}
The sum of product of propagators in this equation is referred to as $G_N$,
\begin{eqnarray}
G_N
(\overline{m}_{\tilde{l}_1}^2,\overline{m}_{\tilde{l}_2}^2,\cdots,\overline{m}_{\tilde{l}_N}^2) 
&\equiv& 
\frac1{(k+p)^2-\overline{m}_{\tilde{l}_1}^2} 
\frac1{k^2-\overline{m}_{\tilde{l}_1}^2} 
\frac1{k^2-\overline{m}_{\tilde{l}_2}^2}
\cdots 
\frac1{k^2-\overline{m}_{\tilde{l}_N}^2}
\nonumber\\
&+& \frac1{(k+p)^2-\overline{m}_{\tilde{l}_1}^2} 
\frac1{(k+p)^2-\overline{m}_{\tilde{l}_2}^2} 
\frac1{k^2-\overline{m}_{\tilde{l}_2}^2} 
\cdots 
\frac1{k^2-\overline{m}_{\tilde{l}_N}^2}
\nonumber\\
&+& 
\cdots 
\nonumber\\
&+&
\frac1{(k+p)^2-\overline{m}_{\tilde{l}_1}^2} 
\frac1{(k+p)^2-\overline{m}_{\tilde{l}_2}^2} 
\cdots 
\frac1{(k+p)^2-\overline{m}_{\tilde{l}_N}^2} 
\frac1{k^2-\overline{m}_{\tilde{l}_N}^2}.
\nonumber\\
\end{eqnarray}
These functions $F_N$ and $G_N$ can be given as a finite difference of 
$F_{N-1}$ and $G_{N-1}$,
\begin{eqnarray}
F_N(\overline{m}_{\tilde{l}_1}^2,\overline{m}_{\tilde{l}_2}^2,\cdots,\overline{m}_{\tilde{l}_N}^2) 
&=& 
D[F_{N-1}(\overline{m}_{\tilde{l}}^2,\overline{m}_{\tilde{l}_3}^2,\cdots,\overline{m}_{\tilde{l}_N}^2);\overline{m}_{\tilde{l}}^2]
(\overline{m}_{\tilde{l}_1}^2,\overline{m}_{\tilde{l}_2}^2),
\label{df}
\\
G_N(\overline{m}_{\tilde{l}_1}^2,\overline{m}_{\tilde{l}_2}^2,\cdots,\overline{m}_{\tilde{l}_N}^2) 
&=& 
D[G_{N-1}(\overline{m}_{\tilde{l}}^2,\overline{m}_{\tilde{l}_3}^2,\cdots,\overline{m}_{\tilde{l}_N}^2);\overline{m}_{\tilde{l}}^2]
(\overline{m}_{\tilde{l}_1}^2,\overline{m}_{\tilde{l}_2}^2),
\label{dg}
\end{eqnarray}
where
\begin{equation}
D[f(x);x](x_1,x_2)
\equiv
\frac{1}{x_1-x_2}\left[f(x_1)-f(x_2)\right].
\end{equation}
Then, by taking finite difference in sequence, each scalar line can be 
represented as a linear combination of flavor conserving scalar lines, 
$F_1$ or $G_1$,
\begin{eqnarray}
F_{N}(\overline{m}_{\tilde{l}_1}^2,\overline{m}_{\tilde{l}_2}^2,\cdots,\overline{m}_{\tilde{l}_N}^2)
&=& 
D^{N-1}[F_1(m^2);m^2](\overline{m}_{\tilde{l}_1}^2,\overline{m}_{\tilde{l}_2}^2,\cdots,\overline{m}_{\tilde{l}_N}^2),
\nonumber\\
G_{N}(\overline{m}_{\tilde{l}_1}^2,\overline{m}_{\tilde{l}_2}^2,\cdots,\overline{m}_{\tilde{l}_N}^2)
&=&
D^{N-1}[G_1(m^2);m^2](\overline{m}_{\tilde{l}_1}^2,\overline{m}_{\tilde{l}_2}^2,\cdots,\overline{m}_{\tilde{l}_N}^2),
\end{eqnarray}
where
\begin{eqnarray}
D^{N-1}[f(x);x](x_1,x_2,\cdots,x_N)  
&\equiv&
\sum^{N}_{i=1} \left(\prod_{j \ne i} \frac{1}{x_i -x_j}\right) f(x_i).
\end{eqnarray}
Due to this fact, we can get amplitudes with any masses inserted from the 
corresponding flavor-conserving diagrams. 

Before we derive amplitudes of $e_i^+ \to e_j^+ \gamma$ $(i> j)$, 
we introduce a mass function $I_{N}^{M}$ as
\begin{eqnarray}
i I_N^M(m_1^2,m_2^2,\cdots,m_N^2) 
&=& 
\int \frac{d^4 k}{(\pi)^2}
k^{2M} \prod_i^N \frac{1}{k^2-m_i^2}.
\end{eqnarray}
This function $I_N^M$ can be reduced to  $I_{N-1}^M$ or $I_{N}^{M-1}$ 
by following rules,
\begin{eqnarray}
I_N^M(m_1^2,m_2^2,\cdots,m_N^2) 
&=& 
D[I_{N-1}^M(m^2,m_3^2,\cdots,m_N^2);m^2](m_1^2,m_2^2),
\label{rule1}
\end{eqnarray}
\begin{eqnarray}
I_N^M(m_1^2,m_2^2,\cdots,m_N^2) &=& m_N^2 I_N^{M-1}(m_1^2,m_2^2,\cdots,m_N^2) 
+ I_{N-1}^{M-1} (m_1^2,m_2^2,\cdots,m_{N-1}^2),
%
\label{rule2}
\end{eqnarray}
and then, all $I_N^M$ for $N>2+M$ can be derived from $I_1^0$,
\begin{eqnarray}
I^0_1(m^2)
&=& 
 \left(1-\log\frac{m^2}{\Lambda^2}\right) m^2,
\end{eqnarray}
where $\Lambda$ is a renormalization point. 
Also, the signs of $I_N^M$ and $D^L[I_N^M]$ are definitely determined as
\begin{eqnarray}
&(-1)^{N+M} I_N^M (m_1^2,m_2^2,\cdots,m_N^2) > 0,&
\label{sign_i}
\\
&(-1)^{L+M+N} D^L[I_N^M(
\underbrace{m^2,m^2,\cdots,m^2}_{N-P},m^2_{1},\cdots,m^2_P);m^2]
(M^2_1,M^2_2,\cdots,M^2_{L+1})>0.&
%
\label{sign_di}
\end{eqnarray}
Then, we can discuss about relative signs between several diagrams definitely
by using $I_N^M$ and the finite differences. Furthermore, since the mass dimension 
of $I_N^M$ is $(4+2M-2N)$, we can derive a following relation,
\begin{eqnarray}
(2+M-N)I_N^M(m_1^2,m_2^2,\cdots,m_N^2) 
&=&\sum^N_{i=1} m_i^2 I_{N+1}^M(m_1^2,\cdots,m_i^2,m_i^2,\cdots,m_N^2) 
%
\label{rule3}
\end{eqnarray}
for $N>2+M$, since
\begin{eqnarray}
\frac{d}{d x} \left\{x^{-2-M+N} I_N^M(x m_1^2,x m_2^2,\cdots,x m_N^2)\right\} &=&  0.
\end{eqnarray}
Due to Eqs.~(\ref{rule1},\ref{rule2},\ref{rule3}) there are some ways to 
represent one function, for example,
\begin{eqnarray}
I_5^2(m^2,m^2,m^2,M^2,M^2)&=& - 3 m^2 I_5^1(m^2,m^2,m^2,m^2,M^2).
\end{eqnarray} 
    
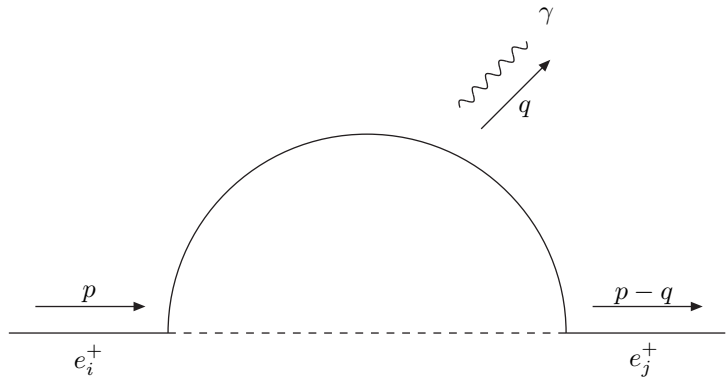
\begin{figure}

\begin{center} 
\begin{picture}(150,150)(0,0)

\CArc(75,25)(75,0,180)

\Text(-30,40)[]{$p$}
\LongArrow(-50,35)(-10,35)
\Text(-30,15)[]{$e^+_i$}

\Line(0,25)(-60,25)
\Line(150,25)(210,25)

\DashLine(0,25)(150,25){3}

\Text(180,15)[]{$e^+_j$}
\LongArrow(160,35)(200,35)
\Text(180,40)[]{$p-q$}

\Photon(110,110)(135,135){2}{5}
\LongArrow(118,102)(143,127)
\Text(143,145)[]{$\gamma$}
\Text(135,110)[]{$q$}

\end{picture} 

\caption{Assignment of the momenta to the external leptons 
and the external photon in a lepton flavor violating diagram. 
An anti-charged lepton $e^+_i$ going into
the left vertex with momentum $p$ is annihilated there,
and a photon with an outgoing momentum $q$ and an anti-charged lepton
$e^+_j$ with an outgoing momentum $p-q$ are emitted.}

\label{fig:pqassigndiag}
\end{center}
\end{figure}

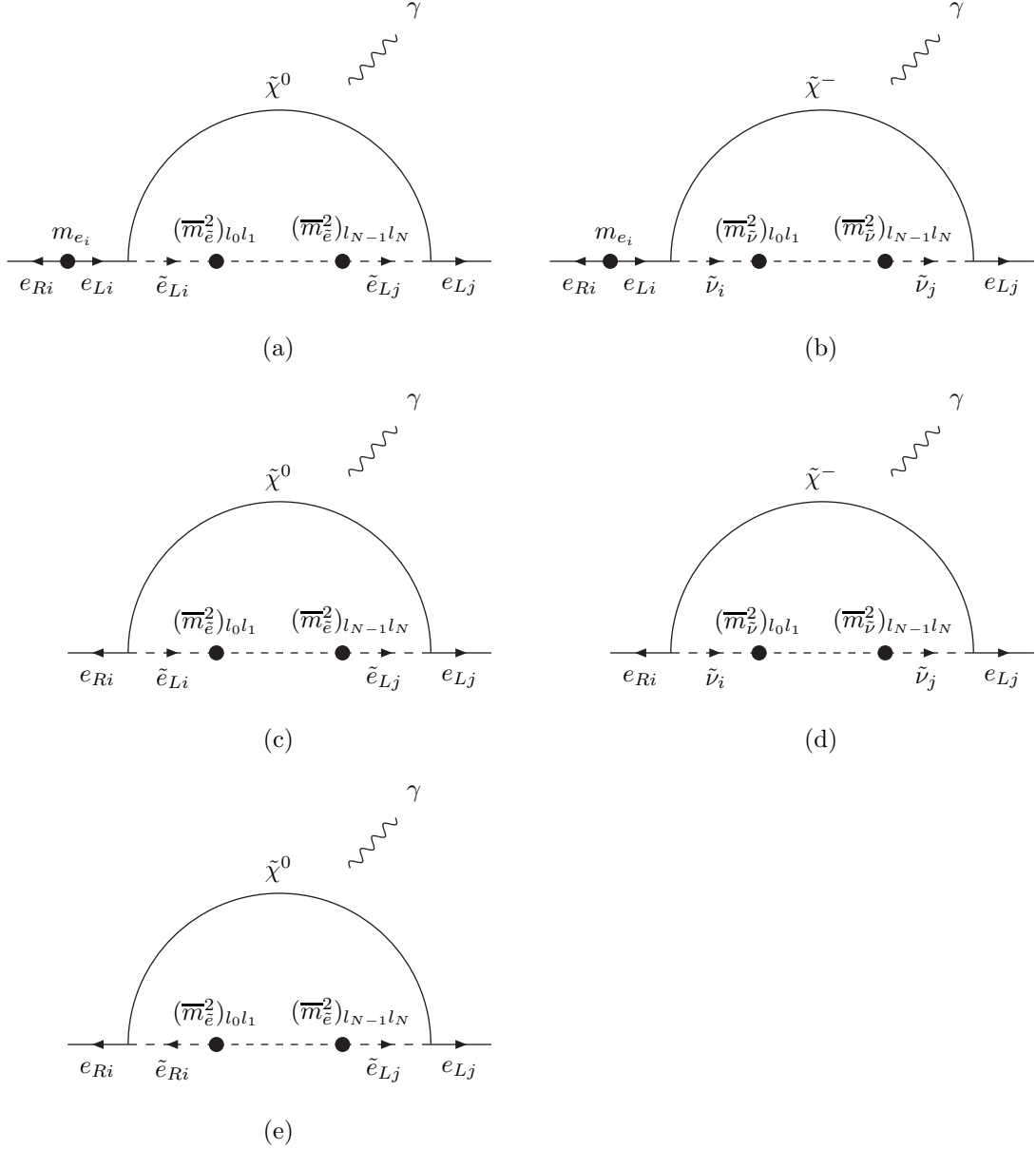
\begin{figure}
\begin{center} 
\begin{picture}(432,155)(0,-20)
\CArc(108,25)(60,0,180)
\Text(108,95)[]{$\tilde{\chi}^0$}

\ArrowLine(24,25)(0,25)                      \Vertex(24,25){3}
\ArrowLine(24,25)(48,25)
\DashArrowLine(48,25)(83,25){3}              \Vertex(83,25){3}
\DashLine(83,25)(133,25){3}                  \Vertex(133,25){3}
\DashArrowLine(133,25)(168,25){3}        
\ArrowLine(168,25)(192,25)

\Text(12,15)[]{$e_{Ri}$}
\Text(26,35)[]{$m_{e_i}$}
\Text(36,15)[]{$e_{Li}$}
\Text(66,15)[]{$\tilde{e}_{Li}$}
\Text(150,15)[]{$\tilde{e}_{Lj}$}
\Text(180,15)[]{$e_{Lj}$}

\Text(83,38)[]{$(\overline{m}^2_{\tilde{e}})_{l_0 l_1}$}
\Text(137,38)[]{$(\overline{m}^2_{\tilde{e}})_{l_{N-1} l_N}$}

\Photon(136,95)(154,115){2}{4}
\Text(162,125)[]{$\gamma$}

\Text(108,-10)[]{(a)}

\SetOffset(216,0)
\CArc(108,25)(60,0,180)
\Text(108,95)[]{$\tilde{\chi}^-$}

\ArrowLine(24,25)(0,25)                      \Vertex(24,25){3}
\ArrowLine(24,25)(48,25)
\DashArrowLine(48,25)(83,25){3}              \Vertex(83,25){3}
\DashLine(83,25)(133,25){3}                  \Vertex(133,25){3}
\DashArrowLine(133,25)(168,25){3}        
\ArrowLine(168,25)(192,25)

\Text(12,15)[]{$e_{Ri}$}
\Text(26,35)[]{$m_{e_i}$}
\Text(36,15)[]{$e_{Li}$}
\Text(66,15)[]{$\tilde{\nu}_{i}$}
\Text(150,15)[]{$\tilde{\nu}_j$}
\Text(180,15)[]{$e_{Lj}$}

\Text(83,38)[]{$(\overline{m}^2_{\tilde{\nu}})_{l_0 l_1}$}
\Text(137,38)[]{$(\overline{m}^2_{\tilde{\nu}})_{l_{N-1} l_N}$}

\Photon(136,95)(154,115){2}{4}
\Text(162,125)[]{$\gamma$}

\Text(108,-10)[]{(b)}

\end{picture} 

\begin{picture}(432,155)(0,-20)
\CArc(108,25)(60,0,180)
\Text(108,95)[]{$\tilde{\chi}^0$}

\ArrowLine(48,25)(24,25)
\DashArrowLine(48,25)(83,25){3}             \Vertex(83,25){3}
\DashLine(83,25)(133,25){3}                 \Vertex(133,25){3}
\DashArrowLine(133,25)(168,25){3}        
\ArrowLine(168,25)(192,25)

\Text(83,38)[]{$(\overline{m}^2_{\tilde{e}})_{l_0 l_1}$}
\Text(137,38)[]{$(\overline{m}^2_{\tilde{e}})_{l_{N-1} l_N}$}

\Text(36,15)[]{$e_{Ri}$}
\Text(66,15)[]{$\tilde{e}_{Li}$}
\Text(150,15)[]{$\tilde{e}_{Lj}$}
\Text(180,15)[]{$e_{Lj}$}

\Photon(136,95)(154,115){2}{4}
\Text(162,125)[]{$\gamma$}

\Text(108,-10)[]{(c)}

\SetOffset(216,0)
\CArc(108,25)(60,0,180)
\Text(108,95)[]{$\tilde{\chi}^-$}

\ArrowLine(48,25)(24,25)
\DashArrowLine(48,25)(83,25){3}             \Vertex(83,25){3}
\DashLine(83,25)(133,25){3}                 \Vertex(133,25){3}
\DashArrowLine(133,25)(168,25){3}        
\ArrowLine(168,25)(192,25)

\Text(36,15)[]{$e_{Ri}$}
\Text(66,15)[]{$\tilde{\nu}_{i}$}
\Text(150,15)[]{$\tilde{\nu}_j$}
\Text(180,15)[]{$e_{Lj}$}

\Text(83,38)[]{$(\overline{m}^2_{\tilde{\nu}})_{l_0 l_1}$}
\Text(137,38)[]{$(\overline{m}^2_{\tilde{\nu}})_{l_{N-1} l_N}$}

\Photon(136,95)(154,115){2}{4}
\Text(162,125)[]{$\gamma$}

\Text(108,-10)[]{(d)}

\end{picture} 

\begin{picture}(432,155)(0,-20)
\CArc(108,25)(60,0,180)
\Text(108,95)[]{$\tilde{\chi}^0$}

\ArrowLine(48,25)(24,25)
\DashArrowLine(83,25)(48,25){3}             \Vertex(83,25){3}
\DashLine(83,25)(133,25){3}                 \Vertex(133,25){3}
\DashArrowLine(133,25)(168,25){3}
\ArrowLine(168,25)(192,25)

\Text(36,15)[]{$e_{Ri}$}
\Text(66,15)[]{$\tilde{e}_{Ri}$}
\Text(150,15)[]{$\tilde{e}_{Lj}$}
\Text(180,15)[]{$e_{Lj}$}

\Text(83,38)[]{$(\overline{m}^2_{\tilde{e}})_{l_0 l_1}$}
\Text(137,38)[]{$(\overline{m}^2_{\tilde{e}})_{l_{N-1} l_N}$}

\Photon(136,95)(154,115){2}{4}
\Text(162,125)[]{$\gamma$}

\Text(108,-10)[]{(e)}

\SetOffset(216,0)

\end{picture} 

\caption{Patterns of the chirality flips in the lepton flavor 
violating diagrams in the $e^+_i \to e^+_j \gamma$ decay $(i > j)$ .
In the diagrams (a) and (b) the lepton chirality is flipped on the external
lines, while in (c) and (d) it is flipped at a vertex of 
lepton-slepton-neutralino (-chargino).
In (e) it flips on the internal slepton line.
Chirality flip on the internal slepton line does not occur in the
diagram with a virtual chargino because of the absence of the right-handed
sneutrino at the low energy region.}

\label{fig:massinsdiag}

\end{center}
\end{figure}

We will derive amplitudes of $e_i^+ \to e_j^+ \gamma$ 
$(i > j)$ by the mass-insertion technique.  The amplitude is generally 
written as
\begin{eqnarray}
T=e~ \epsilon^{\alpha*}~ \bar{v}_i (p)~ 
i \sigma_{\alpha \beta} q^\beta ~(A_L^{(ij)} P_L + A_R^{(ij)} P_R)~
v_j(p-q).
\end{eqnarray}
Here, $e$ is the electric charge, $\epsilon^*$ the photon polarization
vector, $v_i$ and $v_j$ the wave functions for the external leptons.
Assignment of momenta of the external fields is shown in 
Fig.~(\ref{fig:pqassigndiag}). 
Since above term is violating the lepton chirality, it is convenient
to decompose $A_L^{(ij)}$ and $A_R^{(ij)}$ as
\begin{eqnarray}
A_{L}^{(ij)}
&=&
 A^{(ij)}_{L}|_{c1}+A^{(ij)}_{L}|_{c2}
+A^{(ij)}_{L}|_{n1}+A^{(ij)}_{L}|_{n2}+A^{(ij)}_{L}|_{n3}, 
\nonumber \\
A_{R}^{(ij)}
&=&
 A^{(ij)}_{R}|_{c1}+A^{(ij)}_{R}|_{c2}
+A^{(ij)}_{R}|_{n1}+A^{(ij)}_{R}|_{n2}+A^{(ij)}_{R}|_{n3}.
\end{eqnarray}
$A^{(ij)}_{L, R}|_{n1,c1}$ come from diagrams in which the lepton 
chirality is flipped on the external lines 
(Figs.~(\ref{fig:massinsdiag})(a) and (b)). 
$A^{(ij)}_{L, R}|_{n2,c2}$ are contributions
from the diagrams in which the chirality is flipped on a vertex 
of lepton-slepton-neutralino (-chargino) 
[Figs.~(\ref{fig:massinsdiag})(c) and (d)]. 
$A^{(ij)}_{L, R}|_{n3}$ are contributions from those with 
the chirality flip on the internal slepton line
(Fig.~(\ref{fig:massinsdiag})(e)).
Subscripts $c$ and $n$ represent that they are from chargino and  neutralino 
diagrams, respectively.

The neutralino contribution to $A^{(ij)}_L|_{n1}$, 
derived from a diagram where 
the chirality of lepton is flipped in the external lepton line, is given by
\begin{eqnarray}
A_L^{(ij)} |_{n1}
&=&
-\frac{1}{6(4\pi)^2}~ m_{e_i}~
N^{A(i)*}_{LL} N^{A(j)}_{LL}~ 
\sum_{N=1}^{\infty}~ \sum_{l_1,\cdots, l_{N-1}}
(\overline{m}_{\tilde{e}}^2)_{l_0 l_1}
(\overline{m}_{\tilde{e}}^2)_{l_1 l_2}
\cdots
(\overline{m}_{\tilde{e}}^2)_{l_{N-1} l_N}
\nonumber\\
& &
 \times D^{N}[
I_5^2(\overline{m}^2_{\tilde{e}},\overline{m}^2_{\tilde{e}},
    \overline{m}^2_{\tilde{e}},M^2_{\tilde{\chi}^0_A},
    M^2_{\tilde{\chi}^0_A}),\overline{m}^2_{\tilde{e}}]
(\overline{m}^2_{\tilde{e}_{l_0}},\overline{m}^2_{\tilde{e}_{l_1}},\cdots,
\overline{m}^2_{\tilde{e}_{l_N}}) 
\nonumber\\
&&~~~~~~~~~~~~~~~~~~(l_0=i,l_N=j),
\nonumber\\
A_R^{(ij)}|_{n1} &=& 
\left( A_L^{(ij)}|_{n1} \right)_{L\leftrightarrow R,l_0=i+3,l_N=j+3},
\label{a_n1}
\end{eqnarray}
where $m_{e_i}$ is the $i$-th generation charged lepton mass, 
and the explicit form of $I_5^1$ is 
\begin{equation}
I_5^2(\overline{m}^2_{\tilde{e}},\overline{m}^2_{\tilde{e}},
      \overline{m}^2_{\tilde{e}},M^2_{\tilde{\chi}^0},
      M^2_{\tilde{\chi}^0})
= -
  \frac1{\overline{m}^2_{\tilde{e}}} \frac1{2(1-x)^4}
   (1-6x +3 x^2 + 2 x^3 - 6 x^2 \log x)
\end{equation}
with $x=M^2_{\tilde{\chi}^0}/\overline{m}^2_{\tilde{e}}$. 
$N^{A(i)}_{LL}$ 
is a coupling constant between slepton and neutralino, and our definition 
of couplings of slepton to neutralino (or chargino) is 
\begin{eqnarray}
   - {\cal L}_{\rm int}& = & ~~
     \tilde \nu_i^{\dagger}~ \overline{\tilde \chi^-_A}
     (C^{A(i)}_{LR} P_R+ C^{A(i)}_{LL} P_L)~e_i 
\nonumber \\
& &
 +   \tilde e_{Li}^{\dagger}~ \overline{\tilde \chi^0_A}  
     (N^{A(i)}_{LR} P_R +N^{A(i)}_{LL} P_L)~e_i  
\nonumber \\
& &
 +   \tilde e_{Ri}^{\dagger}~ \overline{\tilde \chi^0_A}  
     (N^{A(i)}_{RR} P_R +N^{A(i)}_{RL} P_L)~e_i
   +h.c..
\end{eqnarray}
Here, $C^{A(i)}_{LL}$, $N^{A(i)}_{LL}$, and $N^{A(i)}_{RR}$ correspond to 
the (lepton-chirality conserving) gaugino interactions, and  $C^{A(i)}_{LR}$, 
$N^{A(i)}_{RL}$, and $N^{A(i)}_{LR}$ are the (lepton-chirality violating) 
Higgsino interactions. These coupling constants are represented 
by the MSSM parameters in Appendix~\ref{mssm}. 

Sign of the term with $N$ off-diagonal inserted-masses in 
$A_L^{(ij)}|_{n1}$ is 
definitely determined by signs of coupling constants of slepton to neutralino 
and the inserted masses, and is independent of the 
diagonal slepton masses and neutralino masses since 
\begin{eqnarray}
(-1)^{N+1} D^{N}[
I_5^2(\overline{m}^2_{\tilde{e}},\overline{m}^2_{\tilde{e}},
      \overline{m}^2_{\tilde{e}},M^2_{\tilde{\chi}^0}
    M^2_{\tilde{\chi}^0}),\overline{m}^2_{\tilde{e}}]
(\overline{m}^2_{\tilde{e}_0},\overline{m}^2_{\tilde{e}_1},\cdots,
\overline{m}^2_{\tilde{e}_N}) &>& 0.
\label{sign1}
\end{eqnarray}

A contribution from a neutralino diagram where the lepton chirality
is flipped in the vertex of slepton-lepton-neutralino is given as
\begin{eqnarray}
A_L^{(ij)} |_{n2}
&=&
- \frac{1}{2(4\pi)^2}~M_{\tilde{\chi}^0_A}~
 N^{A(i)*}_{LR} N^{A(j)}_{LL}~
\sum_{N=1}^{\infty}~ \sum_{l_1,\cdots, l_{N-1}}
(\overline{m}_{\tilde{e}}^2)_{l_0 l_1}
(\overline{m}_{\tilde{e}}^2)_{l_1 l_2}
\cdots
(\overline{m}_{\tilde{e}}^2)_{l_{N-1} l_N}
\nonumber\\
& & \times
D^{N}[
I_4^1(\overline{m}^2_{\tilde{e}},\overline{m}^2_{\tilde{e}},
      M^2_{\tilde{\chi}^0_A}, M^2_{\tilde{\chi}^0_A});
      \overline{m}^2_{\tilde{e}}]
(\overline{m}^2_{\tilde{e}_{l_0}},\overline{m}^2_{\tilde{e}_{l_1}},\cdots,
\overline{m}^2_{\tilde{e}_{l_N}}) 
\nonumber\\
&&~~~~~~~~~~~~~~~~~~(l_0=i,l_N=j),
\nonumber\\
A_R^{(ij)}|_{n2} &=& 
\left( A_L^{(ij)}|_{n2} \right)_{L\leftrightarrow R,l_0=i+3,l_N=j+3}.
\label{a_n2}
\end{eqnarray}
The explicit form of $I_4^1$ in Eq.~(\ref{a_n2}) is 
\begin{equation}
I_4^1(\overline{m}^2_{\tilde{e}},\overline{m}^2_{\tilde{e}},
      M^2_{\tilde{\chi}^0}, M^2_{\tilde{\chi}^0})
= -  \frac1{\overline{m}^2_{\tilde{e}}} \frac1{(1-x)^3}
   (1-x^2+2 x \log x )
\end{equation}
with $x=M^2_{\tilde{\chi}^0}/\overline{m}^2_{\tilde{e}}$, and 
the sign of $D^{N}[I_4^1]$ is $(-1)^{N+1}$ from Eq.~(\ref{sign_di}).

A contribution from a neutralino diagram where the lepton chirality
is flipped in the internal slepton line is given as
\begin{eqnarray}
A_L^{(ij)} |_{n3}
&=&
- \frac{1}{2(4\pi)^2}~M_{\tilde{\chi}^0_A}~
 N^{A(i)*}_{RR} N^{A(j)}_{LL}~
\sum_{N=1}^{\infty}~ \sum_{l_1,\cdots, l_{N-1}}
(\overline{m}_{\tilde{e}}^2)_{l_0 l_1}
(\overline{m}_{\tilde{e}}^2)_{l_1 l_2}
\cdots
(\overline{m}_{\tilde{e}}^2)_{l_{N-1} l_N}
\nonumber\\
& & \times
D^{N}[
I_4^1(\overline{m}^2_{\tilde{e}},\overline{m}^2_{\tilde{e}},
      M^2_{\tilde{\chi}^0_A} ,M^2_{\tilde{\chi}^0_A});
      \overline{m}^2_{\tilde{e}}]
(\overline{m}^2_{\tilde{e}_{l_0}},\overline{m}^2_{\tilde{e}_{l_1}},\cdots,
\overline{m}^2_{\tilde{e}_{l_N}}) 
\nonumber\\
&&~~~~~~~~~~~~~~~~~~(l_0=i+3,l_N=j),
\nonumber\\
A_R^{(ij)}|_{n3} &=& 
\left(A_L^{(ij)}|_{n3} \right)_{L\leftrightarrow R,l_0=i,l_N=j+3 } .
\label{a_n3}
\end{eqnarray}

A contribution from a chargino diagram where the lepton chirality
is flipped in the external lepton line is given as
\begin{eqnarray}
A_L^{(ij)} |_{c1}
&=&
 \frac{1}{6(4\pi)^2}~ 
  m_{e_i} ~C^{A(i)*}_{LL} C^{A(j)}_{LL}~
\sum_{N=1}^{\infty}~ \sum_{l_1,\cdots, l_{N-1}}
(\overline{m}_{\tilde{\nu}}^2)_{l_0 l_1}
(\overline{m}_{\tilde{\nu}}^2)_{l_1 l_2}
\cdots
(\overline{m}_{\tilde{\nu}}^2)_{l_{N-1} l_N}
\nonumber\\
& & \times
D^{N}[
I_5^2(\overline{m}^2_{\tilde{\nu}},\overline{m}^2_{\tilde{\nu}},
      M^2_{\tilde{\chi}^{-}_A},M^2_{\tilde{\chi}^{-}_A}
      ,M^2_{\tilde{\chi}^{-}_A});\overline{m}^2_{\tilde{\nu}}]
(\overline{m}^2_{\tilde{\nu}_{l_0}},\overline{m}^2_{\tilde{\nu}_{l_1}},\cdots,
\overline{m}^2_{\tilde{\nu}_{l_N}}) 
\nonumber\\
&&~~~~~~(l_0=i,l_N=j),
\nonumber\\
A_R^{(ij)}|_{c1} &=& {\cal O}(m_{e_j}) .
\label{a_c1}
\end{eqnarray}
The explicit form of $I_5^2$ in Eq.~(\ref{a_c1}) is 
\begin{equation}
I_5^2(\overline{m}^2_{\tilde{\nu}},\overline{m}^2_{\tilde{\nu}},
      M^2_{\tilde{\chi}^{-}},M^2_{\tilde{\chi}^{-}},
      M^2_{\tilde{\chi}^{-}})
=  -\frac1{\overline{m}^2_{\tilde{\nu}}} \frac1{2(1-x)^4}
   (2+3x -6x^2+x^3+6x\log x)
\end{equation}
with $x=M^2_{\tilde{\chi}^{-}}/\overline{m}^2_{\tilde{\nu}}$, and 
the sign of $D^{N}[I_5^2]$ is $(-1)^{N+1}$. 

A contribution from a chargino diagram where the lepton chirality
is flipped in the vertex of lepton-sneutrino-chargino is given as
\begin{eqnarray}
A_L^{(ij)} |_{c2}
&=&
  \frac1{(4\pi)^2} M_{\tilde{\chi}^{-}_A}
 C^{A(i)*}_{LR} C^{A(j)}_{LL}~ 
\sum_{N=1}^{\infty}~ \sum_{l_1,\cdots, l_{N-1}}
(\overline{m}_{\tilde{\nu}}^2)_{l_0 l_1}
(\overline{m}_{\tilde{\nu}}^2)_{l_1 l_2}
\cdots
(\overline{m}_{\tilde{\nu}}^2)_{l_{N-1} l_N}
\nonumber\\
& & \times
D^{N}[
I_4^1(\overline{m}^2_{\tilde{\nu}},M^2_{\tilde{\chi}^{-}_A},
      M^2_{\tilde{\chi}^{-}_A},M^2_{\tilde{\chi}^{-}_A});
      \overline{m}^2_{\tilde{\nu}}]
(\overline{m}^2_{\tilde{\nu}_{l_0}},\overline{m}^2_{\tilde{\nu}_{l_1}},\cdots,
\overline{m}^2_{\tilde{\nu}_{l_N}}) 
\nonumber\\
&&~~~~~~(l_0=i,l_N=j),
\nonumber\\
A_R^{(ij)}|_{c2} &=& {\cal O}(m_{e_j}) .
\label{a_c2}
\end{eqnarray}
The explicit form of $I_4^1$ in Eq.~(\ref{a_c2}) is 
\begin{equation}
I_4^1(\overline{m}^2_{\tilde{\nu}},M^2_{\tilde{\chi}^{-}},
      M^2_{\tilde{\chi}^{-}},M^2_{\tilde{\chi}^{-}})
=  \frac1{\overline{m}^2_{\tilde{\nu}}} \frac1{2(1-x)^3}
   (3-4x+x^2+2 \log x)
\end{equation}
with $x=M^2_{\tilde{\chi}^{-}}/\overline{m}^2_{\tilde{\nu}}$, and 
the sign of $D^{N}[I_4^1]$ is $(-1)^{N+1}$.

\section{Application of Mass Insertion Formulas to 
$\tau^+ \to \mu^+\gamma$ and $\mu^+ \to e^+\gamma$}
\label{section:applimassins}
In this section we apply the formulas derived in the previous section to 
$\tau^+ \to \mu^+ \gamma$ and $\mu^+ \to e^+ \gamma$ processes. 
We neglect $m_\mu$ ($m_e$) in the calculation of $\tau^+\to\mu^+\gamma$ 
($\mu^+\to e^+\gamma$). 
First we consider $\tau^+\to\mu^+\gamma$. 
The whole contribution to $\tau^+\to\mu^+\gamma$ can be written as
\begin{eqnarray}
A_L^{(\tau\mu)} &=&  A_L^{(\tau\mu)}|_{c1} +A_L^{(\tau\mu)}|_{c2}
                   + A_L^{(\tau\mu)}|_{n1} +A_L^{(\tau\mu)}|_{n2}
                   + A_L^{(\tau\mu)}|_{n3}                       , \nonumber \\
A_R^{(\tau\mu)} &=&  A_R^{(\tau\mu)}|_{c1} +A_R^{(\tau\mu)}|_{c2}
                   + A_R^{(\tau\mu)}|_{n1} +A_R^{(\tau\mu)}|_{n2} 
                   + A_R^{(\tau\mu)}|_{n3}                       ,
\end{eqnarray}
in the same notation as the previous section.
In many models the dominant contributions to $\tau \to \mu \gamma$
are from the diagrams with single insertion of 
$(m_{\tilde L}^2)_{32}$, $(m_{\tilde e}^2)_{32}$, $A_e^{32}$, or 
$A_e^{23}$. 
Among these lepton flavor violating coupling constants
only $(m_{\tilde L}^2)_{32}$ and  $(m_{\tilde e}^2)_{32}$ are 
important for $\tan\beta\gsim 1$.
We here show explicit expressions of the diagrams 
with single off-diagonal slepton mass matrix element insertion
for $\tan\beta\gsim 1$.
From the formula we derived in the previous Appendix, 
the contributions from diagrams with the lepton 
chirality flipped in the external line 
(Figs.~(\ref{fig:massinsdiag})(a) and (b)) are given as 
\begin{eqnarray}
A_L^{(\tau\mu)}|_{c1} &=& - m_\tau \frac16 
                          \frac{\alpha_2}{4 \pi} (O_R)_{A1}^2    \nonumber\\
&& 
\times \left[  \frac{(m^2_{\tilde{L}})_{32}}
                    {m_{\tilde{\nu}_\mu}^2 - m_{\tilde{\nu}_\tau}^2}
      \right]~
      \left\{  \frac{1}{m^2_{\tilde{\nu}_\mu }} f_{c1} (x_{A \tilde{\nu}_\mu})
              -\frac{1}{m^2_{\tilde{\nu}_\tau}} f_{c1} (x_{A \tilde{\nu}_\tau})
      \right\}   ,
\label{eq:tmLc1}   \\
A_R^{(\tau\mu)}|_{c1} &=& {\cal O}(m_\mu) , \\
A_L^{(\tau\mu)}|_{n1}&=& ~ m_\tau \frac1{12}
                         \frac{\alpha_2}{4 \pi}~
                         \left((O_N)_{A2} + (O_N)_{A1} \tan\theta_W \right)^2  
\nonumber\\
& & 
\times \left[  \frac{(m^2_{\tilde{L}})_{32}}
                    {m_{\tilde{\mu}_L}^2 - m_{\tilde{\tau}_L}^2}
       \right]~
       \left\{   \frac{1}{m^2_{\tilde{\mu }_L}} f_{n1} (x_{A \tilde{\mu }_L})
                -\frac{1}{m^2_{\tilde{\tau}_L}} f_{n1} (x_{A \tilde{\tau}_L})
       \right\},
\label{eq:tmLn1}   \\
A_R^{(\tau\mu)}|_{n1} &=& ~  m_\tau \frac13
                            \frac{\alpha_Y}{4 \pi}~
                            (O_N)_{A1}^2 
\nonumber\\
& & 
\times \left[  \frac{(m^2_{\tilde{e}})_{32}}
                    {m_{\tilde{\mu}_R}^2 - m_{\tilde{\tau}_R}^2}
       \right]~
       \left\{   \frac{1}{m^2_{\tilde{\mu }_R}} f_{n1} (x_{A \tilde{\mu }_R})
                -\frac{1}{m^2_{\tilde{\tau}_R}} f_{n1} (x_{A \tilde{\tau}_R})
       \right\},
\label{eq:tmRn1}   
\end{eqnarray}
where $x_{A \tilde{l}}=M^2_{\tilde{\chi}_A}/m^2_{\tilde{l}}$ with 
$l = \nu_\mu$, $\nu_\tau$, $\mu_{L}$, $\tau_{L}$, $\mu_{R}$, and 
$\tau_{R}$ and $\tilde{\chi}_A=\tilde{\chi}_A^-$ and $\tilde{\chi}_A^0$. 
The coefficient functions $f_{c1}(x)$ and $f_{n1}(x)$ are given as
\begin{eqnarray}
f_{c1}(x) &\equiv& \frac{1}{2(1-x)^4} (2 + 3x - 6 x^2 + x^3 + 6 x \log x),  \\
f_{n1}(x) &\equiv& \frac{1}{2(1-x)^4} (1 - 6x + 3 x^2 + 2 x^3 - 6 x^2 \log x),
\end{eqnarray}
which are positive definite and monotonically decreasing.  Following 
coefficient functions are also defined to be positive definite and 
monotonically decreasing.
 
The diagrams where the lepton chirality is flipped in the vertices 
give the following contributions,
\begin{eqnarray}
A_L^{(\tau\mu)}|_{c2} &=&  m_\tau
      \frac{\alpha_2}{4 \pi}~
      \frac{M_{\tilde{\chi}_A^-}}{\sqrt{2}m_W \cos\beta} (O_R)_{A1}(O_L)_{A2}  
\nonumber\\
&& 
\times \left[ \frac{(m^2_{\tilde{L}})_{32}}
                   {m_{\tilde{\nu}_\mu}^2 - m_{\tilde{\nu}_\tau}^2}
       \right]
       \left\{ \frac{1}{m^2_{\tilde{\nu}_\mu }} f_{c2}(x_{A \tilde{\nu}_\mu })
              -\frac{1}{m^2_{\tilde{\nu}_\tau}} f_{c2}(x_{A \tilde{\nu}_\tau})
       \right\},  
\label{eq:tmLc2}   \\
A_R^{(\tau\mu)}|_{c2} &=& {\cal O}(m_\mu) , \\
A_L^{(\tau\mu)}|_{n2} &=& -  m_\tau \frac14
                              \frac{\alpha_2}{4 \pi}
                \frac{M_{\tilde{\chi}_A^0}}{m_Z \cos\theta_W  \cos\beta} 
                (O_N)_{A3} 
                \left((O_N)_{A2} + (O_N)_{A1} \tan\theta_W \right)  
\nonumber\\
&& \times \left[ \frac{(m^2_{\tilde{L}})_{32}}
                      {m_{\tilde{\mu}_L}^2 - m_{\tilde{\tau}_L}^2}
          \right]
          \left\{ \frac{1}{m^2_{\tilde{\mu }_L}} f_{n2} (x_{A \tilde{\mu }_L})
                 -\frac{1}{m^2_{\tilde{\tau}_L}} f_{n2} (x_{A \tilde{\tau}_L})
          \right\}, 
\label{eq:tmLn2}   \\
A_R^{(\tau\mu)}|_{n2} &=&  m_\tau \frac12
                \frac{\alpha_Y}{4 \pi}
                \frac{M_{\tilde{\chi}_A^0}}{m_Z \sin\theta_W \cos\beta} 
                (O_N)_{A3} (O_N)_{A1}
\nonumber\\
& & \times \left[ \frac
                    {(m^2_{\tilde{e}})_{32}}
                    {m_{\tilde{\mu}_R}^2 - m_{\tilde{\tau}_R}^2}
           \right]
           \left\{ \frac{1}{m^2_{\tilde{\mu }_R}} f_{n2} (x_{A \tilde{\mu }_R})
                  -\frac{1}{m^2_{\tilde{\tau}_R}} f_{n2} (x_{A \tilde{\tau}_R})
           \right\}, 
\label{eq:tmRn2}   
\end{eqnarray}
where the functions $f_{c2}(x)$ and $f_{n2}(x)$ are defined as
\begin{eqnarray}
f_{c2}(x) &\equiv& - \frac1{2(1-x)^3} (3 - 4 x + x^2 + 2 \log x), \\
f_{n2}(x) &\equiv&   \frac1{ (1-x)^3} (1 - x^2 + 2 x \log x)   .
\end{eqnarray}

Finally the contribution from the diagrams in which the lepton chirality
is flipped on the internal slepton lines are
\begin{eqnarray}
A_L^{(\tau\mu)}|_{n3}&=& -\frac12 \frac{\alpha_2}{4\pi}
                              (O_N)_{A1} 
                        \left((O_N)_{A2} + (O_N)_{A1} \tan\theta_W \right)
                  \tan\theta_W  
                  (m^2_{LR})_{33} (m^2_{\tilde{L}})_{32}  
                  M_{\tilde{\chi}_A^0}  \nonumber \\     
&& \times \left\{ \phantom{+}
  \frac{1}{m^2_{\tilde{\tau}_R}} 
      \frac{1}{m^2_{\tilde{\tau}_R}-m^2_{\tilde{\tau}_L}}
      \frac{1}{m^2_{\tilde{\tau}_R}-m^2_{\tilde{\mu }_L}} 
                 f_{n2}(x_{A \tilde{\tau}_R})             \right. \nonumber \\
&& \phantom{\times \left\{ \right.}
+ \frac{1}{m^2_{\tilde{\tau}_L}} 
      \frac{1}{m^2_{\tilde{\tau}_L}-m^2_{\tilde{\tau}_R}}
      \frac{1}{m^2_{\tilde{\tau}_L}-m^2_{\tilde{\mu }_L}} 
                 f_{n2}(x_{A \tilde{\tau}_L})                     \nonumber \\
&& \phantom{\times \left\{ \right.} \left. 
+ \frac{1}{m^2_{\tilde{\mu}_L}} 
      \frac{1}{m^2_{\tilde{\mu}_L}-m^2_{\tilde{\tau}_R}}
      \frac{1}{m^2_{\tilde{\mu}_L}-m^2_{\tilde{\tau}_L}} 
                 f_{n2}(x_{A \tilde{\mu }_L})             \right\} , \\
A_R^{(\tau\mu)}|_{n3}&=& -\frac12 \frac{\alpha_2}{4\pi}
                              (O_N)_{A1} 
                        \left((O_N)_{A2} + (O_N)_{A1} \tan\theta_W \right)
                  \tan\theta_W  
                  (m^2_{LR})_{33} (m^2_{\tilde{e}})_{32} 
                 M_{\tilde{\chi}_A^0}  \nonumber \\     
&& \times \left\{ \phantom{+}
  \frac{1}{m^2_{\tilde{\tau}_L}} 
      \frac{1}{m^2_{\tilde{\tau}_L}-m^2_{\tilde{\tau}_R}}
      \frac{1}{m^2_{\tilde{\tau}_L}-m^2_{\tilde{\mu }_R}} 
                 f_{n2}(x_{A \tilde{\tau}_L})             \right. \nonumber \\
&& \phantom{\times \left\{ \right.}
+ \frac{1}{m^2_{\tilde{\tau}_R}} 
      \frac{1}{m^2_{\tilde{\tau}_R}-m^2_{\tilde{\tau}_L}}
      \frac{1}{m^2_{\tilde{\tau}_R}-m^2_{\tilde{\mu }_R}} 
                 f_{n2}(x_{A \tilde{\tau}_R})                     \nonumber \\
&& \phantom{\times \left\{ \right.} \left. 
+ \frac{1}{m^2_{\tilde{\mu}_R}} 
      \frac{1}{m^2_{\tilde{\mu}_R}-m^2_{\tilde{\tau}_L}}
      \frac{1}{m^2_{\tilde{\mu}_R}-m^2_{\tilde{\tau}_R}} 
                 f_{n2}(x_{A \tilde{\mu }_R})             \right\} .
\end{eqnarray} 

Next we present the mass insertion formula for $\mu^+ \to e^+\gamma$.
Here we assume mass degeneracy between selectron and smuon, then
\begin{eqnarray}
m_{\tilde{\nu}_e} &=& m_{\tilde{\nu}_\mu} ~\equiv~~ m_{\tilde{\nu}}, 
\nonumber\\
m_{\tilde{e}_L}   &=& m_{\tilde{\mu}_L} ~\equiv~~ m_{\tilde{l}_L}, 
\nonumber\\
m_{\tilde{e}_R}   &=& m_{\tilde{\mu}_R} ~\equiv~~ m_{\tilde{l}_R} .
\nonumber
\end{eqnarray}

The contributions from diagrams with the lepton chirality flipped 
in the external
line (Figs.~(\ref{fig:massinsdiag})(a) and (b)) are given as 
\begin{eqnarray}
A_L^{(\mu e)}|_{c1} &=& ~~~  m_\mu \frac16
                \frac{\alpha_2}{4 \pi} (O_R)^2_{A1}
\nonumber\\
&&
\times \left\{ 
       \left[ (m^2_{\tilde{L}})_{21}
         +   \frac{(m^2_{\tilde{L}})_{23}(m^2_{\tilde{L}})_{31}}
                  {m_{\tilde{\nu}}^2 - m_{\tilde{\nu}_\tau}^2}
       \right]
             \frac{1}{m^4_{\tilde{\nu}}} g_{c1} (x_{A \tilde{\nu}})
                 \right. 
\nonumber\\
&& ~~-~\left. 
       \left[  \frac{(m^2_{\tilde{L}})_{23}(m^2_{\tilde{L}})_{31}}
                    {(m_{\tilde{\nu}}^2 - m_{\tilde{\nu}_\tau}^2)^2}
       \right]   
                       \frac{1}{m^2_{\tilde{\nu}_\tau}} 
                       f_{c1}(x_{A \tilde{\nu}_\tau})
       \right\} ,                                           \\
A_R^{(\mu e)}|_{c1} &=& {\cal O}(m_e) ,  \\
A_L^{(\mu e)}|_{n1} &=&  -  m_\mu \frac{1}{12}
                \frac{\alpha_2}{4 \pi}
                \left( (O_N)_{A2} + (O_N)_{A1} \tan\theta_W \right)^2  
\nonumber\\
       & & \times \left\{     
       \left[    (m^2_{\tilde{L}})_{21}
               + \frac{(m^2_{\tilde{L}})_{23}(m^2_{\tilde{L}})_{31}}
                      {m_{\tilde{l}_L}^2 - m_{\tilde{\tau}_L}^2}
       \right]
                 \frac{1}{m^4_{\tilde{l}_L}} g_{n1} (x_{A \tilde{l}_L})
                 \right. 
\nonumber\\
&& ~~-~\left.  \left[ 
                 \frac
                    {(m^2_{\tilde{L}})_{23}(m^2_{\tilde{L}})_{31}}
                    {(m_{\tilde{l}_L}^2 - m_{\tilde{\tau}_L}^2)^2}
                 \right]   
                       \frac{1}{m^2_{\tilde{\tau}_L}} 
                       f_{n1} (x_{A \tilde{\tau}_L})
                  \right\}   ,                          \\
A_R^{(\mu e)}|_{n1} &=& ~- \frac13 m_\mu 
                \frac{\alpha_Y}{4 \pi}
                (O_N)_{A1}^2                   \nonumber\\
& & 
\times \left\{  \left[ 
                 (m^2_{\tilde{e}})_{21}
               + \frac{(m^2_{\tilde{e}})_{23}(m^2_{\tilde{e}})_{31}}
                      {m_{\tilde{l}_R}^2 - m_{\tilde{\tau}_R}^2}
                 \right]
~                                          
                 \frac{1}{m^4_{\tilde{l}_R}} g_{n1} (x_{A \tilde{l}_R})
                 \right. 
\nonumber\\
             & & ~~-~\left. 
                 \left[ 
                 \frac{(m^2_{\tilde{e}})_{23}(m^2_{\tilde{e}})_{31}}
                      {(m_{\tilde{l}_R}^2 - m_{\tilde{\tau}_R}^2)^2}
                 \right]   
                       \frac{1}{m^2_{\tilde{\tau}_R}} 
                       f_{n1} (x_{A \tilde{\tau}_R})
                  \right\},
\end{eqnarray}
with
\begin{eqnarray}
g_{c1}(x)  &\equiv& f_{c1}(x)  + x f^\prime_{c1}(x),   \nonumber\\
g_{n1}(x)  &\equiv& f_{n1}(x)  + x f^\prime_{n1}(x).
\end{eqnarray}
Here primes on $f_{c1,n1}(x)$ mean differentiation about $x$.

The diagrams where the lepton chirality is flipped in the vertices 
give the following contributions,
\begin{eqnarray}
A_L^{(\mu e)}|_{c2} &=& - m_\mu  \frac{\alpha_2}{4 \pi}
                 \frac{M_{\tilde{\chi}^-_A}}{\sqrt{2}m_W \cos\beta} 
                 (O_R)_{A1} (O_L)_{A2}                          \nonumber\\
&& 
\times \left\{  
       \left[   (m^2_{\tilde{L}})_{21}
             +   \frac{(m^2_{\tilde{L}})_{23}(m^2_{\tilde{L}})_{31}}
                      {m_{\tilde{\nu}}^2 - m_{\tilde{\nu}_\tau}^2}   
       \right]
               \frac{1}{m^4_{\tilde{\nu}}} g_{c2} (x_{A \tilde{\nu}})  \right. 
\nonumber\\
&& ~~-~\left.  \left[ 
             \frac{(m^2_{\tilde{L}})_{23}(m^2_{\tilde{L}})_{31}}
                  {(m_{\tilde{\nu}}^2 - m_{\tilde{\nu}_\tau}^2)^2}  \right]   
             \frac{1}{m^2_{\tilde{\nu}_\tau}} 
                   f_{c2} (x_{A \tilde{\nu}_\tau})   \right\} , \\
A_R^{(\mu e)}|_{c2} &=& {\cal O}(m_e) ,  \\
A_L^{(\mu e)}|_{n2} &=& ~~~m_\mu \frac14 \frac{\alpha_2}{4 \pi}
                \frac{M_{\tilde{\chi}^0_A}}{m_Z \cos\theta_W \cos\beta} 
                (O_N)_{A3} 
                \left((O_N)_{A2} + (O_N)_{A1} \tan\theta_W \right)  
\nonumber\\
&&
\times \left\{
       \left[   (m^2_{\tilde{L}})_{21}
          +     \frac{(m^2_{\tilde{L}})_{23}(m^2_{\tilde{L}})_{31}}
                     {m_{\tilde{l}_L}^2 - m_{\tilde{\tau}_L}^2}
       \right]
                \frac{1}{m^4_{\tilde{l}_L}} g_{n2}(x_{A \tilde{l}_L})   \right. 
\nonumber\\
&& ~~-~\left.     \left[ 
                 \frac{(m^2_{\tilde{L}})_{23}(m^2_{\tilde{L}})_{31}}
                      {(m_{\tilde{l}_L}^2 - m_{\tilde{\tau}_L}^2)^2}
                 \right]   
                      \frac{1}{m^2_{\tilde{\tau}_L}} 
                       f_{n2}(x_{A \tilde{\tau}_L})      \right\} ,   \\
A_R^{(\mu e)}|_{n2} &=& ~~-m_\mu \frac12 \frac{\alpha_Y}{4 \pi}
                \frac{M_{\tilde{\chi}^0_A}}{m_Z \sin\theta_W  \cos\beta} 
                (O_N)_{A3} (O_N)_{A1}
\nonumber\\
&& 
\times \left\{   \left[   (m^2_{\tilde{e}})_{21}
                 +  \frac{(m^2_{\tilde{e}})_{23}(m^2_{\tilde{e}})_{31}}
                         {m_{\tilde{l}_R}^2 - m_{\tilde{\tau}_R}^2}
                 \right]
                 \frac{1}{m^4_{\tilde{l}_R}} g_{n2}(x_{A \tilde{l}_R})
                 \right. 
\nonumber\\
&& - \left. \left[ \frac{(m^2_{\tilde{e}})_{23}(m^2_{\tilde{e}})_{31}}
                        {(m_{\tilde{l}_R}^2 - m_{\tilde{\tau}_R}^2)^2}  
            \right]   
                   \frac{1}{m^2_{\tilde{\tau}_R}} 
                     f_{n2}(x_{A \tilde{\tau}_R})
     \right\} .
\end{eqnarray}
Here the functions $g_{c2}(x)$ and $g_{n2}(x)$ are defined, similarly to
$g_{c1,n1}(x)$, as 
\begin{eqnarray}
g_{c2}(x) &\equiv & f_{c2}(x) + x f^\prime_{c2}(x), \nonumber \\ 
g_{n2}(x) &\equiv & f_{n2}(x) + x f^\prime_{n2}(x). 
\end{eqnarray}

The contributions from Fig.~(\ref{fig:massinsdiag})(e) 
in which the lepton chirality
is flipped in the slepton lines are following,
\begin{eqnarray}
A_L^{(\mu e)}|_{n3}&=& \frac12 \frac{\alpha_2}{4\pi} M_{\tilde{\chi}_A^0}
            (O_N)_{A1} ((O_N)_{A2} + (O_N)_{A1} \tan\theta_W) \tan\theta_W
\nonumber \\
&&
\times \left\{ -(m^2_{\tilde e})_{23}(m^2_{LR})_{33}(m^2_{\tilde L})_{31} 
  \phantom{\frac12} \right.
\nonumber \\
&&
\phantom{\times \left\{ \right. -}
\times \left( \phantom{+}
 \frac{1}{m^2_{\tilde{l}_R}} 
 \frac{1}{m^2_{\tilde{l}_R} - m^2_{\tilde{\tau}_R}} 
 \frac{1}{m^2_{\tilde{l}_R} - m^2_{\tilde{\tau}_L}} 
 \frac{1}{m^2_{\tilde{l}_R} - m^2_{\tilde{l   }_L}} f_{n2}(x_{A \tilde{l}_R})
\right.
\nonumber \\
&& 
\phantom{\times \left\{ \right. - \times \left( \right.}
+\frac{1}{m^2_{\tilde{\tau}_L}} 
 \frac{1}{m^2_{\tilde{\tau}_L} -m^2_{\tilde{l   }_R}} 
 \frac{1}{m^2_{\tilde{\tau}_L} -m^2_{\tilde{\tau}_R}} 
 \frac{1}{m^2_{\tilde{\tau}_L} -m^2_{\tilde{l}_L}} f_{n2}(x_{A \tilde{\tau}_L})
\nonumber \\
&&
\phantom{\times \left\{ \right. - \times \left( \right.}
+\frac{1}{m^2_{\tilde{\tau}_R}} 
 \frac{1}{m^2_{\tilde{\tau}_R} -m^2_{\tilde{l   }_R}} 
 \frac{1}{m^2_{\tilde{\tau}_R} -m^2_{\tilde{\tau}_L}} 
 \frac{1}{m^2_{\tilde{\tau}_R} -m^2_{\tilde{l}_L}} f_{n2}(x_{A \tilde{\tau}_R})
\nonumber \\
&& 
\phantom{\times \left\{ \right. - \times \left( \right.}
\left.
+\frac{1}{m^2_{\tilde{l}_L}} 
 \frac{1}{m^2_{\tilde{l}_L} - m^2_{\tilde{l   }_R}} 
 \frac{1}{m^2_{\tilde{l}_L} - m^2_{\tilde{\tau}_R}} 
 \frac{1}{m^2_{\tilde{l}_L} - m^2_{\tilde{\tau}_L}} f_{n2}(x_{A \tilde{l}_L})
\right)
\nonumber \\
&& \phantom{\times \left\{ \right. }
+(m^2_{LR})_{22}(m^2_{\tilde{L}})_{21} \nonumber \\
&& \phantom{\times \left\{ \right. +}
\times \left( -\frac{1}{ m^2_{\tilde{l}_R}}
               \frac{1}{(m^2_{\tilde{l}_R} - m^2_{\tilde{l}_L})^2} 
                      f_{n2}(x_{A \tilde{l}_R})
+              \frac{1}{ m^4_{\tilde{l}_L}}
               \frac{1}{m^2_{\tilde{l}_L} - m^2_{\tilde{l}_R}} 
                      g_{n2}(x_{A \tilde{l}_L})
       \right)                    \nonumber \\
&&  \phantom{\times \left\{ \right. }
+(m^2_{LR})_{22} (m^2_{\tilde{L}})_{23} (m^2_{\tilde{L}})_{31}
          \nonumber \\
&&  \phantom{\times \left\{ \right. +}
\times \left(
-\frac{1}{ m^2_{\tilde{l}_R}}
 \frac{1}{(m^2_{\tilde{l}_R} - m^2_{\tilde{l   }_L})^2} 
 \frac{1}{ m^2_{\tilde{l}_R} - m^2_{\tilde{\tau}_L}}     
       f_{n2}(x_{A \tilde{l}_R})     \right.  \nonumber \\
&& \phantom{\times \left\{ \right. + \times \left( \right.}
-\frac{1}{ m^2_{\tilde{\tau}_L}}
 \frac{1}{(m^2_{\tilde{\tau}_L} - m^2_{\tilde{l}_L})^2} 
 \frac{1}{ m^2_{\tilde{\tau}_L} - m^2_{\tilde{l}_R}}     
       f_{n2}(x_{A \tilde{\tau}_L})    \nonumber \\
&& \phantom{\times \left\{ \right. + \times \left( \right.} \left. \left.
+\frac{1}{ m^4_{\tilde{l}_L}}
 \frac{1}{ m^2_{\tilde{l}_L} - m^2_{\tilde{l   }_R}} 
 \frac{1}{ m^2_{\tilde{l}_L} - m^2_{\tilde{\tau}_L}}     
       g_{n2}(x_{A \tilde{l}_L})             \right)  \right\}  , \\
A_R^{(\mu e)}|_{n3}&=& \frac12 \frac{\alpha_2}{4\pi} M_{\tilde{\chi}_A^0}
          (O_N)_{A1} ((O_N)_{A2} + (O_N)_{A1} \tan\theta_W) \tan\theta_W
\nonumber \\
&&
\times \left\{ -(m^2_{\tilde L})_{23}(m^2_{LR})_{33}(m^2_{\tilde e})_{31} 
  \phantom{\frac12} \right.
\nonumber \\
&&
\phantom{\times \left\{ \right. -}
\times \left( \phantom{+}
 \frac{1}{m^2_{\tilde{l}_L}} 
 \frac{1}{m^2_{\tilde{l}_L} - m^2_{\tilde{\tau}_L}} 
 \frac{1}{m^2_{\tilde{l}_L} - m^2_{\tilde{\tau}_R}} 
 \frac{1}{m^2_{\tilde{l}_L} - m^2_{\tilde{l   }_R}} f_{n2}(x_{A \tilde{l}_L})
\right.
\nonumber \\
&& 
\phantom{\times \left\{ \right. - \times \left( \right.}
+\frac{1}{m^2_{\tilde{\tau}_R}} 
 \frac{1}{m^2_{\tilde{\tau}_R} -m^2_{\tilde{l   }_L}} 
 \frac{1}{m^2_{\tilde{\tau}_R} -m^2_{\tilde{\tau}_L}} 
 \frac{1}{m^2_{\tilde{\tau}_R} -m^2_{\tilde{l}_R}} f_{n2}(x_{A \tilde{\tau}_R})
\nonumber \\
&&
\phantom{\times \left\{ \right. - \times \left( \right.}
+\frac{1}{m^2_{\tilde{\tau}_L}} 
 \frac{1}{m^2_{\tilde{\tau}_L} -m^2_{\tilde{l   }_L}} 
 \frac{1}{m^2_{\tilde{\tau}_L} -m^2_{\tilde{\tau}_R}} 
 \frac{1}{m^2_{\tilde{\tau}_L} -m^2_{\tilde{l}_R}} f_{n2}(x_{A \tilde{\tau}_L})
\nonumber \\
&& 
\phantom{\times \left\{ \right. - \times \left( \right.}
\left.
+\frac{1}{m^2_{\tilde{l}_R}} 
 \frac{1}{m^2_{\tilde{l}_R} - m^2_{\tilde{l   }_L}} 
 \frac{1}{m^2_{\tilde{l}_R} - m^2_{\tilde{\tau}_L}} 
 \frac{1}{m^2_{\tilde{l}_R} - m^2_{\tilde{\tau}_R}} f_{n2}(x_{A \tilde{l}_R})
\right)
\nonumber \\
&& \phantom{\times \left\{ \right. }
+(m^2_{LR})_{22}(m^2_{\tilde{e}})_{21} \nonumber \\
&& \phantom{\times \left\{ \right. +}
\times \left( -\frac{1}{ m^2_{\tilde{l}_L}}
               \frac{1}{(m^2_{\tilde{l}_L} - m^2_{\tilde{l}_R})^2} 
                      f_{n2}(x_{A \tilde{l}_L})
+              \frac{1}{ m^4_{\tilde{l}_R}}
               \frac{1}{m^2_{\tilde{l}_R} - m^2_{\tilde{l}_L}} 
                      g_{n2}(x_{A \tilde{l}_R})
       \right)                    \nonumber \\
&&  \phantom{\times \left\{ \right. }
+(m^2_{LR})_{22} (m^2_{\tilde{e}})_{23} (m^2_{\tilde{e}})_{31}
          \nonumber \\
&&  \phantom{\times \left\{ \right. +}
\times \left(
-\frac{1}{ m^2_{\tilde{l}_L}}
 \frac{1}{(m^2_{\tilde{l}_L} - m^2_{\tilde{l   }_R})^2} 
 \frac{1}{ m^2_{\tilde{l}_L} - m^2_{\tilde{\tau}_R}}     
       f_{n2}(x_{A \tilde{l}_L})     \right.  \nonumber \\
&& \phantom{\times \left\{ \right. + \times \left( \right.}
-\frac{1}{ m^2_{\tilde{\tau}_R}}
 \frac{1}{(m^2_{\tilde{\tau}_R} - m^2_{\tilde{l}_R})^2} 
 \frac{1}{ m^2_{\tilde{\tau}_R} - m^2_{\tilde{l}_L}}     
       f_{n2}(x_{A \tilde{\tau}_R})    \nonumber \\
&& \phantom{\times \left\{ \right. + \times \left( \right.} \left. \left.
+\frac{1}{ m^4_{\tilde{l}_R}}
 \frac{1}{ m^2_{\tilde{l}_R} - m^2_{\tilde{l   }_L}} 
 \frac{1}{ m^2_{\tilde{l}_R} - m^2_{\tilde{\tau}_R}}     
       g_{n2}(x_{A \tilde{l}_R})             \right)  \right\}  .
\end{eqnarray}

\section{Renormalization Group Equations}
\label{section:RGE}
In this Appendix we show the one-loop level renormalization group 
equations (RGE's) relevant to our discussion in the text.
In the equations below we use a shorthand notation, that is,
for matrices in generation space $A$ and $B$ we define $\{A,B\}$ by
\begin{equation}
\{ A,B \}_{ij} \equiv \sum_k ( A_{ik}B_{kj}+ B_{ik}A_{kj} ) .
\end{equation}

\subsection{RGE's for the MSSMRN Model}
\label{RGEMSSMRN}
First we devote ourselves to the MSSMRN model.
The RGE's for the gauge coupling constants and the gaugino masses are 
unchanged from the MSSM since the right-handed neutrinos are singlet 
under the Standard Model gauge group.
First we list the RGE's of the Yukawa coupling constants.
\begin{eqnarray}
16\pi^2\mu\frac{d}{d\mu} f_{e_{ij}}&=& 
\left \{ -\frac{9}{5} g_1^2 - 3 g_2^2 
  + 3 {\rm Tr}(f_d f_d^{\dagger})
  +   {\rm Tr}(f_e f_e^{\dagger}) \right \} f_{e_{ij}}     \nonumber \\
&& 
+3(f_e f_e^{\dagger} f_e)_{ij} + (f_e f_{\nu}^{\dagger} f_{\nu})_{ij} , \\
16\pi^2\mu\frac{d}{d\mu} f_{\nu_{ij}}&=&
\left \{ -\frac{3}{5} g_1^2 - 3 g_2^2 
  + 3 {\rm Tr}(f_{u} f_{u}^{\dagger})
  +   {\rm Tr}(f_{\nu} f_{\nu}^{\dagger}) \right \} f_{\nu_{ij}}  \nonumber \\
&& 
+3(f_{\nu} f_{\nu}^{\dagger} f_{\nu})_{ij}
+ (f_{\nu} f_{e}^{\dagger} f_{e})_{ij} .
\end{eqnarray}
Next RGE's of the soft masses and A parameters.
\begin{eqnarray}
16\pi^2\mu\frac{d}{d\mu} (m^2_{\tilde{L}})_{ij}&=&   
-\left (\frac{6}{5}g_1^2 \left| M_1 \right|^2
                +6 g_2^2 \left| M_2 \right|^2 \right) \delta_{ij}
-\frac{3}{5} g_1^2 S \delta_{ij}                              \nonumber \\
&& 
+\{ m^2_{\tilde L} , f_e^{\dagger} f_e + f_{\nu}^{\dagger} f_{\nu} \}_{ij}
\nonumber \\
&&
+2 \left( f_e^{\dagger} m^2_{\tilde e} f_e
           +{\tilde m}^2_{h1}f_e^{\dagger}f_e
+A_e^{\dagger} A_e \right)_{ij}                    \nonumber \\
&&
+2 \left( f_{\nu}^{\dagger} m^2_{\tilde \nu} f_{\nu}
+{\tilde m}^2_{h2}f_{\nu}^{\dagger} f_{\nu}
+A_{\nu}^{\dagger} A_{\nu} \right)_{ij} ,              \\
16\pi^2\mu\frac{d}{d\mu} (m^2_{\tilde{e}})_{ij}&=& 
-\frac{24}{5} g_1^2 \left| M_1 \right|^2 \delta_{ij}
  +\frac{6}{5}g_1^2 S \delta_{ij}  \nonumber \\
&& 
+2 \{ m^2_{\tilde e} ,  f_e f_e^{\dagger} \}_{ij}   \nonumber \\
&&
+4 \left( f_e m^2_{\tilde L} f_e^{\dagger}
+{\tilde m}^2_{h1}f_e f_e^{\dagger}
+A_e A_e^{\dagger}\right)_{ij}           ,               \\
16\pi^2\mu\frac{d}{d\mu} (m^2_{\tilde{\nu}})_{ij}&=& 
\phantom{+} 2 \{ m^2_{\tilde \nu} , f_{\nu} f_{\nu}^{\dagger} \}_{ij} 
\nonumber \\
&&
+4 \left( f_{\nu} m^2_{\tilde L} f_{\nu}^{\dagger}
+{\tilde m}^2_{h2}f_{\nu} f_{\nu}^{\dagger}
+A_{\nu} A_{\nu}^{\dagger}\right)_{ij}    ,            \\
16\pi^2\mu\frac{d}{d\mu} A_{e_{ij}}&=&  
 \left\{ -\frac{9}{5} g_1^2 -3 g_2^2
+ 3 {\rm Tr}(f_d^{\dagger} f_d)
+   {\rm Tr}(f_e^{\dagger} f_e) \right \} A_{e_{ij}} \nonumber \\
&&
+2 \left\{
-\frac{9}{5} g_1^2 M_1 -3 g_2^2 M_2 
+ 3 {\rm Tr}(f_d^{\dagger} A_d)
+   {\rm Tr}(f_e^{\dagger} A_e) \right \} f_{e_{ij}} \nonumber \\
&&
+4 (f_e f_e^{\dagger} A_e)_{ij} 
+5 (A_e f_e^{\dagger} f_e)_{ij}                      \nonumber \\
&&
+2 (f_e f_{\nu}^{\dagger} A_{\nu})_{ij} 
+  (A_e f_{\nu}^{\dagger} f_{\nu})_{ij}    ,        \\
16\pi^2\mu\frac{d}{d\mu} A_{\nu_{ij}}&=&  
\left\{ -\frac{3}{5}g_1^2 -3g_2^2 
+3 {\rm Tr}(f_u^{\dagger} f_u)
+  {\rm Tr}(f_{\nu}^{\dagger} f_{\nu}) \right \} A_{\nu_{ij}} \nonumber \\
&&
+2 \left\{-\frac{3}{5} g_1^2 M_1 -3 g_2^2 M_2 
+ 3 {\rm Tr}(f_u^{\dagger} A_u)
+   {\rm Tr}(f_{\nu}^{\dagger} A_{\nu}) \right \} f_{\nu_{ij}} \nonumber \\
&&
+4(f_{\nu} f_{\nu}^{\dagger}A_{\nu})_{ij}
+5(A_{\nu}f_{\nu}^{\dagger} f_{\nu})_{ij} \nonumber \\
&&
+2(f_{\nu} f_e^{\dagger}A_e)_{ij}
+(A_{\nu} f_e^{\dagger} f_e)_{ij} , 
\end{eqnarray}
where
\begin{equation}
S={\rm Tr}(m_{\tilde Q}^2 + m_{\tilde d}^2 -2 m_{\tilde u}^2
-m_{\tilde L}^2 +m_{\tilde e}^2)
- {\tilde m}^2_{h1}+{\tilde m}^2_{h2}.
\end{equation}
Here $g_1$ is the U(1) gauge coupling constant in the GUT convention,
which is related to the U(1)$_Y$ gauge coupling constant $g_Y$ by
$g_Y^2=\frac35 g_1^2$.

\subsection{RGE's for the SU(5)RN Model}
Next we list the RGE's for the SU(5)RN model.
The superpotential of the matter sector is
\begin{eqnarray}
W&=&\frac14 f_{u_{ij}} \psi_i^{AB} \psi_j^{CD} H^E \epsilon_{ABCDE} 
  + \sqrt{2} f_{d_{ij}} \psi_i^{AB} \phi_{jA} \overline{H}_B  \nonumber  \\
 &&+ f_{\nu_{ij}} \eta_i \phi_{jA} H^A 
  + \frac12 M_{\eta_i \eta_j} \eta_i \eta_j , \nonumber 
\end{eqnarray}
and the soft SUSY breaking terms are
\begin{eqnarray}
-{\cal L}_{\rm SUSY~breaking} 
&=&
 (m_{\psi}^2)_{ij}  \tilde{\psi}_i^{\dagger} \tilde{\psi}_j
+(m_{\phi}^2)_{ij}  \tilde{\phi}_i^{\dagger} \tilde{\phi}_{j} 
+(m_{\eta}^2)_{ij}  \tilde{\eta}_i^{\dagger} \tilde{\eta}_{j} 
+m_{h}^2 h^\dagger h +  m_{\bar{h}}^2 \bar{h}^\dagger \bar{h} 
\nonumber\\
&&
+\left\{
 \frac14  A_{u_{ij}} \tilde{\psi}_i \tilde{\psi}_j h
+ \sqrt{2}A_{d_{ij}} \tilde{\psi}_i \tilde{\phi}_{j} \bar{h}
+ A_{\nu_{ij}} \tilde{\eta}_i \tilde{\phi}_{j} h + h.c.
\right\} 
\nonumber\\
&&
+\frac12M_5 \lambda_{5L} \lambda_{5L} + h.c. .
\end{eqnarray} 
We denote the SU(5)$_{\rm GUT}$ gauge coupling constant as $g_5$,
the SU(5)$_{\rm GUT}$ gaugino $\lambda_5$, and its soft Majorana mass $M_5$.
We neglect a coupling $\lambda H \Sigma \overline{H}$, 
as stated in the text.

First is the RGE's for the dimensionless coupling constants.
\begin{eqnarray}
16\pi^2\mu\frac{d}{d\mu} g_5 &=&  -3 g_5^3 , \\
16\pi^2\mu\frac{d}{d\mu} f_{d_{ij}} &=& 
\left[ -\frac{84}{5} g_5^2 + 4{\rm Tr}(f_d^\dagger f_d) \right] f_{d_{ij}}
\nonumber \\
&&
+ 6 (f_d f_d^\dagger f_d)_{ij} 
+ 3 (f_u f_u^\dagger f_d)_{ij}
+   (f_d f_\nu^\dagger f_\nu)_{ij}   , \\
16\pi^2\mu\frac{d}{d\mu} f_{u_{ij}} &=& 
\left[ -\frac{96}{5} g_5^2 
       + 3{\rm Tr}(f_u^\dagger f_u) 
       +  {\rm Tr}(f_\nu^\dagger f_\nu) 
\right] f_{u_{ij}} \nonumber \\
&&
+ 6 (f_u f_u^\dagger f_u)_{ij}
+ 2 (f_d f_d^\dagger f_u)_{ij} 
+ 2 (f_u f_d^\ast f_d^\top)_{ij} ,  \\
16\pi^2\mu\frac{d}{d\mu} f_{\nu_{ij}} &=&  
\left[ -\frac{48}{5} g_5^2 
       + 3{\rm Tr}(f_u^\dagger f_u) 
       +  {\rm Tr}(f_\nu^\dagger f_\nu) 
\right] f_{\nu_{ij}} \nonumber \\
&&
+ 6 (f_\nu f_\nu^\dagger f_\nu)_{ij} 
+ 4 (f_\nu f_d^\dagger f_d)_{ij}  .
\end{eqnarray}
Next is for the soft masses.
\begin{eqnarray}
16\pi^2\mu\frac{d}{d\mu} M_5 &=&  -6 g_5^2 M_5 , \\
16\pi^2\mu\frac{d}{d\mu} m^2_{h} &=&  
    -\frac{96}{5} g_5^2 M_5^2                    \nonumber \\
&&
+6{\rm Tr}(f_u f_u^\dagger) m^2_{h}               \nonumber \\ 
&&
+6{\rm Tr}(f_u m^2_\psi f_u^\dagger 
 +         f_u^\dagger m^{2\top}_\psi f_u
 +         A_u A_u^\dagger           )            \nonumber \\
&&
+2{\rm Tr}(f_\nu f_\nu^\dagger) m^2_{h}     \nonumber \\ 
&& 
+2{\rm Tr}(f_\nu m^2_\phi f_\nu^\dagger 
 +         f_\nu^\top m^{2}_\eta f_\nu^\ast
 +         A_\nu A_\nu^\dagger       )           ,     \\
16\pi^2\mu\frac{d}{d\mu} m^2_{\overline{h}} &=&  
-\frac{96}{5} g_5^2 M_5^2                                  \nonumber \\
&&
+8{\rm Tr}(f_d f_d^\dagger) m^2_{\overline{h}}             \nonumber \\ 
&&
+8{\rm Tr}(f_d m^2_\phi f_d^\dagger 
 +         f_d^\dagger m^{2\top}_\psi f_d
 +         A_d A_d^\dagger           )            ,   \\
16\pi^2\mu\frac{d}{d\mu} (m^2_{\psi})_{ij} &=&  
-\frac{144}{5} g_5^2 M_5^2 \delta_{ij}
+ \{ m^2_{\psi}, 2f_d^\ast f_d^\top + 3f_u^\ast f_u^\top \}_{ij} \nonumber \\
&& +4 \left[ (f_d^\ast f_d^\top)_{ij} m^2_{\overline{h}} 
      +      (f_d^\ast m^{2\top}_\phi f_d^\top 
             +A_d^\ast A_d^\top               )_{ij} \right] 
\nonumber \\
&& +6 \left[ (f_u^\ast f_u^\top)_{ij} m^2_h + 
      (f_u^\ast m_\psi^{2\top} f_u^\top + A_u^\ast A_u^\top )_{ij} \right] ,
 \\
16\pi^2\mu\frac{d}{d\mu} (m^2_{\phi})_{ij} &=&  
-\frac{96}{5} g_5^2 M_5^2 \delta_{ij}
+\{ m^2_{\phi}, 4f_d^\dagger f_d + f_{\nu}^\dagger f_{\nu} \}_{ij} 
\nonumber \\
&& +8 \left[ (f_d^\dagger f_d)_{ij} m^2_{\overline{h}} + 
         (f_d^\dagger m^{2\top}_\psi f_d + A_d^\dagger A_d )_{ij} \right] 
\nonumber \\
&& +2 \left[ (f_{\nu}^\dagger f_{\nu})_{ij} m^2_h + 
 (f_{\nu}^\dagger m_\eta^{2\top} f_{\nu} + A_{\nu}^\dagger A_{\nu})_{ij} 
      \right] ,  \\
16\pi^2\mu\frac{d}{d\mu} (m^2_{\eta})_{ij} &=&  
\phantom{+} 5 \{ m^2_{\eta}, f_\nu^\ast f_\nu^\top \}_{ij} \nonumber \\
&& +10 \left[ 
   (f_\nu^\ast f_\nu^\top)_{ij} m^2_{h}
 + (f_\nu^\ast m^{2\top}_\phi f_\nu^\top 
 +  A_\nu^\ast A_\nu^\top                 )_{ij}    \right] .
\end{eqnarray}
Finally A terms.
\begin{eqnarray}
16\pi^2\mu\frac{d}{d\mu} A_{d_{ij}} &=& 
\left[ -\frac{84}{5} g_5^2 + 4{\rm Tr}(f_d^\dagger f_d) \right] A_{d_{ij}}
\nonumber \\
&&
+2 \left[ -\frac{84}{5} g_5^2 M_5 + 4{\rm Tr}(f_d^\dagger A_d) \right] 
f_{d_{ij}} \nonumber \\
&&
+ 10 (f_d f_d^\dagger A_d)_{ij} 
+ 3 (f_u f_u^\dagger A_d)_{ij} \nonumber \\
&&
+ 8 (A_d f_d^\dagger f_d)_{ij} 
+ 6 (A_u f_u^\dagger f_d)_{ij} \nonumber \\
&&
+   (A_d f_\nu^\dagger f_\nu)_{ij} 
+ 2 (f_d f_\nu^\dagger A_\nu)_{ij}  , \\
16\pi^2\mu\frac{d}{d\mu} A_{u_{ij}} &=& 
\left[ -\frac{96}{5} g_5^2 
       + 3{\rm Tr}(f_u^\dagger f_u) 
       +  {\rm Tr}(f_\nu^\dagger f_\nu) 
\right] A_{u_{ij}} \nonumber \\
&&
+2\left[ -\frac{96}{5} g_5^2 M_5
       + 3{\rm Tr}(f_u^\dagger A_u) 
       +  {\rm Tr}(f_\nu^\dagger A_\nu) 
\right] f_{u_{ij}} \nonumber \\
&&
+ 2( f_d f_d^\dagger A_u)_{ij}   + 9(f_u f_u^\dagger A_u)_{ij}
+ 2( A_u f_d^\ast f_d^\top)_{ij} 
 \nonumber \\ 
&&
+ 4 (A_d f_d^\dagger f_u)_{ij} + 9( A_u f_u^\dagger f_u)_{ij} 
+ 4 (f_u f_d^\ast A_d^\top)_{ij} , \\
16\pi^2\mu\frac{d}{d\mu} A_{\nu_{ij}} &=&  
\left[ -\frac{48}{5} g_5^2 
       + 3{\rm Tr}(f_u^\dagger f_u) 
       +  {\rm Tr}(f_\nu^\dagger f_\nu) 
\right] A_{\nu_{ij}} \nonumber \\
&&
+2\left[ -\frac{48}{5} g_5^2 M_5
       + 3{\rm Tr}(f_u^\dagger A_u) 
       +  {\rm Tr}(f_\nu^\dagger A_\nu) 
\right] f_{\nu_{ij}} \nonumber \\
&&
+ 7 (f_\nu f_\nu^\dagger A_\nu)_{ij} 
+ 4 (A_\nu f_d^\dagger f_d)_{ij}        \nonumber \\
&&
+11 (A_\nu f_\nu^\dagger f_\nu)_{ij} 
+ 8 (f_\nu f_d^\dagger A_d)_{ij} .
\end{eqnarray}

\newpage
\newcommand{\Journal}[4]{{\it #1} {\bf #2} {(#3)} {#4}}
\newcommand{\APJ}{Ap. J.}
\newcommand{\CJP}{Can. J. Phys.}
\newcommand{\NC}{Nuovo Cimento}
\newcommand{\NP}{Nucl. Phys.}
\newcommand{\PL}{Phys. Lett.}
\newcommand{\PR}{Phys. Rev.}
\newcommand{\PRep}{Phys. Rep.}
\newcommand{\PRL}{Phys. Rev. Lett.}
\newcommand{\PTP}{Prog. Theor. Phys.}
\newcommand{\RMP}{Rev. Mod. Phys.}
\newcommand{\SJNP}{Sov. J. Nucl. Phys.}
\newcommand{\ZP}{Z. Phys.}




\newpage


\end{document}